\documentclass[preprint, showpacs]{revtex4}
\usepackage{graphicx}
\usepackage[bf,SL,BF]{subfigure}
\usepackage{psfrag}
\usepackage{color}
\usepackage{amssymb}
\usepackage{amsmath}
\usepackage{epstopdf}
\usepackage{bbold}

\newcommand{\be}{\begin{eqnarray}}
\newcommand{\ee}{\end{eqnarray}}

\newcommand\tab[1][1cm]{\hspace*{#1}}
\newcommand{\bfq}{{\bf q}_{\perp}}

\newcommand{\bfk}{{\bf k}_{\perp}}

\newcommand{\bfb}{{\bf b}_{\perp}}

\newcommand{\bfp}{{\bf p}_{\perp}}

\newcommand{\bfS}{{\bf S}_{\perp}}

\begin{document}
\title{Transverse structure of proton in a light-front quark-diquark model}
\author{ Tanmay Maji}
\author{Dipankar Chakrabarti }
\affiliation{ Department of Physics, 
Indian Institute of Technology Kanpur,
Kanpur 208016, India}
\date{\today}

\begin{abstract}
We present all the leading twist T-even TMDs in the light-front quark diquark model(LFQDM) and study the relations among them. The model contains both the scalar and vector diquark with the light front wave functions modeled from the soft-wall AdS/QCD prediction.  The $x-p^2_\perp$ factorization used in phenomenological extraction for TMDs is observed in this model. We present the results for the quark densities and the transverse shape of proton. The shape of the transversely polarized proton is shown to be non-spherical for nonzero transverse momentum. The scale evolution of  both integrated and unintegrated  TMDs  are also presented in this paper.


\end{abstract}
\pacs{14.20.Dh, 12.39.-x,12.38.Aw}
\maketitle
\section{Introduction\label{intro}}
Transverse momentum dependent parton distributions(TMDs)  encode three dimensional structure as well as angular momentum information and hence have attracted lot of attentions in recent time to unravel the three dimensional structure of the proton.  Being nonperturbative in nature, the  TMDs are very difficult to be calculated in full QCD. So, they have been  studied in different QCD inspired models to understand the spin and three dimensional structure of the proton in parton level. 
The TMDs (see \cite{collins} and references therein) are required to describe the  Semi-Inclusive Deep Inelastic Scattering(SIDIS) or Drell-Yan processes where a final state particle  with transverse momentum is observed. The collinear picture of DIS  cannot explain the single or double spin asymmetries in SIDIS or Drell-Yan processes.  
At leading twist, there are eight TMDs, three of them  $f_1(x,p_\perp),~g_{1L}(x,p_\perp),~h_1(x,p_\perp)$ are generalization of the three PDFs and when integrated over transverse momentum, reduce to PDFs, namely,  the unpolarized distribution $f_1(x)$, helicity distribution function $g_{1}(x)$ and the transversity distribution $h_1(x)$. 
 TMDs are  also rich in information about the spin-orbit correlations at the parton level.
 
The TMDs provide interesting insights into the proton structure. From the TMDs, one can extract the quark densities for different proton and quark polarization. 
 In the recent time, the transversity TMD $h_1(x,\bfp)$ has drawn a lot of attention for its contribution to the Collins asymmetry in the leading order QCD parton models\cite{Ans07,Ans11,Ans13}.  Phenomenological restrictions suggest that the transversity distribution should be positive for $u$ and negative for $d$ quarks.  When integrated over $x,\bfp$, it reduces to the tensor charge.
The distribution $g_{1T}(x,\bfp)$ encodes the information of longitudinally polarized quark in a transversely polarized proton. $\bfp^2$ weighted moment of $g_{1T}(x,\bfp)$ contributes to the double spin asymmetry $A_{LT}$ \cite{DSA}. The distribution corresponding to the transverse quark in a longitudinal proton $h^\perp_{1L}(x,\bfp)$ is found to be negative for $u$ and positive for $d$ quarks in some model calculations\cite{Pasquini08}. The transverse moment of $h^\perp_{1L}(x,\bfp)$ can be connected with the higher twist TMDs using the Wandzura-Wilczek-type approximation on the basis of available data from HERMESS\cite{Ava08}. The pretzelosity TMD, $h^\perp_{1T}(x,\bfp)$ contributes to the single spin asymmetry(SSA) $A^{\sin(3\phi_h-\phi_S)}_{UT}$\cite{Proku15,Ava08,Ans11}. It is also related to the orbital angular momentum(OAM) of quarks\cite{She09,bag,Efr10}. The non-vanishing $h^\perp_{1T}(x,\bfp)$ indicates that the polarized proton is not spherically symmetric\cite{Miller07}. Most of the models predict a negative distribution for $u$ quarks and a positive distributions for $d$ quark\cite{bag,Jakob97} whereas model extraction from experimental data shows opposite behavior with large error corridor\cite{Proku15}. Some models calculation shows that the difference between helicity and transversity distributions is related to pretzelosity distributions\cite{Ava08}. 

TMDs have been investigated in several QCD inspired models, e.g., in a diquark spectator model\cite{Jakob97, Bacc08}, in MIT bag model\cite{bag}, in a covariant  parton model\cite{parton}. The power counting rule of $h^\perp_{1T}(x,\bfp)$ compared with unpolarized distribution, for large $x$ regime,  is discussed in\cite{Bur08}. TMDs satisfy different relations with PDFs and GPDs. These relations are model dependent and it is not guaranteed that they should hold in QCD. A model independent derivation of the relations is not yet possible. Nevertheless, from phenomenological point of view, these relations may provide additional constraints on model predictions.   The model dependent relations among TMDs and GPDs have been investigated in Ref.\cite{meissner, MC_rel} and the relations with PDFs have been investigated in Ref.\cite{efremov}. 

In this work, we study the T-even TMDs in a light front quark-diquark model\cite{TM_VD} where the wave functions are constructed from the AdS/QCD prediction.  The TMDs in our model satisfy certain inequalities, specially, the unpolarized, helicity and transversity TMDs satisfy a Soffer bound type inequality.  In many phenomenological models, the unpolarized TMD is modeled as the unpolarized PDF with a Gaussian transverse momentum dependence. In our model, this $x-p_\perp^2$  factorization is not apparent, but interestingly, numerical analysis support the phenomenological assumption.
 The model is defined at an initial scale $\mu_0=0.8$ GeV and the TMDs at the energy scales accessible to different experiments are evaluated using the evolution scheme proposed in\cite{Aybat11,Anselmino12}.  
 
 In Sec.\ref{model}, we introduce the quark-diquark model of the proton. In Sec.\ref{TMDS} the TMDs are defined and the results for the TMDs in our model are given in Sec.\ref{results}.    The TMD inequality relations are discussed in Sec.\ref{rel} and in Sec.\ref{densities} we present the results for quark densities. The TMD evolution and results for integrated TMDs are  presented in Sec.\ref{evol} and \ref{integrated}. The distortion in  the transverse shape of the proton due to the pretzelosity TMD  is  discussed in Sec.\ref{proton_shape}. Finally, a brief conclusion and summary is presented in Sec.\ref{concl}.
 A discussion and the values of the  parameters in the model are given in  Appendix \ref{AppA} and some details of the quark correlator calculations are given in Appendix \ref{AppB}.

\section{light-front quark diquark model for nucleon\label{model}}
Here, we consider the light-front quark-diquark model proposed in \cite{TM_VD}.  In this model, the proton is written 
as a sum of isoscalar-scalar diquark singlet $|u~ S^0\rangle$, isoscalar-vector diquark $|u~ A^0\rangle$ and isovector-vector diquark $|d~ A^1\rangle$ states\cite{Jakob97,Bacc08} having a spin-flavor $SU(4)$ structure 
\be 
|P; \pm\rangle = C_S|u~ S^0\rangle^\pm + C_V|u~ A^0\rangle^\pm + C_{VV}|d~ A^1\rangle^\pm. \label{PS_state}
\ee
Where $S$ and $A$ represent the scalar and vector diquark and their superscripts  represent the isospin of that diquark.  

We use the light-cone convention $x^\pm=x^0 \pm x^3$ and choose a frame where the transverse momentum of proton vanishes i,e. $P \equiv \big(P^+,\frac{M^2}{P^+},\textbf{0}_\perp\big)$. Where the momentum of struck quark and diquark are   
$p\equiv (xP^+, \frac{p^2+|\bfp|^2}{xP^+},\bfp)$
 and  $P_X\equiv ((1-x)P^+,P^-_X,-\bfp)$ respectively. The longitudinal momentum fraction carried by the struck quark is denoted by $x=p^+/P^+$. The two particle Fock-state expansion for $J^z =\pm1/2$ for scalar diquark  is given by
\be
|u~ S\rangle^\pm & =& \int \frac{dx~ d^2\bfp}{2(2\pi)^3\sqrt{x(1-x)}} \bigg[ \psi^{\pm(u)}_{+}(x,\bfp)|+\frac{1}{2}~s; xP^+,\bfp\rangle \nonumber \\
 &+& \psi^{\pm(u)}_{-}(x,\bfp)|-\frac{1}{2}~s; xP^+,\bfp\rangle\bigg],\label{fock_SD}
\ee
and the light front wave functions for scalar diquark are given by\cite{Lepa80}
\be 
\psi^{+(u)}_+(x,\bfp)&=& N_S~ \varphi^{(u)}_{1}(x,\bfp),\nonumber \\
\psi^{+(u)}_-(x,\bfp)&=& N_S\bigg(- \frac{p^1+ip^2}{xM} \bigg)\varphi^{(u)}_{2}(x,\bfp),\label{LFWF_S}\\
\psi^{-(u)}_+(x,\bfp)&=& N_S \bigg(\frac{p^1-ip^2}{xM}\bigg) \varphi^{(u)}_{2}(x,\bfp),\nonumber \\
\psi^{-(u)}_-(x,\bfp)&=&  N_S~ \varphi^{(u)}_{1}(x,\bfp),\nonumber
\ee
where $|\lambda_q~\lambda_S; xP^+,\bfp\rangle$ represents the two particle state having struck quark of helicity $\lambda_q$ and a scalar diquark having helicity $\lambda_S=s$(spin-0 singlet diquark helicity is denoted by s to distinguish from triplet diquark). Similarly the two particle fock-state expansion for vector diquark is given as \cite{Ellis08}
\be
|\nu~ A \rangle^\pm & =& \int \frac{dx~ d^2\bfp}{2(2\pi)^3\sqrt{x(1-x)}} \bigg[ \psi^{\pm(\nu)}_{++}(x,\bfp)|+\frac{1}{2}~+1; xP^+,\bfp\rangle \nonumber\\
 &+& \psi^{\pm(\nu)}_{-+}(x,\bfp)|-\frac{1}{2}~+1; xP^+,\bfp\rangle +\psi^{\pm(\nu)}_{+0}(x,\bfp)|+\frac{1}{2}~0; xP^+,\bfp\rangle \nonumber \\
 &+& \psi^{\pm(\nu)}_{-0}(x,\bfp)|-\frac{1}{2}~0; xP^+,\bfp\rangle + \psi^{\pm(\nu)}_{+-}(x,\bfp)|+\frac{1}{2}~-1; xP^+,\bfp\rangle \nonumber\\
 &+& \psi^{\pm(\nu)}_{--}(x,\bfp)|-\frac{1}{2}~-1; xP^+,\bfp\rangle  \bigg].\label{fock_VD}
\ee
Where $|\lambda_q~\lambda_D; xP^+,\bfp\rangle$ is the two-particle state with a quark of helicity $\lambda_q=\pm\frac{1}{2}$ and a vector diquark of helicity $\lambda_D=\pm 1,0(triplet)$.
The light front wave functions for vector diquark are given as, for $J^z=+1/2$
\be 
\psi^{+(\nu)}_{+~+}(x,\bfp)&=& N^{(\nu)}_1 \sqrt{\frac{2}{3}} \bigg(\frac{p^1-ip^2}{xM}\bigg) \varphi^{(\nu)}_{2}(x,\bfp),\nonumber \\
\psi^{+(\nu)}_{-~+}(x,\bfp)&=& N^{(\nu)}_1 \sqrt{\frac{2}{3}} \varphi^{(\nu)}_{1}(x,\bfp),\nonumber \\
\psi^{+(\nu)}_{+~0}(x,\bfp)&=& - N^{(\nu)}_0 \sqrt{\frac{1}{3}} \varphi^{(\nu)}_{1}(x,\bfp),\label{LFWF_Vp}\\
\psi^{+(\nu)}_{-~0}(x,\bfp)&=& N^{(\nu)}_0 \sqrt{\frac{1}{3}} \bigg(\frac{p^1+ip^2}{xM} \bigg)\varphi^{(\nu)}_{2}(x,\bfp),\nonumber \\
\psi^{+(\nu)}_{+~-}(x,\bfp)&=& 0,\nonumber \\
\psi^{+(\nu)}_{-~-}(x,\bfp)&=&  0, \nonumber 
\ee
and for $J^z=-1/2$
\be 
\psi^{-(\nu)}_{+~+}(x,\bfp)&=& 0,\nonumber \\
\psi^{-(\nu)}_{-~+}(x,\bfp)&=& 0,\nonumber \\
\psi^{-(\nu)}_{+~0}(x,\bfp)&=& N^{(\nu)}_0 \sqrt{\frac{1}{3}} \bigg( \frac{p^1-ip^2}{xM} \bigg) \varphi^{(\nu)}_{2}(x,\bfp),\label{LFWF_Vm}\\
\psi^{-(\nu)}_{-~0}(x,\bfp)&=& N^{(\nu)}_0\sqrt{\frac{1}{3}} \varphi^{(\nu)}_{1}(x,\bfp),\nonumber \\
\psi^{-(\nu)}_{+~-}(x,\bfp)&=& - N^{(\nu)}_1 \sqrt{\frac{2}{3}} \varphi^{(\nu)}_{1}(x,\bfp),\nonumber \\
\psi^{-(\nu)}_{-~-}(x,\bfp)&=& N^{(\nu)}_1 \sqrt{\frac{2}{3}} \bigg(\frac{p^1+ip^2}{xM}\bigg) \varphi^{(\nu)}_{2}(x,\bfp),\nonumber
\ee
having flavour index $\nu=u,d$.
We adopt a generic ansatz of LFWFs $\varphi^{(\nu)}_i(x,\bfp)$ from the soft-wall AdS/QCD prediction\cite{BT1,BT2} and introduce the parameters $a^\nu_i,~b^\nu_i$ and $\delta^\nu$ as
\be
\varphi_i^{(\nu)}(x,\bfp)=\frac{4\pi}{\kappa}\sqrt{\frac{\log(1/x)}{1-x}}x^{a_i^\nu}(1-x)^{b_i^\nu}\exp\bigg[-\delta^\nu\frac{\bfp^2}{2\kappa^2}\frac{\log(1/x)}{(1-x)^2}\bigg].
\label{LFWF_phi}
\ee
The wave functions $\varphi_i^\nu ~(i=1,2)$ reduce to the AdS/QCD prediction for the parameters $a_i^\nu=b_i^\nu=0$  and $\delta^\nu=1.0$.
 We use the AdS/QCD scale parameter $\kappa =0.4~GeV$ as determined in \cite{CM1} and the quarks are  assumed  to be  massless.  The parameters of the model and the pdf scale evolution are discussed in appendix \ref{AppA}.

\section{TMDs}\label{TMDS}
In the light front formalism, the unintegrated quark-quark correlator for semi inclusive deep inelastic scattering(SIDIS) is defined as 
\be
\Phi^{\nu [\Gamma]}(x,\textbf{p}_{\perp};S)&=&\frac{1}{2}\int \frac{dz^- d^2z_T}{2(2\pi)^3} e^{ip.z} \langle P; S|\overline{\psi}^\nu (0)\Gamma \mathcal{W}_{[0,z]} \psi^\nu (z) |P;S\rangle\Bigg|_{z^+=0} \label{TMD_cor}
\ee
 at equal light-front time $z^+=0$. The summations over the color indicates of quarks are implied. $x ~(x=p^+/P^+)$ is the momentum fraction carried by the struck quark of helicity $\lambda$ and $P$ is the momentum of the proton with helicity $\lambda_N$. 
We choose the light cone gauge $A^+=0$ and a frame where the nucleon momentum $ P\equiv (P^+,\frac{M^2}{P^+},\textbf{0} ),$ and virtual photon momentum $q\equiv (x_B P^+, \frac{Q^2}{x_BP^+},\textbf{0})$, where $x_B= \frac{Q^2}{2P.q}$ is the Bjorken variable and $Q^2 = -q^2$. The nucleon with helicity $\lambda$ has  spin components $S^+ = \lambda \frac{P^+}{M},~ S^- = \lambda\frac{P^-}{M},$ and $ S_T $. The Wilson line $\mathcal{W}_{[0,z]}$ goes along $[0, 0, 0_\perp] \to [0,1, 0_\perp] \to
[0,1, z_\perp] \to [0, z^-, z_\perp]$ \cite{Bacc08, Boer03} and at light-cone gauge it does not contribute to the T-even TMDs.  In this work, we concentrate only on the T-even TMDs and the Wilson line is taken to be unity.
In leading twist, the TMDs are defined as
\begin{eqnarray}
\Phi^{\nu [\gamma^+]}(x,\textbf{p}_{\perp};S)&=& f_1^\nu (x,\textbf{p}_{\perp}^2) - \frac{\epsilon^{ij}_Tp^i_\perp S^j_T}{M}f^{\perp  \nu} _{1T}(x,\textbf{p}_{\perp}^2),\label{Phi_1}\\
\Phi^{\nu [\gamma^+ \gamma^5]}(x,\textbf{p}_{\perp};S) &=&  \lambda g_{1L}^\nu (x,\textbf{p}_{\perp}^2) + \frac{\textbf{p}_{\perp}.\textbf{S}_T}{M} g^\nu _{1T}(x,\textbf{p}_{\perp}^2),\label{Phi_2}\\
\Phi^{\nu [i \sigma^{j +}\gamma^5]}(x,\textbf{p}_{\perp};S)& = & S^j_T h_1^\nu (x,\textbf{p}_{\perp}^2) + \lambda\frac{p^j_\perp}{M}h^{\perp  \nu} _{1L}(x,\textbf{p}_{\perp}^2)\nonumber\\
&&+ \frac{2 p^j_\perp \textbf{p}_{\perp}.\textbf{S}_T - S^j_T \textbf{p}^2_{\perp}}{2M^2} h^{\perp  \nu} _{1T}(x,\textbf{p}_{\perp}^2) - \frac{\epsilon_T^{ij}p^i_{\perp}}{M}h^{\perp  \nu} _1(x,\textbf{p}_{\perp}^2).\label{Phi_3}
\end{eqnarray}
The $p_\perp$ integrated function of $f_1^\nu (x,p_\perp^2)$ gives the unpolarized distribution $f^{\nu }_1(x)$ and that of $g_{1L}^\nu (x,p_\perp^2)$ ($=g_1^\nu (x,p_\perp^2)$) gives the helicity distribution $g^{\nu }_1(x)$. The transversity TMD $h^{\nu }_1 (x,\textbf{p}_{\perp}^2)$ is defined  as 
\begin{eqnarray}
h^{\nu }_1 (x,\textbf{p}_{\perp}^2) &=& h^{\nu }_{1T} (x,\textbf{p}_{\perp}^2) + \frac{\textbf{p}^2_{\perp}}{2M^2}h^{\perp  \nu} _{1T} (x,\textbf{p}_{\perp}^2)\label{h1q}\\
h^{\nu }_1 (x) &=& \int d^2p_\perp h^{\nu }_1 (x,\textbf{p}_{\perp}^2).
\end{eqnarray} 
 $h^{\nu }_1 (x)$ is called the transversity distribution.
There are altogether six T-even TMDs and two T-odd TMDs at the leading twist. In this work, we concentrate  on the T-even TMDs only.

\section{Results}\label{results}


Using the Eqs.(\ref{fock_SD},\ref{fock_VD}) in the correlator Eq.(\ref{TMD_cor}), we calculate the TMDs for different polarization from Eqs.(\ref{Phi_1},\ref{Phi_2},\ref{Phi_3}). The transverse momentum dependent parton distributions, in this model, can be written in terms of LFWFs as\\
for scalar diquark:
\be
{f}^{\nu(S)}_1(x,\textbf{p}_{\perp}^2)&=&\frac{1}{16\pi^3}\bigg[|\psi ^{+\nu}_+(x,\textbf{p}_{\perp})|^2+|\psi ^{ + \nu}_-(x,\textbf{p}_{\perp})|^2\bigg], \nonumber \\ 
{g}^{\nu (S)}_{1L}(x,\textbf{p}_{\perp}^2)&=&\frac{1}{16\pi^3}\bigg[|\psi ^{ + \nu}_+(x,\textbf{p}_{\perp})|^2 - |\psi ^{ + \nu}_-(x,\textbf{p}_{\perp})|^2\bigg],\nonumber \\
\frac{\textbf{p}_{\perp}.\textbf{S}_T}{M}~ {g}^{\nu (S) }_{1T}(x,\textbf{p}_{\perp}^2)&=&  \frac{1}{16\pi^3}\bigg[\psi ^{+\nu \dagger}_+(x,\textbf{p}_{\perp})\psi^{- \nu}_+(x,\textbf{p}_{\perp})
 - {\psi ^{+\nu \dagger}_-}(x,\textbf{p}_{\perp}){\psi^{- \nu}_-}(x,\textbf{p}_{\perp})\nonumber\\
 &&+\psi ^{-\nu \dagger}_+(x,\textbf{p}_{\perp})\psi^{+ \nu}_+(x,\textbf{p}_{\perp})
 - {\psi ^{-\nu \dagger}_-}(x,\textbf{p}_{\perp}){\psi^{+ \nu}_-}(x,\textbf{p}_{\perp}\bigg],\label{TMD_SD}\\
\frac{\textbf{p}_{\perp}.\textbf{s}^\nu_{ T}}{M} ~{h}^{\nu\perp (S) }_{1L}(x,\textbf{p}_{\perp}^2)&=&\frac{1}{16\pi^3}\bigg[\psi ^{ + \nu\dagger}_+(x,\textbf{p}_{\perp})\psi ^{ + \nu}_-(x,\textbf{p}_{\perp}) + \psi ^{ + \nu\dagger}_-(x,\textbf{p}_{\perp})\psi ^{ + \nu}_+(x,\textbf{p}_{\perp})\bigg], \nonumber\\
\textbf{S}_T . \textbf{s}^\nu_{ T}~ {h}^{\nu (S) }_{1}(x,\textbf{p}_{\perp}^2) &+& \frac{2 \bfp . \textbf{S}_{T}~\bfp . \textbf{s}^\nu_{T}-\textbf{S}_{T}.\textbf{s}^\nu_{T} \bfp^2}{2M^2} {h}^{\perp q (S) }_{1T}(x,\textbf{p}_{\perp}^2)
\nonumber\\  ~~
&=& \frac{1}{16\pi^3}\bigg[\psi ^{+\nu \dagger}_+(x,\textbf{p}_{\perp})\psi^{- \nu}_-(x,\textbf{p}_{\perp})
 + {\psi ^{+\nu \dagger}_-}(x,\textbf{p}_{\perp}){\psi^{- \nu}_+}(x,\textbf{p}_{\perp})\nonumber\\
 &&+\psi ^{-\nu \dagger}_+(x,\textbf{p}_{\perp})\psi^{+ \nu}_-(x,\textbf{p}_{\perp})
 - {\psi ^{-\nu \dagger}_-}(x,\textbf{p}_{\perp}){\psi^{+ \nu}_+}(x,\textbf{p}_{\perp}\bigg].\nonumber
\ee
for vector diquark:
\be
{f}^{\nu(A)}_1(x,\textbf{p}_{\perp}^2)&=&\sum_{\lambda_A} \frac{1}{16\pi^3}\bigg[|\psi ^{+\nu}_{+\lambda_A}(x,\textbf{p}_{\perp})|^2+|\psi ^{ + \nu}_{-\lambda_A}(x,\textbf{p}_{\perp})|^2\bigg],\nonumber\\ 
{g}^{\nu (A)}_{1L}(x,\textbf{p}_{\perp}^2)&=&\sum_{\lambda_A}\frac{1}{16\pi^3}\bigg[|\psi ^{ + \nu}_{+\lambda_A}(x,\textbf{p}_{\perp})|^2 - |\psi ^{ + \nu}_{-\lambda_A}(x,\textbf{p}_{\perp})|^2\bigg],\nonumber\\
\frac{\textbf{p}_{\perp}.\textbf{S}_T}{M}~ {g}^{\nu (A) }_{1T}(x,\textbf{p}_{\perp}^2)&=& \sum_{\lambda_A} \frac{1}{16\pi^3}\bigg[\psi ^{+\nu \dagger}_{+\lambda_A}(x,\textbf{p}_{\perp})\psi^{- \nu}_{+\lambda_A}(x,\textbf{p}_{\perp})
 - \psi ^{+\nu \dagger}_{-\lambda_A}(x,\textbf{p}_{\perp})\psi^{- \nu}_{-\lambda_A}(x,\textbf{p}_{\perp})\nonumber\\
 &&+\psi ^{-\nu \dagger}_{+\lambda_A}(x,\textbf{p}_{\perp})\psi^{+ \nu}_{+\lambda_A}(x,\textbf{p}_{\perp})
 - \psi ^{-\nu \dagger}_{-\lambda_A}(x,\textbf{p}_{\perp})\psi^{+ \nu}_{-\lambda_A}(x,\textbf{p}_{\perp}\bigg],\label{TMD_VD}\\
\frac{\textbf{p}_{\perp}.\textbf{s}^\nu_{ T}}{M} ~{h}^{\nu\perp (A) }_{1L}(x,\textbf{p}_{\perp}^2)&=&\sum_{\lambda_A}\frac{1}{16\pi^3}\bigg[\psi ^{ + \nu\dagger}_{+\lambda_A}(x,\textbf{p}_{\perp})\psi ^{ + \nu}_{-\lambda_A}(x,\textbf{p}_{\perp}) + \psi ^{ + \nu\dagger}_{-\lambda_A}(x,\textbf{p}_{\perp})\psi ^{ + \nu}_{+\lambda_A}(x,\textbf{p}_{\perp})\bigg], \nonumber\\
\textbf{S}_T . \textbf{s}^\nu_{ T}~ {h}^{\nu (A) }_{1}(x,\textbf{p}_{\perp}^2) &+& \frac{2 \bfp . \textbf{S}_{T}~\bfp . \textbf{s}^\nu_{T}-\textbf{S}_{T}.\textbf{s}^\nu_{T} \bfp^2}{2M^2} {h}^{\perp q (A) }_{1T}(x,\textbf{p}_{\perp}^2)
\nonumber\\  ~~
&=&\sum_{\lambda_A} \frac{1}{16\pi^3}\bigg[\psi ^{+\nu \dagger}_{+\lambda_A}(x,\textbf{p}_{\perp})\psi^{- \nu}_{-\lambda_A}(x,\textbf{p}_{\perp})
 + \psi ^{+\nu \dagger}_{-\lambda_A}(x,\textbf{p}_{\perp})\psi^{- \nu}_{+\lambda_A}(x,\textbf{p}_{\perp})\nonumber\\
 &&+\psi ^{-\nu \dagger}_{+\lambda_A}(x,\textbf{p}_{\perp})\psi^{+ \nu}_{-\lambda_A}(x,\textbf{p}_{\perp})
 - \psi ^{-\nu \dagger}_{-\lambda_A}(x,\textbf{p}_{\perp})\psi^{+ \nu}_{+\lambda_A}(x,\textbf{p}_{\perp}\bigg].\nonumber
\ee
Where the summation is taken over helicity of the vector diquark, $\lambda_A=0,\pm$. 
Using the light-front wave functions from Eqs. (\ref{LFWF_S}) and Eqs.(\ref{LFWF_Vp},\ref{LFWF_Vm}), the  explicit expressions for the TMDs can be written as:
\be
{f}^{\nu  }_1(x,\textbf{p}^2_{\perp})&=&\bigg(C^2_SN^{\nu 2}_S+C^2_V\big(\frac{1}{3}N^{\nu 2}_0+\frac{2}{3}N^{\nu 2}_1\big)\bigg)\frac{\ln(1/x)}{\pi\kappa^2}\bigg[T^\nu_1(x) +\frac{\textbf{p}^2_{\perp}}{M^2} T^\nu_2(x)\bigg]\exp\big[-R^\nu(x)\bfp^2\big], \nonumber\\ 
\label{TMD_f1}\\
{g}^{\nu  }_{1L}(x,\textbf{p}^2_{\perp})&=&\bigg(C^2_SN^{\nu 2}_S+C^2_V\big(\frac{1}{3}N^{\nu 2}_0 - \frac{2}{3}N^{\nu 2}_1\big)\bigg)\frac{\ln(1/x)}{\pi\kappa^2}\bigg[T^\nu_1(x)-
 \frac{\textbf{p}^2_{\perp}}{M^2} T^\nu_2(x)\bigg]\exp\big[-R^\nu(x)\bfp^2\big], \nonumber\\
 \label{TMD_g1L}\\
{h}^{\nu  }_1(x,\textbf{p}^2_{\perp}) &=& \bigg(C^2_SN^{\nu 2}_S-C^2_V\frac{1}{3}N^{\nu 2}_0\bigg) \frac{\ln(1/x)}{\pi\kappa^2}T^\nu_1(x)\exp\big[-R^\nu(x)\bfp^2\big], \label{TMD_h1}\\
{g}^{\nu}_{1T}(x,\textbf{p}^2_{\perp})&=&\bigg(C^2_SN^{\nu 2}_S-C^2_V\frac{1}{3}N^{\nu 2}_0\bigg) \frac{2\ln(1/x)}{\pi\kappa^2}T^\nu_3(x)\exp\big[-R^\nu(x)\bfp^2\big],\label{TMD_g1T}\\
{h}^{\nu\perp}_{1L}(x,\textbf{p}^2_{\perp})&=& -\bigg(C^2_SN^{\nu 2}_S+C^2_V\big(\frac{1}{3}N^{\nu 2}_0 - \frac{2}{3}N^{\nu 2}_1\big)\bigg)\frac{2\ln(1/x)}{\pi\kappa^2} T^\nu_3(x)\exp\big[-R^\nu(x)\bfp^2\big],\label{TMD_h1Lp}\\
{h}^{\nu}_{1T}(x,\textbf{p}^2_{\perp})&=& \bigg(C^2_SN^{\nu 2}_S-C^2_V\frac{1}{3}N^{\nu 2}_0\bigg) \frac{\ln(1/x)}{\pi\kappa^2}\bigg[T^\nu_1(x)+ \frac{\textbf{p}^2_{\perp}}{M^2}  T^\nu_2(x)\bigg]\exp\big[-R^\nu(x)\bfp^2\big],\label{TMD_h1T}\\
{h}^{\nu\perp}_{1T}(x,\textbf{p}^2_{\perp})&=& - \bigg(C^2_SN^{\nu 2}_S-C^2_V\frac{1}{3}N^{\nu 2}_0\bigg) \frac{2\ln(1/x)}{\pi\kappa^2}T^\nu_2(x)\exp\big[-R^\nu(x)\bfp^2\big],\label{TMD_h1Tp}
\ee
where 
\be
T^\nu_1(x)&=& x^{2a^{\nu}_1}(1-x)^{2b^{\nu}_1-1}, \nonumber\\
T^\nu_2(x)&=& x^{2a^{\nu}_2 -2}(1-x)^{2b^{\nu}_2-1},\label{Fx}\\
T^\nu_3(x)&=& x^{a^{\nu}_1 +a^{\nu}_2-1}(1-x)^{b^{\nu}_1+b^{\nu}_2-1}.\nonumber\\
R^\nu(x)&=&\delta^\nu \frac{\ln(1/x)}{\kappa^2 (1-x)^2},\nonumber
\ee

\begin{figure}[htbp]
\begin{minipage}[c]{0.98\textwidth}
\small{(a)}\includegraphics[width=7.5cm,clip]{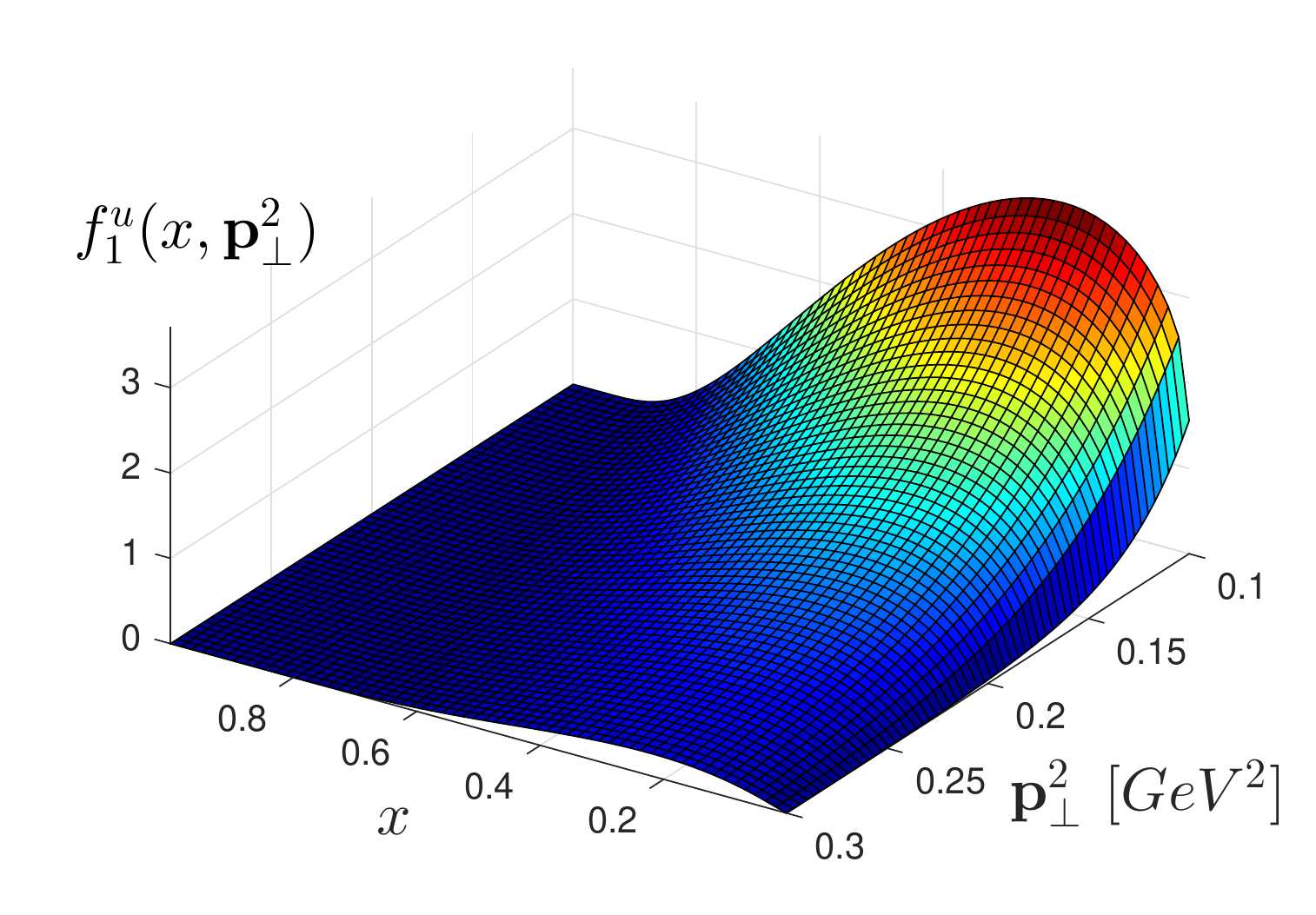}
\small{(b)}\includegraphics[width=7.5cm,clip]{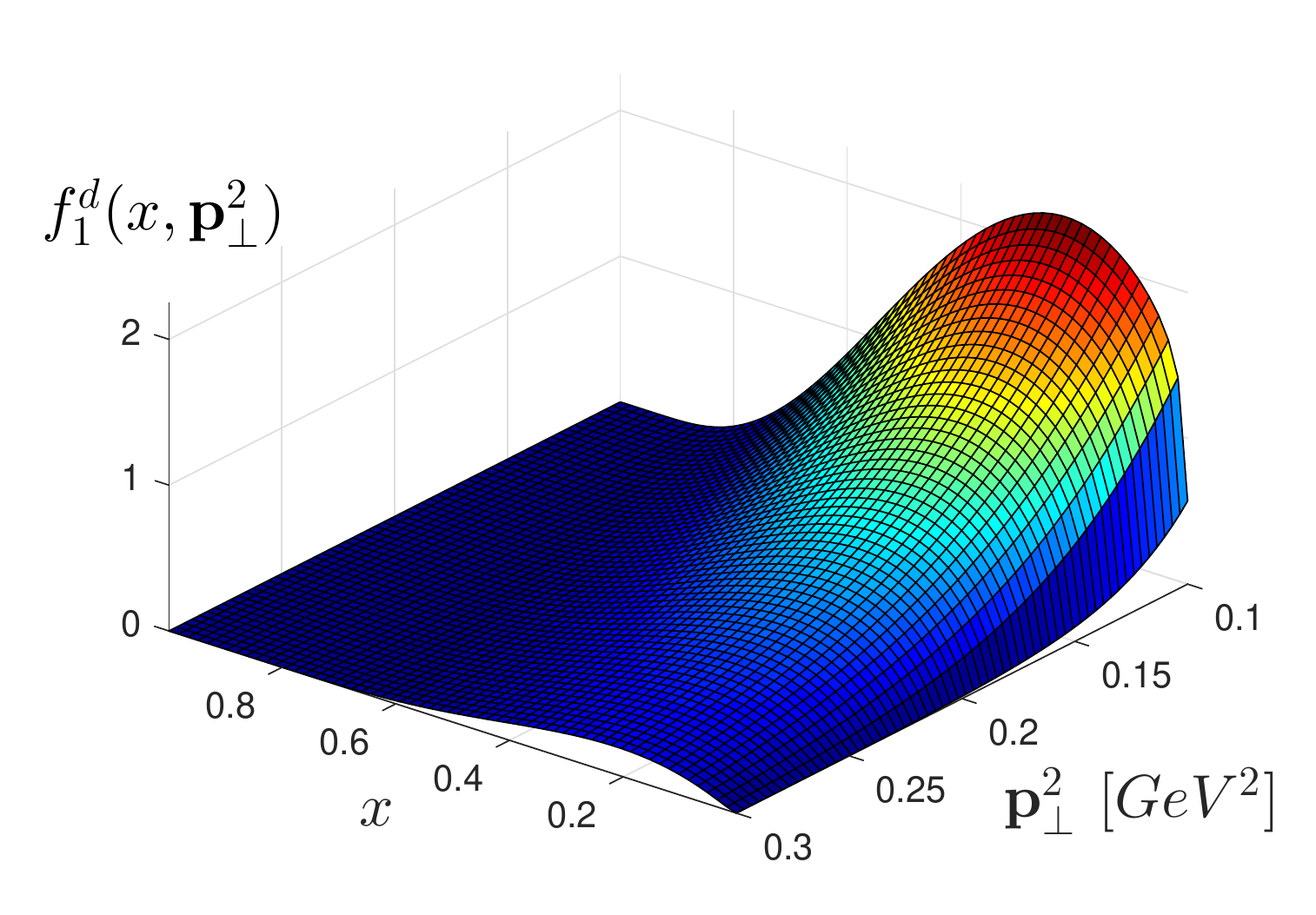}
\end{minipage}
\begin{minipage}[c]{0.98\textwidth}
\small{(c)}\includegraphics[width=7.5cm,clip]{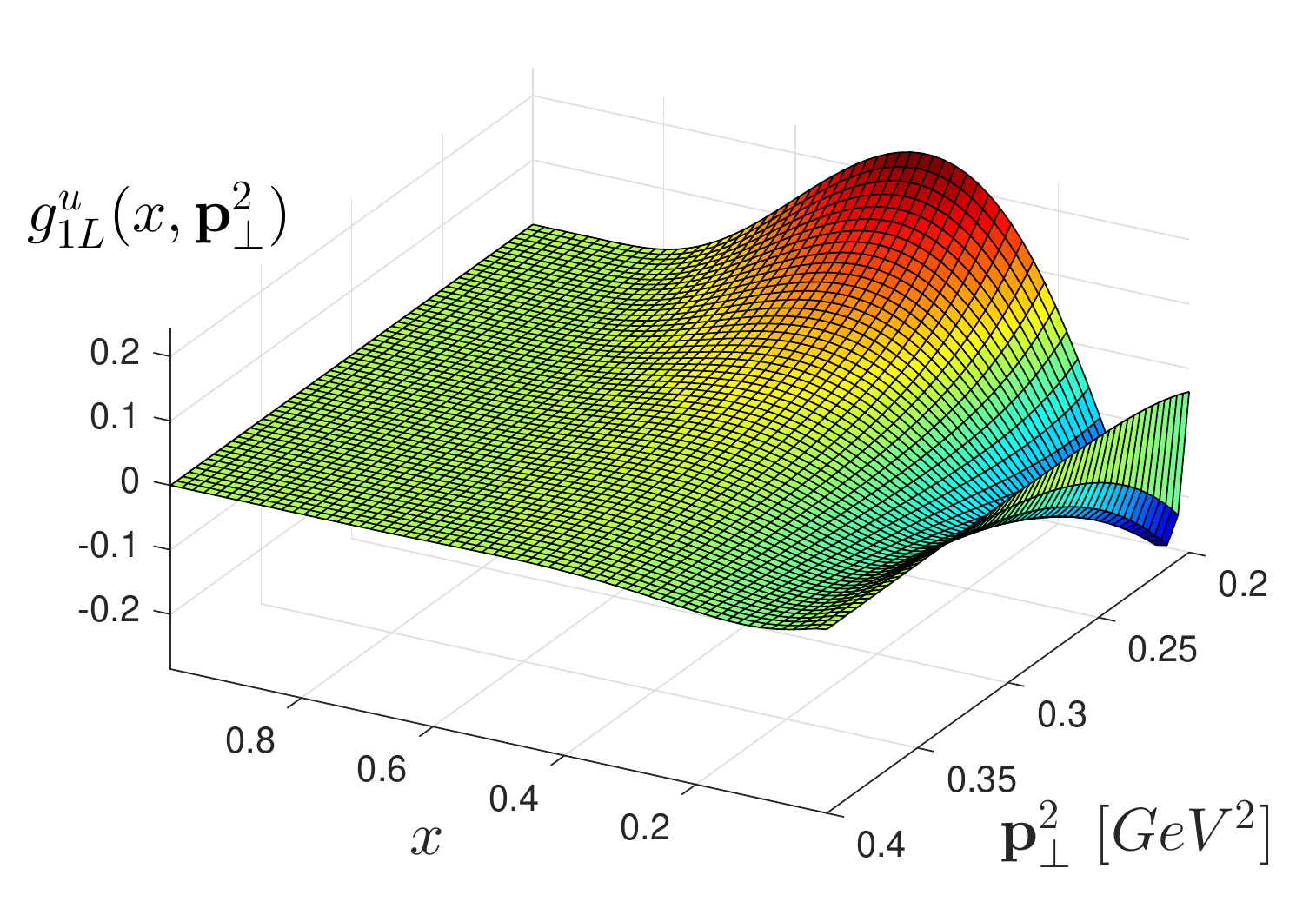}
\small{(d)}\includegraphics[width=7.5cm,clip]{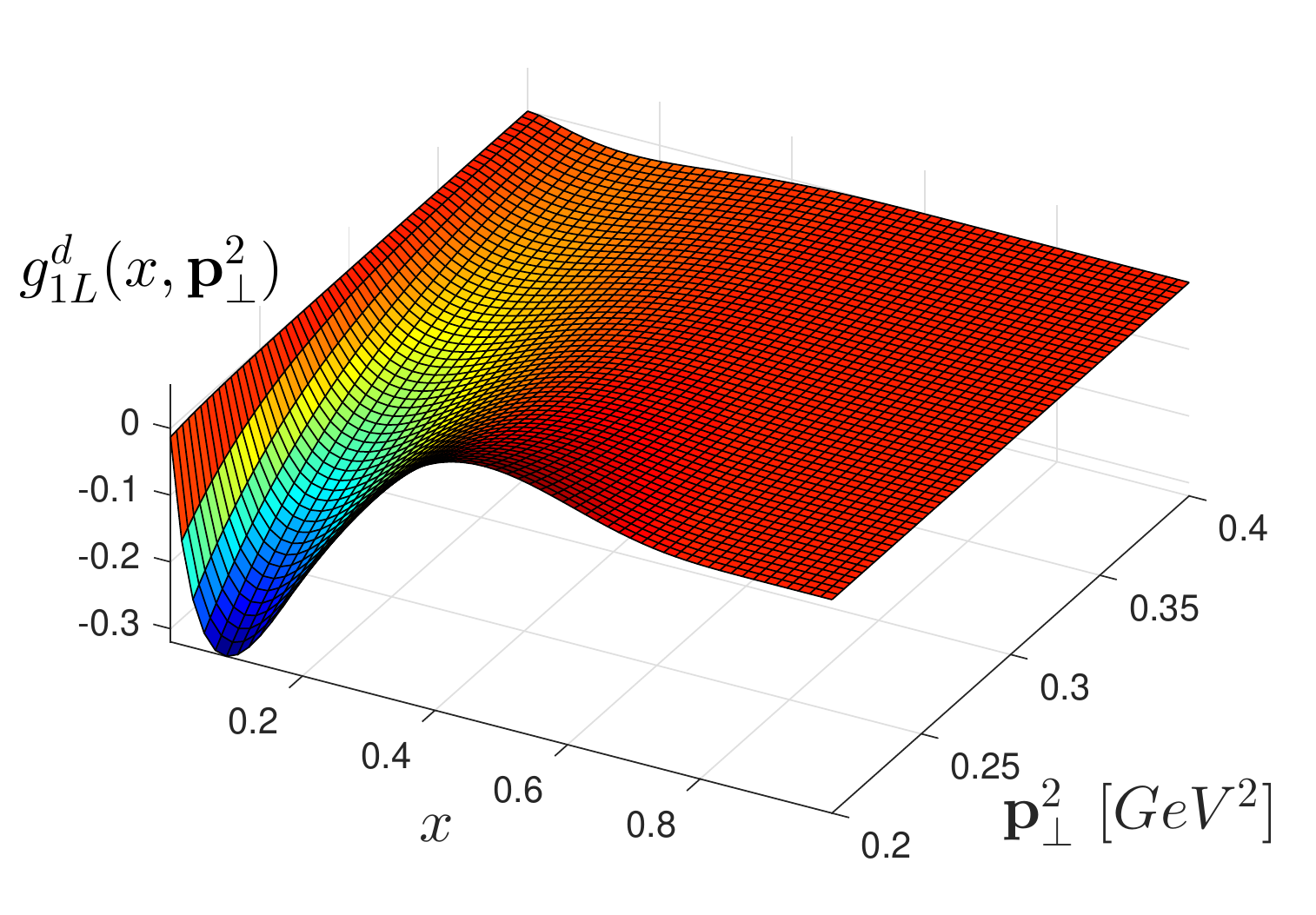} 
\end{minipage}
\begin{minipage}[c]{0.98\textwidth}
\small{(a)}\includegraphics[width=7.5cm,clip]{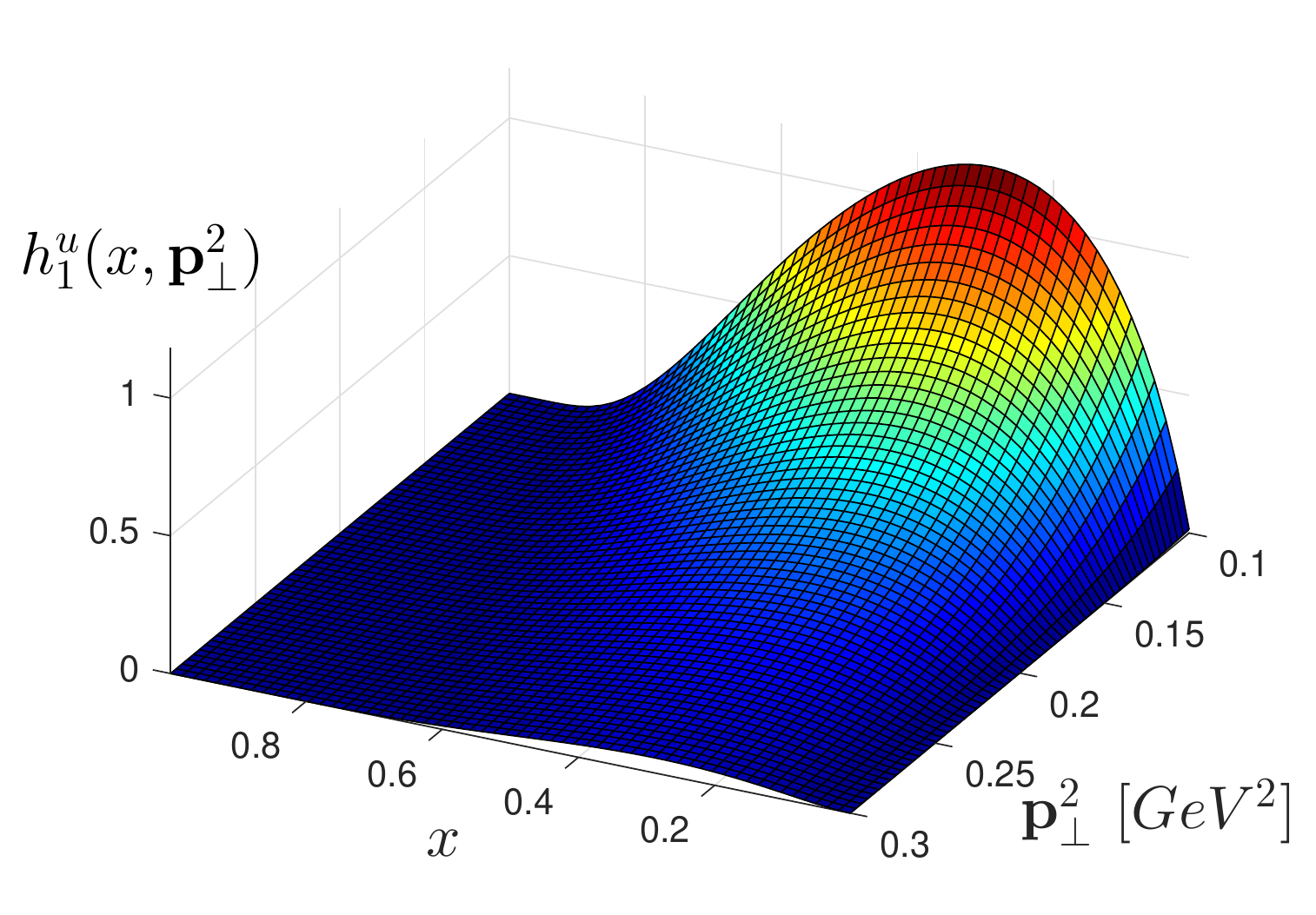}
\small{(b)}\includegraphics[width=7.5cm,clip]{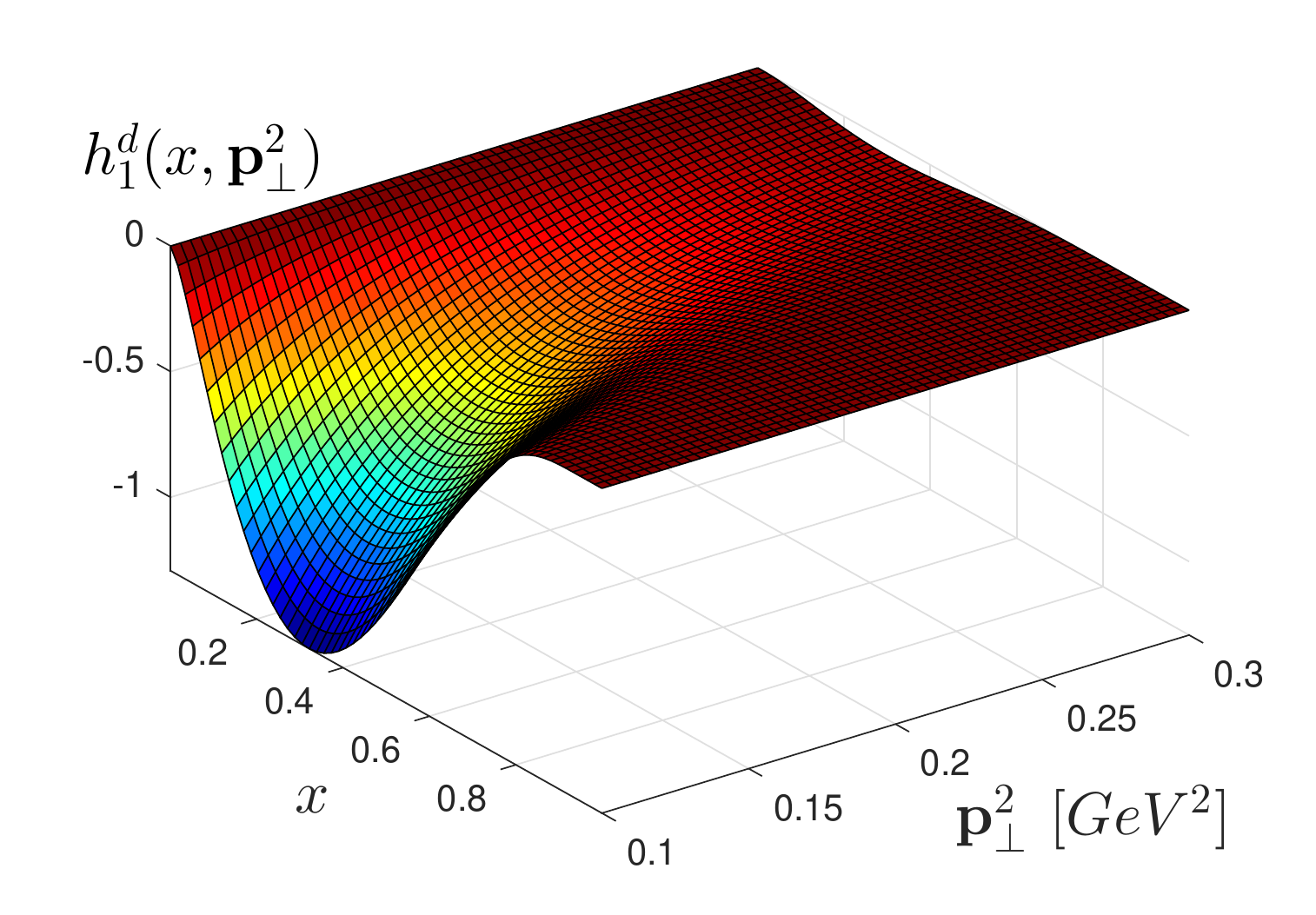}
\end{minipage}
\caption{\label{fig_TMD1}The transverse momentum dependent parton distributions $f^\nu_{1}(x,\bfp^2),~g^\nu_{1L}(x,\bfp^2),~h^\nu_{1}(x,\bfp^2)$ are shown for u quark(left column) and d quark(right column) at the initial scale $\mu_0$.} 
\end{figure} 

\begin{figure}[htbp]
\begin{minipage}[c]{0.98\textwidth}
\small{(a)}\includegraphics[width=7.5cm,clip]{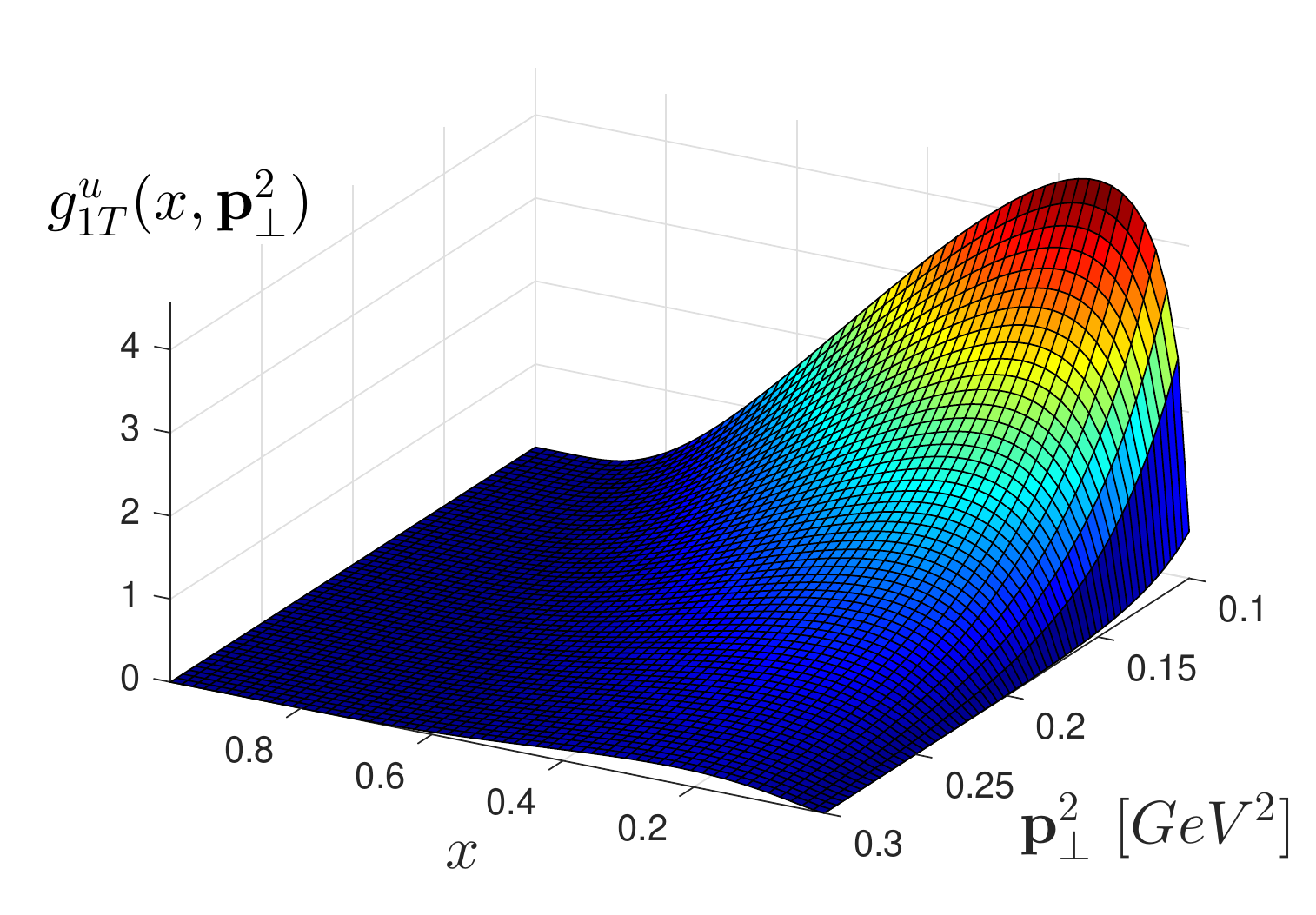}
\small{(b)}\includegraphics[width=7.5cm,clip]{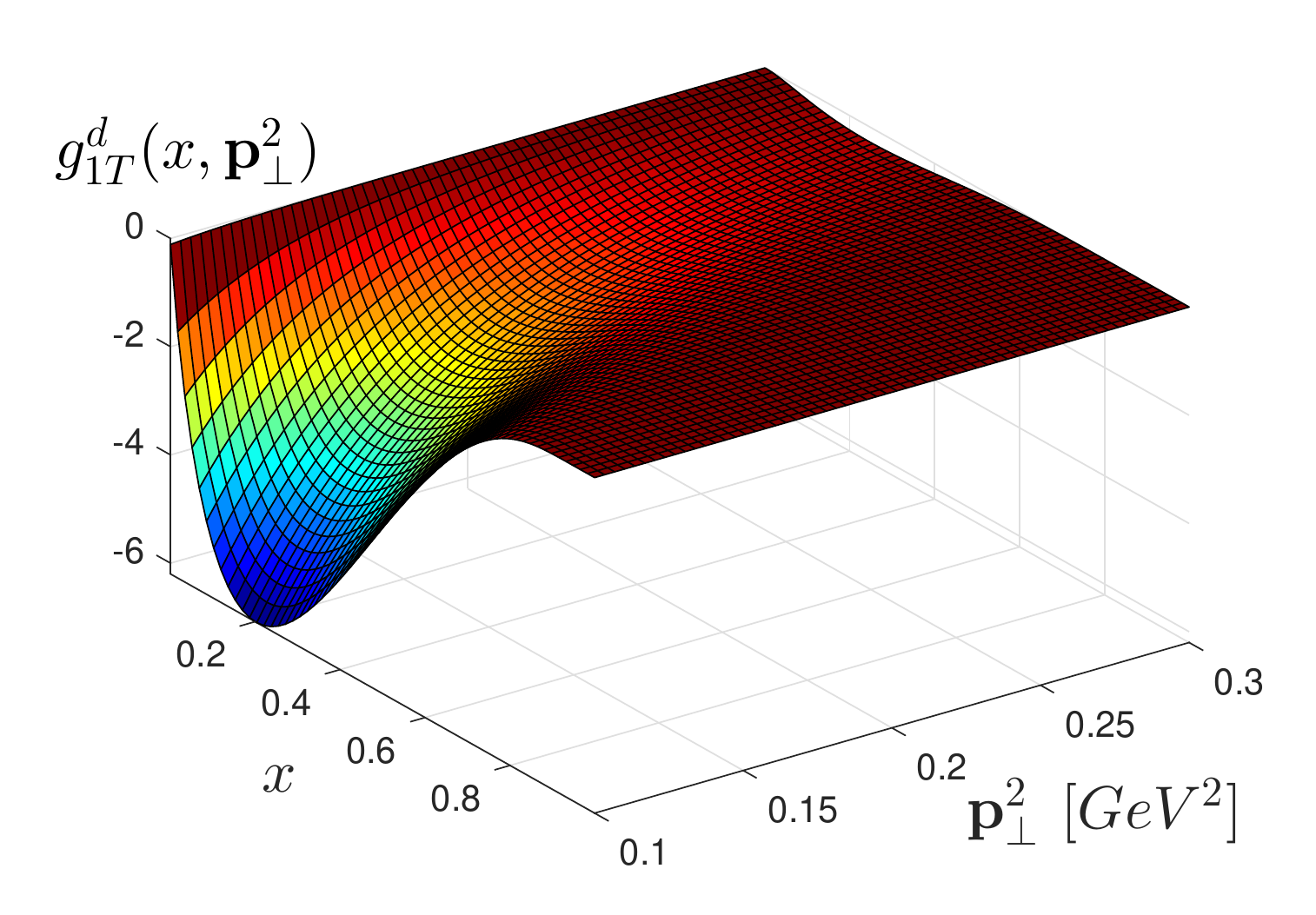}
\end{minipage}
\begin{minipage}[c]{0.98\textwidth}
\small{(c)}\includegraphics[width=7.5cm,clip]{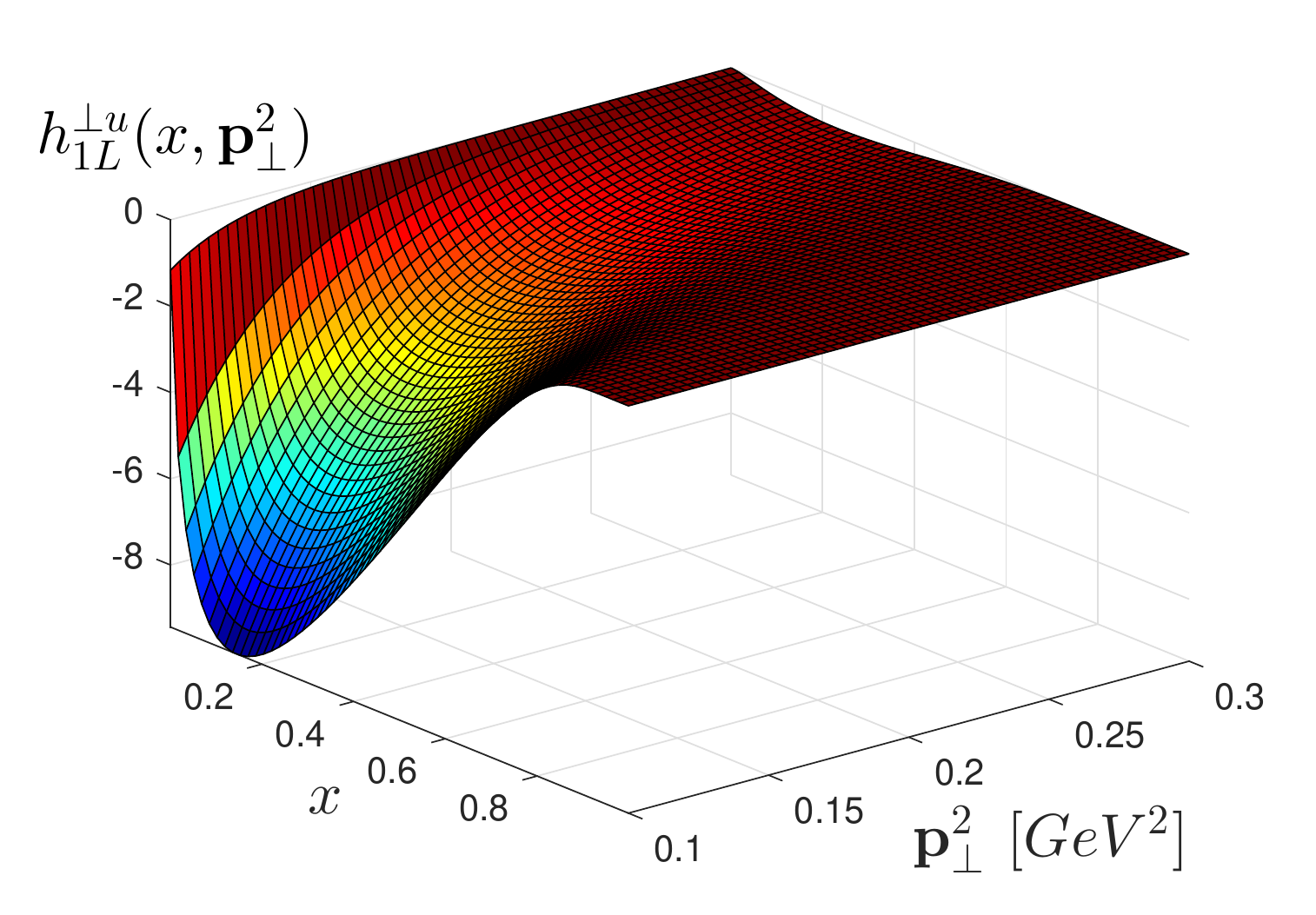}
\small{(d)}\includegraphics[width=7.5cm,clip]{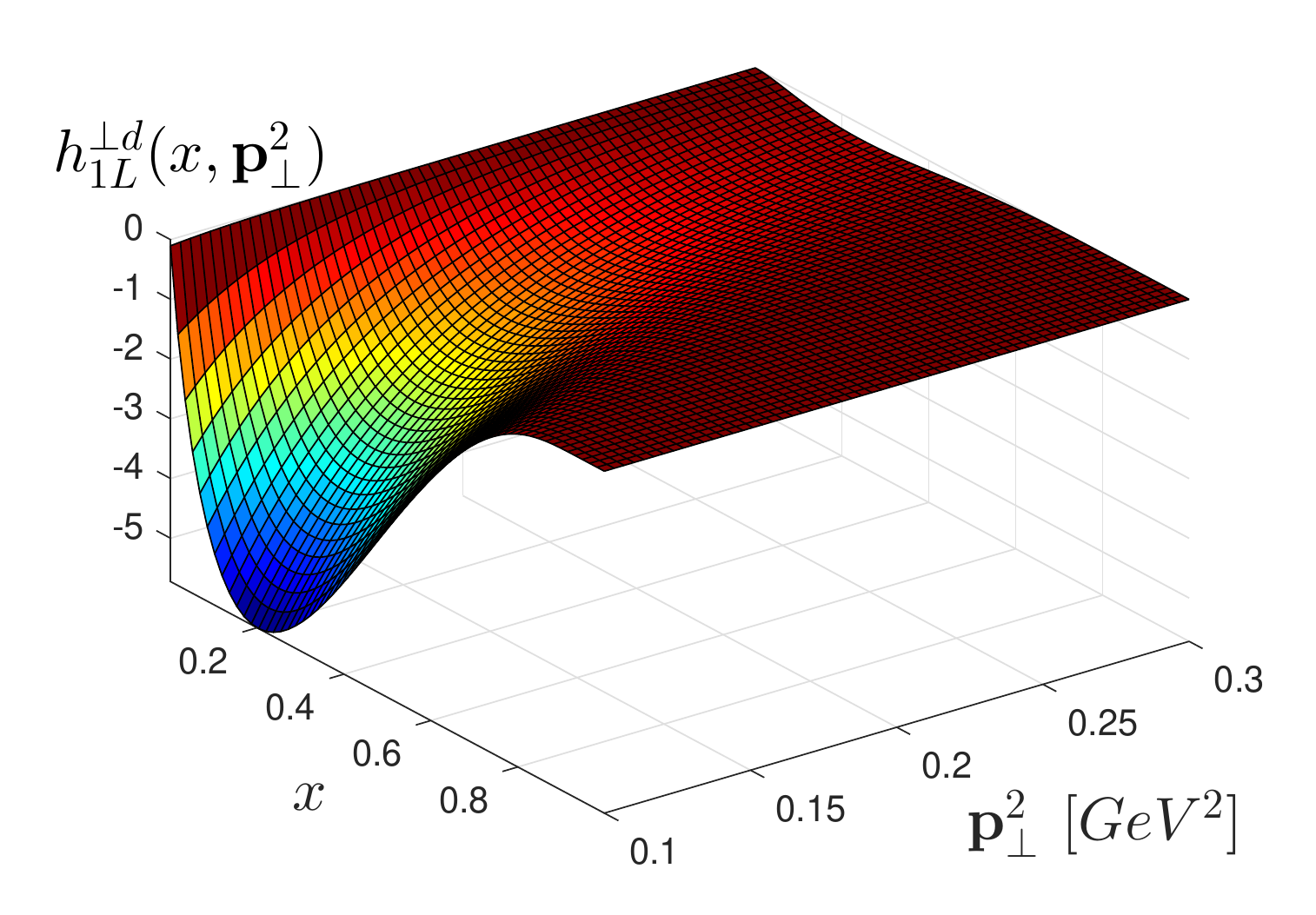} 
\end{minipage}
\begin{minipage}[c]{0.98\textwidth}
\small{(e)}\includegraphics[width=7.5cm,clip]{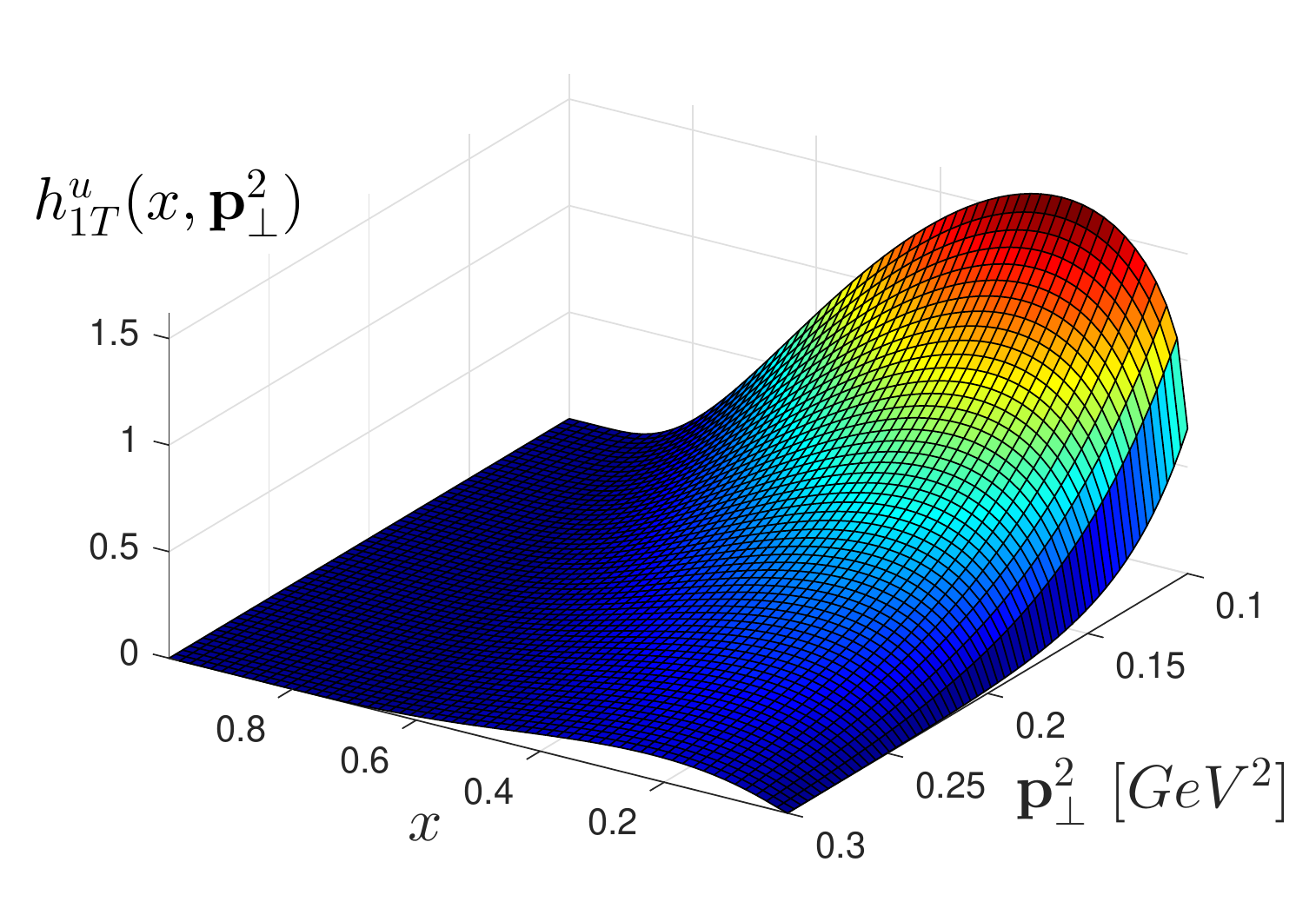}
\small{(f)}\includegraphics[width=7.5cm,clip]{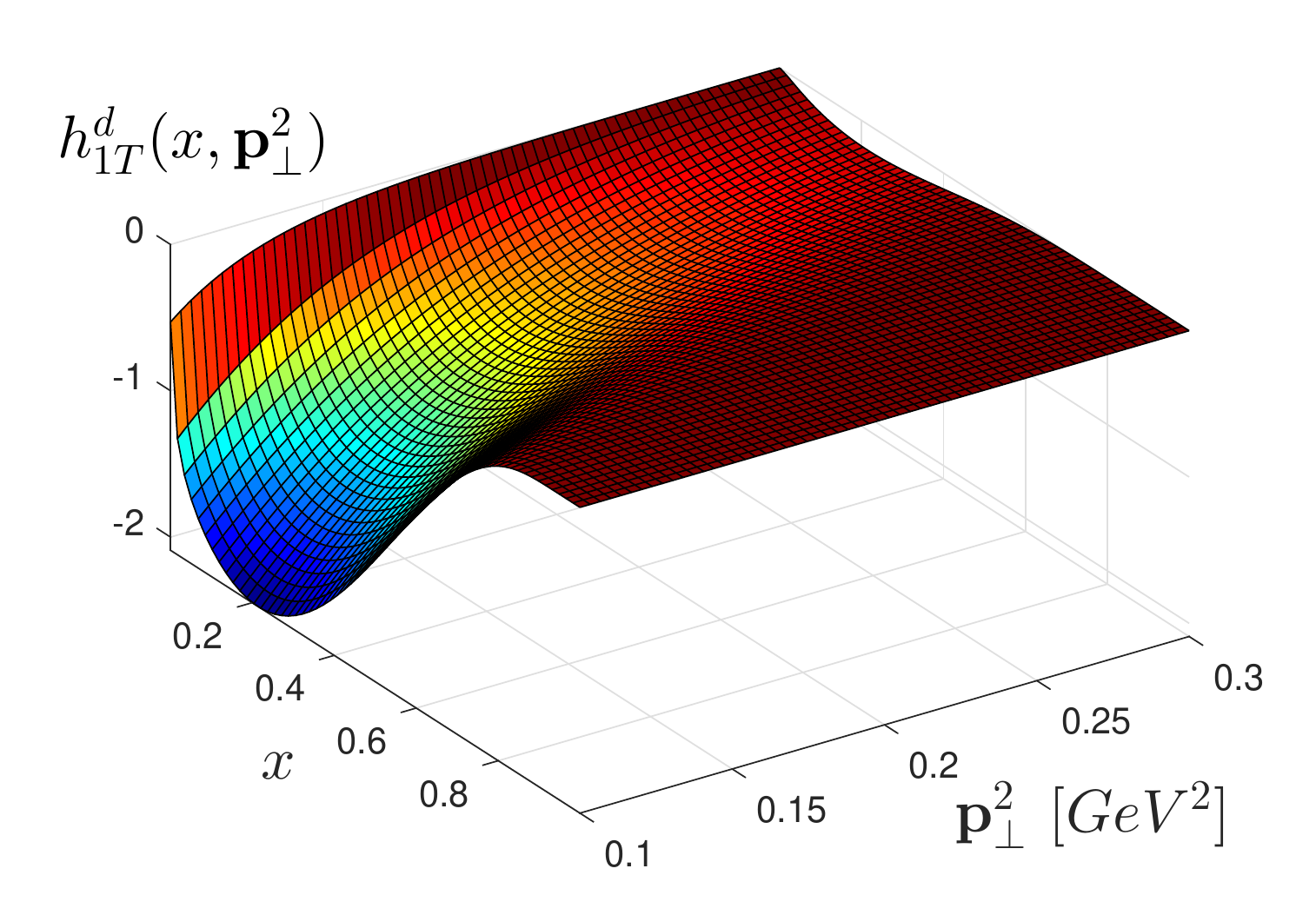}
\end{minipage}
\begin{minipage}[c]{0.98\textwidth}
\small{(g)}\includegraphics[width=7.5cm,clip]{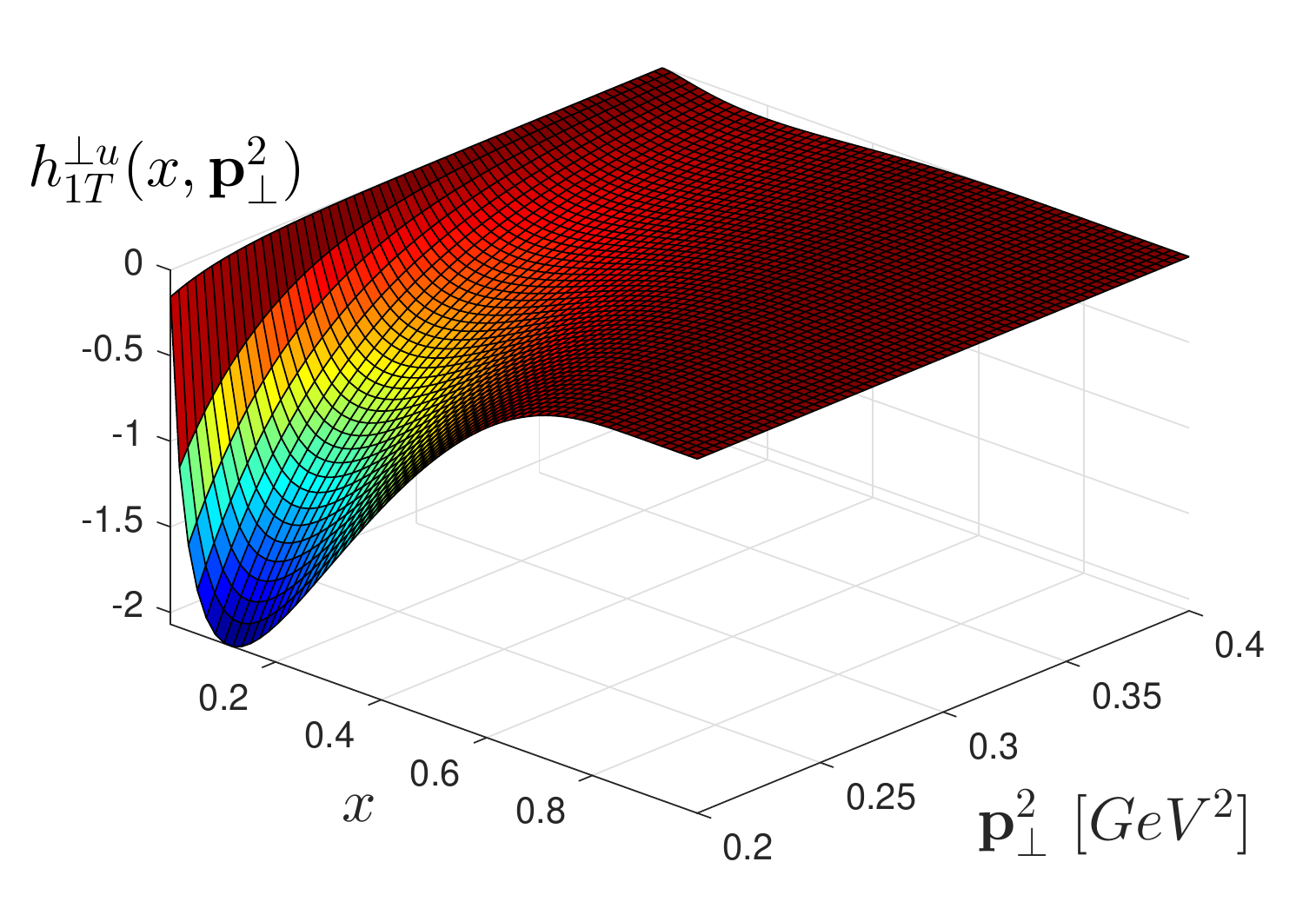}
\small{(h)}\includegraphics[width=7.5cm,clip]{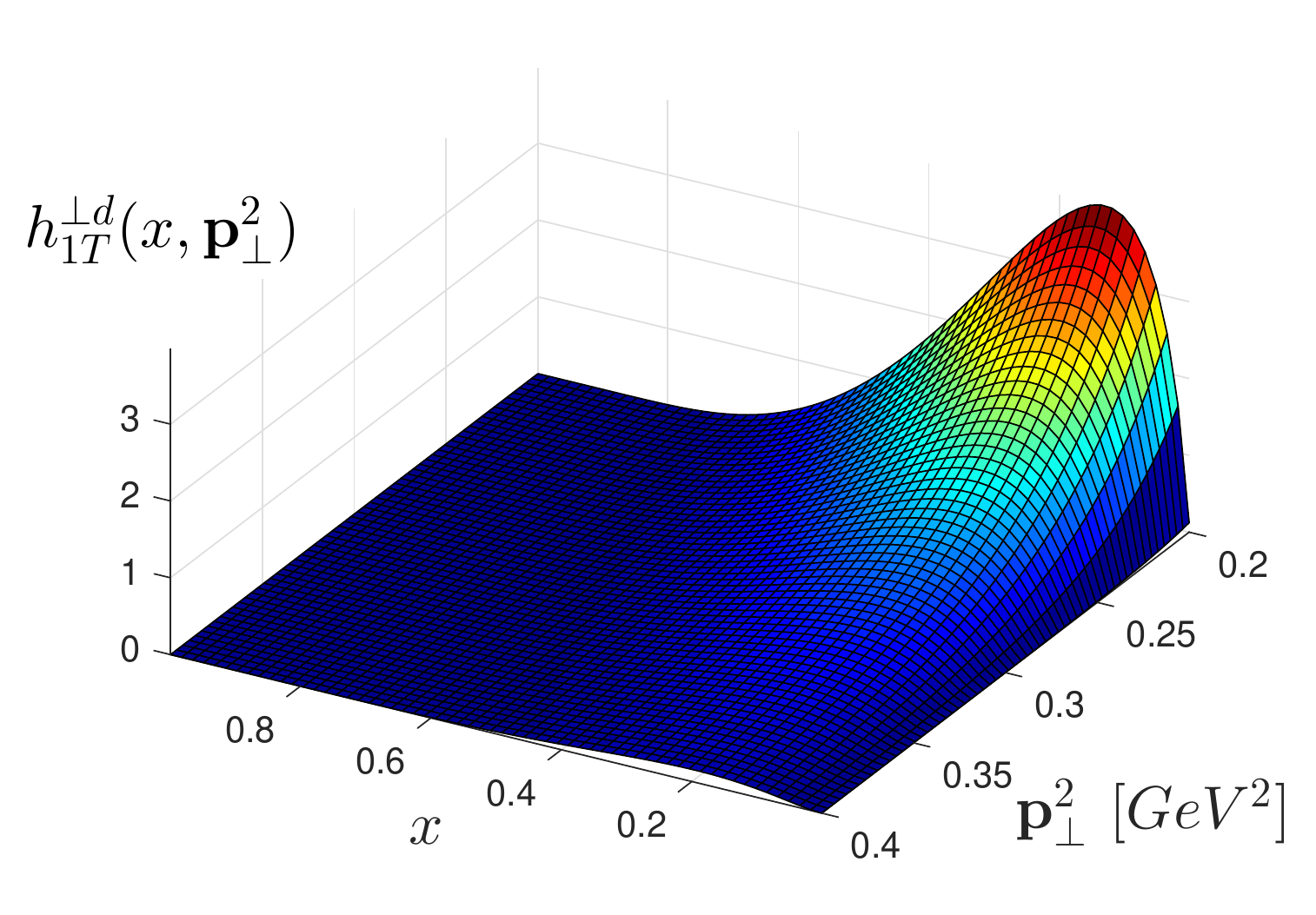}
\end{minipage}
\caption{\label{fig_TMD2}The transverse momentum dependent parton distributions $g^\nu_{1T}(x,\bfp^2),~h^{\perp\nu}_{1L}(x,\bfp^2),~h^\nu_{1T}(x,\bfp^2)$ and $~h^{\perp\nu}_{1T}(x,\bfp^2)$ for u and d quarks at the initial scale $\mu_0$. }. 
\end{figure} 

The three dimensional variation of $f^\nu_{1}(x,\bfp^2),~g^\nu_{1L}(x,\bfp^2),~h^\nu_{1}(x,\bfp^2)$ with $x$ and $\bfp^2$ are shown in Fig.\ref{fig_TMD1} for u and d quarks at the initial scale $\mu_0=0.8~GeV$(see appendix-B). We found $f^\nu_{1}(x,\bfp^2)$ have positive peaks for both u and d quarks, as expected from quark counting rules. In case of  helicity TMD, $g^\nu_{1L}(x,\bfp^2)$, the distribution is mostly positive(except near x=0) for u quark and negative for $d$ quark which indicates the sign difference in the experimental values of axial charges $g^\nu_{A}=\int dx d^2\bfp g^\nu_{1L}(x,\bfp^2)$, for $u$ to $d$ quarks. The axial charge in this model is calculated in \cite{TM_VD} and compared with the experimental data. 
The transversity TMD, $h^\nu_{1}(x,\bfp^2)$, is shown in Fig.\ref{fig_TMD1}(c,d) for u and d quarks respectively. We observed a positive distribution for u quarks and negative distribution for d quarks. Therefore it indicates that the tensor charge, for u quark is positive and negative for d quark. The flavor dependent tensor charge is defined as $g^\nu_{T}=\int dx d^2\bfp h^\nu_{1}(x,\bfp^2)$. A detail discussion on tensor charge is included in \cite{TM_VD}.
A similar behavior of $f^\nu_{1}(x,\bfp^2),~g^\nu_{1L}(x,\bfp^2)$ and $h^\nu_{1}(x,\bfp^2)$ are found in light-cone constituent quark model(LCCQM)\cite{Pasquini08}.

The TMDs $g^\nu_{1T}(x,\bfp^2)$ as a function of $x$ and $\bfp^2$ are shown in Fig.\ref{fig_TMD2}(a,b) for u and d quarks respectively. The distributions for d quark is opposite in sign of the $u$ quark distribution. The TMDs associated with a transversely polarized quarks in a longitudinally polarized proton, $h^{\perp\nu}_{1L}(x,\bfp^2)$, are shown in Fig.\ref{fig_TMD2}(c,d) for u and d quarks respectively. In this model, the distributions are negative for both the quarks. Whereas in LCCQM the $h^{\perp\nu}_{1L}(x,\bfp^2)$ is positive for d quark\cite{Pasquini08}. From Eq.(\ref{TMD_g1T}) and Eq.(\ref{TMD_h1Lp}), we see that the $x$ and $\bfp$ variation of the distributions $g^\nu_{1T}(x,\bfp^2)$ and $h^{\perp\nu}_{1L}(x,\bfp^2)$ are same and they differ by the normalizations factors only. In this model $|g^\nu_{1T}(x,\bfp^2)|<|h^{\perp\nu}_{1L}(x,\bfp^2)|$.
 The TMDs $h^\nu_{1T}(x,\bfp^2)$ is shown in Fig.\ref{fig_TMD2}(e,f) for u and d quarks respectively. We find a positive distribution for u quark and negative distribution for d quark. From Eq.(\ref{TMD_f1}) and Eq.(\ref{TMD_h1T}), we see that the variation of the distributions $f^\nu_{1}(x,\bfp^2)$ and $h^\nu_{1T}(x,\bfp^2)$ are similar. The difference in the peaks are because of the different normalization factors and $|f^\nu_{1}(x,\bfp^2)|>|h^\nu_{1T}(x,\bfp^2)|$. The pretzelosity TMDs, $h^{\perp\nu}_{1T}(x,\bfp^2)$, are shown in Fig.\ref{fig_TMD2}(g,h) for u and d quarks respectively. The model predicts  a negative distribution for u quark and positive distribution for d quark  consistent with the findings of other models e.g,  LCCQM\cite{Pasquini08}, MIT Bag model\cite{bag, Avakian08}, Spectaror model\cite{Avakian08} etc. The pretzelosity distribution extracted by Lefky and Prokudin\cite{Proku15} shows the opposite behavior with a large error corridor.

\subsection{$x-\bfp^2$ factorization}\label{facto}
\begin{figure}[htbp]
\begin{minipage}[c]{0.98\textwidth}
\small{(a)}\includegraphics[width=7.5cm,clip]{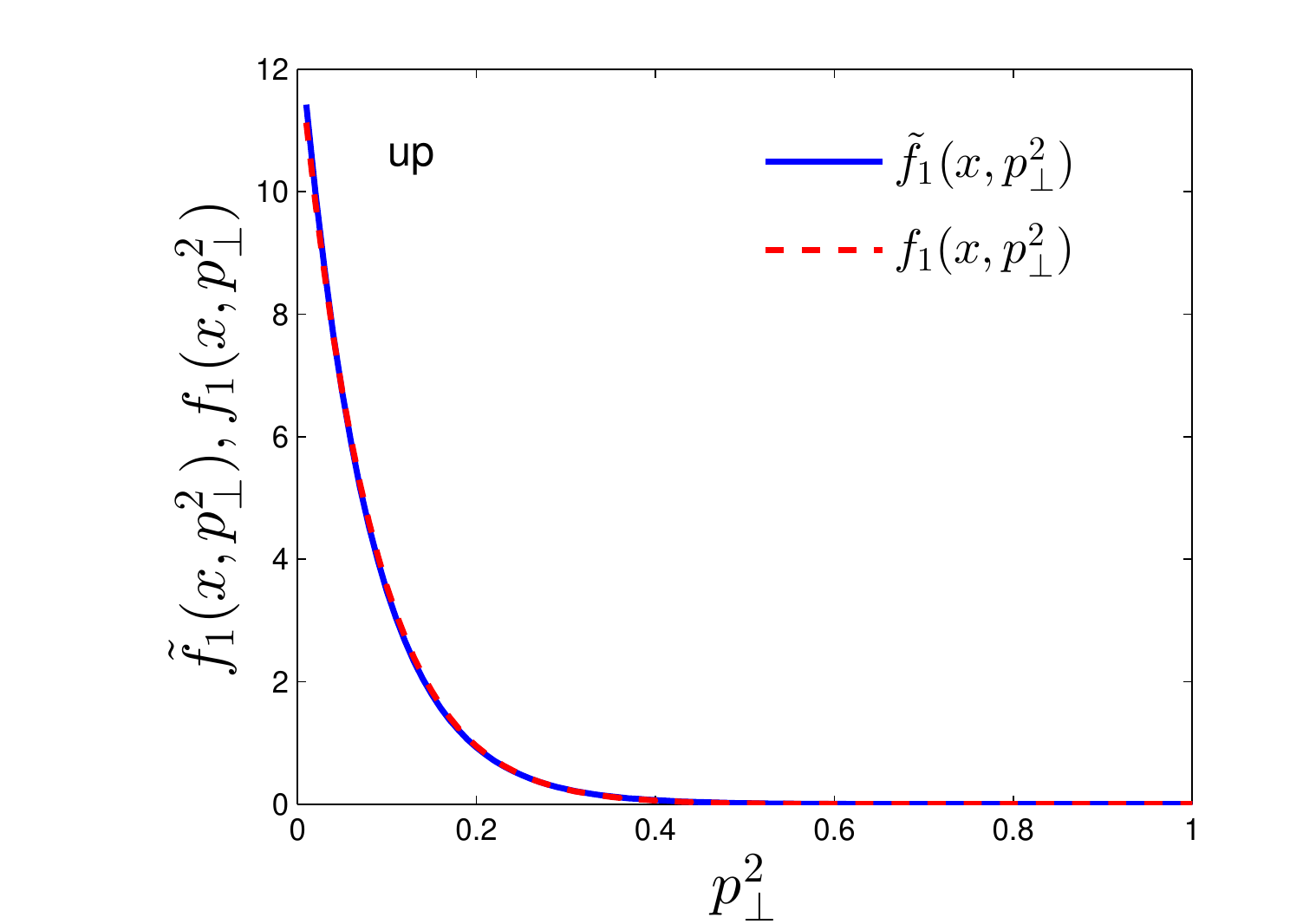}
\small{(a)}\includegraphics[width=7.5cm,clip]{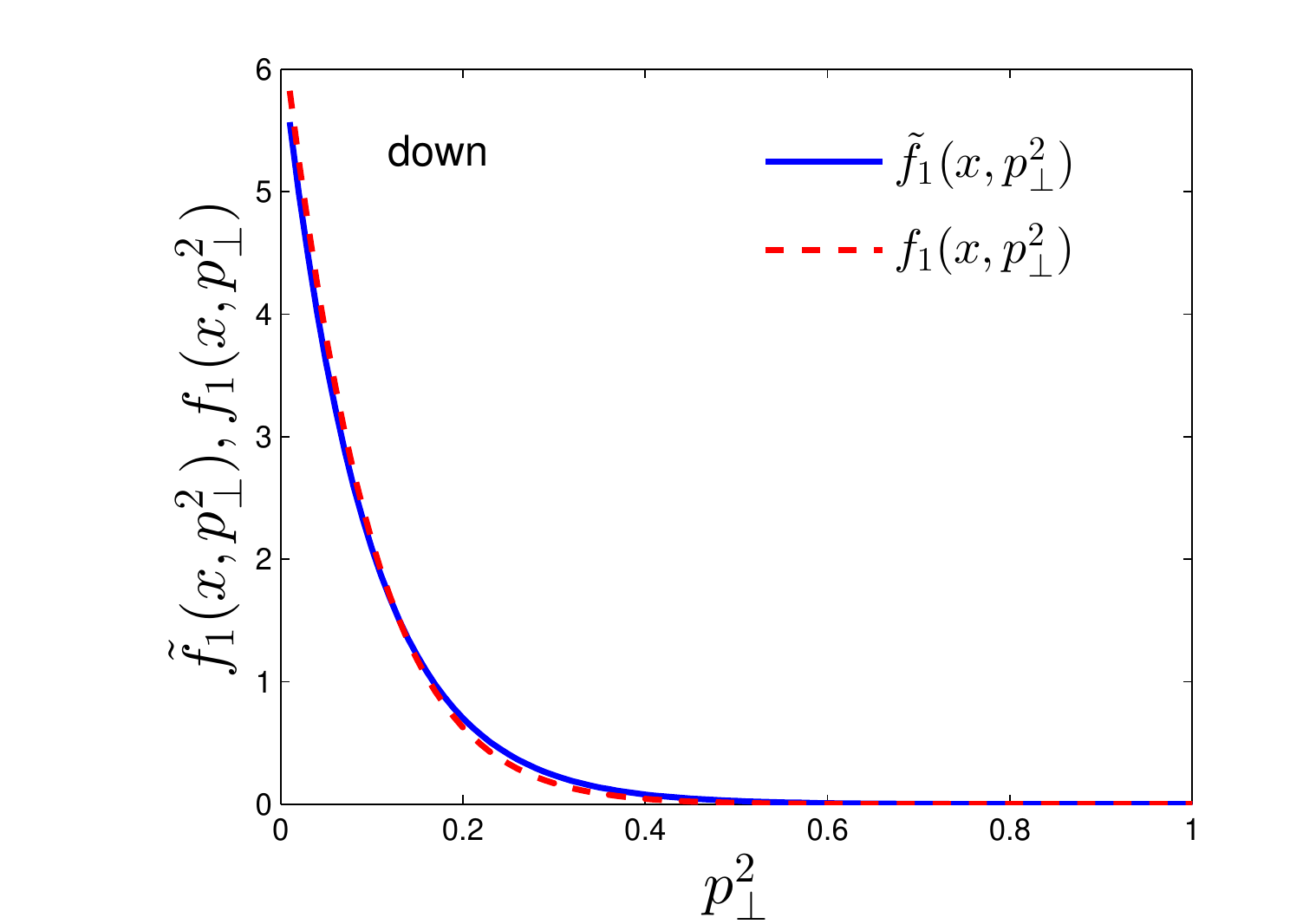}
\end{minipage}
\caption{\label{fig_facto} $x-\bfp^2$ factorization check: $f_1(x,\bfp^2)$ is from Eq.(\ref{TMD_f1}) and $\tilde{f}_1(x,\bfp^2)$ is from Eq.(\ref{f1_tilde}) for $u$ and $d$ quarks at $x=0.3$.} 
\end{figure}

A $x-\bfp^2$ factorization in TMDs are assumed in many places e,g. phenomenological extraction\cite{Anselmino12}, Lattice QCD\cite{Hagler09,Musch09} etc. 
In the Gaussian ansatz the unpolarized TMDs are written as
\be 
\tilde{f}^\nu_1(x,\bfp^2)=f^\nu_1(x) \frac{e^{-\bfp^2/\langle\bfp^2(f_1)\rangle^\nu}}{\pi \langle\bfp^2(f_1)\rangle^\nu}. \label{f1_tilde}
\ee
Where, the averaged $\bfp^2$ is defined as 
\be 
\langle\bfp^2(f_1)\rangle^\nu= \frac{\int dx\int d^2p_\perp p^2_\perp f_1(x,\bfp^2)}{\int dx\int d^2p_\perp f^\nu_1(x,\bfp^2)}
\ee 
To check whether our results satisfy $x-\bfp^2$ factorization in TMDs, we compare $\tilde{f}_1(x,\bfp^2)$ and  $f_1(x,\bfp^2)$(Eq.\ref{TMD_f1}), as shown in Fig.\ref{fig_facto}.  The agreement of these two results  shows that though the $x-\bfp^2$ factorization is not explicit in our model, but  numerically the factorization holds. 

\section{Relations}\label{rel}
It is interesting to study interrelations among the T-even TMDs at the leading twist.
The the transversity TMD have a uppercut specified by the unpolarized TMD and helicity TMD as  
\be 
| {h}^{\nu  }_1(x,\textbf{p}^2_{\perp})| &<& \frac{1}{2} \bigg| {f}^{\nu  }_1(x,\textbf{p}^2_{\perp}) + {g}^{\nu  }_{1L}(x,\textbf{p}^2_{\perp}) \bigg|.\label{soffer_bound}
\ee
This represents the Soffer bound \cite{soffer} for TMDs. 
The leading twist TMDs in this model also satisfy the inequality relations which are valid in QCD and all models\cite{bounds,bag}:
\be
{f}^{\nu  }_1(x,\textbf{p}^2_{\perp}) & > & 0 ,\\
\mid {g}^{\nu  }_{1L}(x,\textbf{p}^2_{\perp})\mid & < & \mid {f}^{\nu  }_1(x,\textbf{p}^2_{\perp})\mid.
\ee
Other inequalities are 
\be
\frac{\bfp}{2M^2}| {h}^{\nu \perp }_{1T}(x,\textbf{p}^2_{\perp})| &<& \frac{1}{2} \bigg| {f}^{\nu  }_1(x,\textbf{p}^2_{\perp}) - {g}^{\nu  }_{1L}(x,\textbf{p}^2_{\perp}) \bigg|,\\
|f^{\nu  }_{1}(x,\textbf{p}^2_{\perp})| &>&   |h^{\nu}_{1}(x,\textbf{p}^2_{\perp})|,\\ 
|f^{\nu  }_{1}(x,\textbf{p}^2_{\perp})| &>&   |h^{\nu}_{1T}(x,\textbf{p}^2_{\perp})|.
\ee 
The above relations are consistent with the relations found in other models like Bag model\cite{bag}, LCCQM  and are proved to be generic for diquark models\cite{Lorce11}. Note that all the  relations listed above are independent of the parameters of our model.

 In this model, we observe a generic relation between TMDs and GPDs 
\be 
\frac{\partial}{\partial |t|}\ln[\rm{GPD}^\nu(x,t)] = \frac{(1-x)^2}{4}\frac{\partial}{\partial p^2_\perp}\ln[\rm{TMD}^\nu(x,\bfp^2)],
\ee
as found in quark-scalar-diquark model\cite{MC_rel}. Where $\rm{GPD}^\nu(x,t)$ represents the $H$ and $E$ GPDs and the $\rm{TMD}^\nu(x,\bfp^2)$ stands for all the leading twist T-even TMDs. The $\bfp^2$ and $|t|$ are treated in same footing. We observe that contribution of vector diquark does not effect the relation. 
A detail discussion on GPDs in this model are given in \cite{GPD_LFQDM}. Note that the above equation is not exact for all the distributions, however the deviation is found to be negligible at high transverse momentum.

\section{Quark densities}\label{densities}
\begin{figure}[htbp]
\begin{minipage}[c]{0.98\textwidth}
\small{(a)}\includegraphics[width=7.5cm,clip]{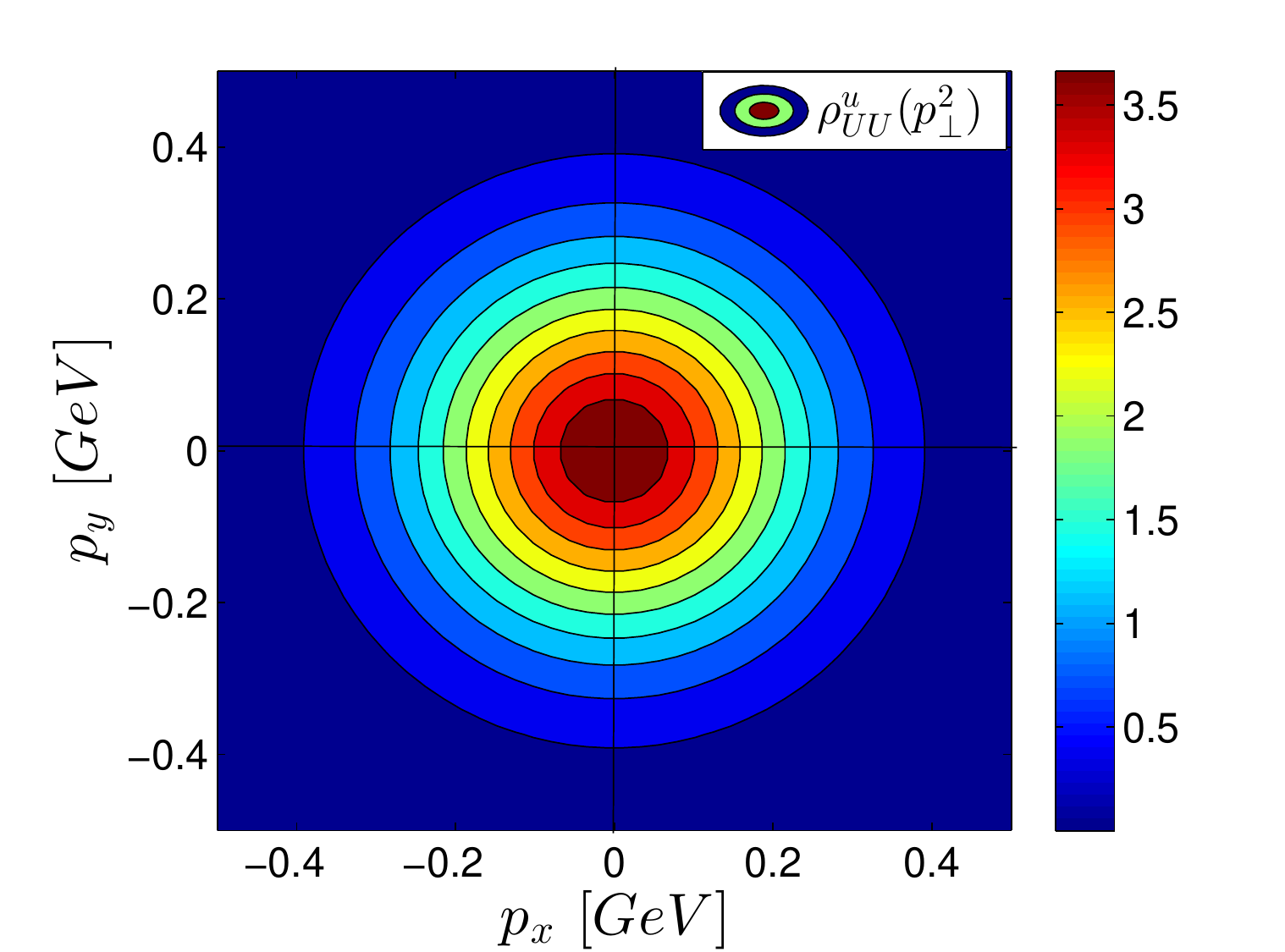}
\small{(a)}\includegraphics[width=7.5cm,clip]{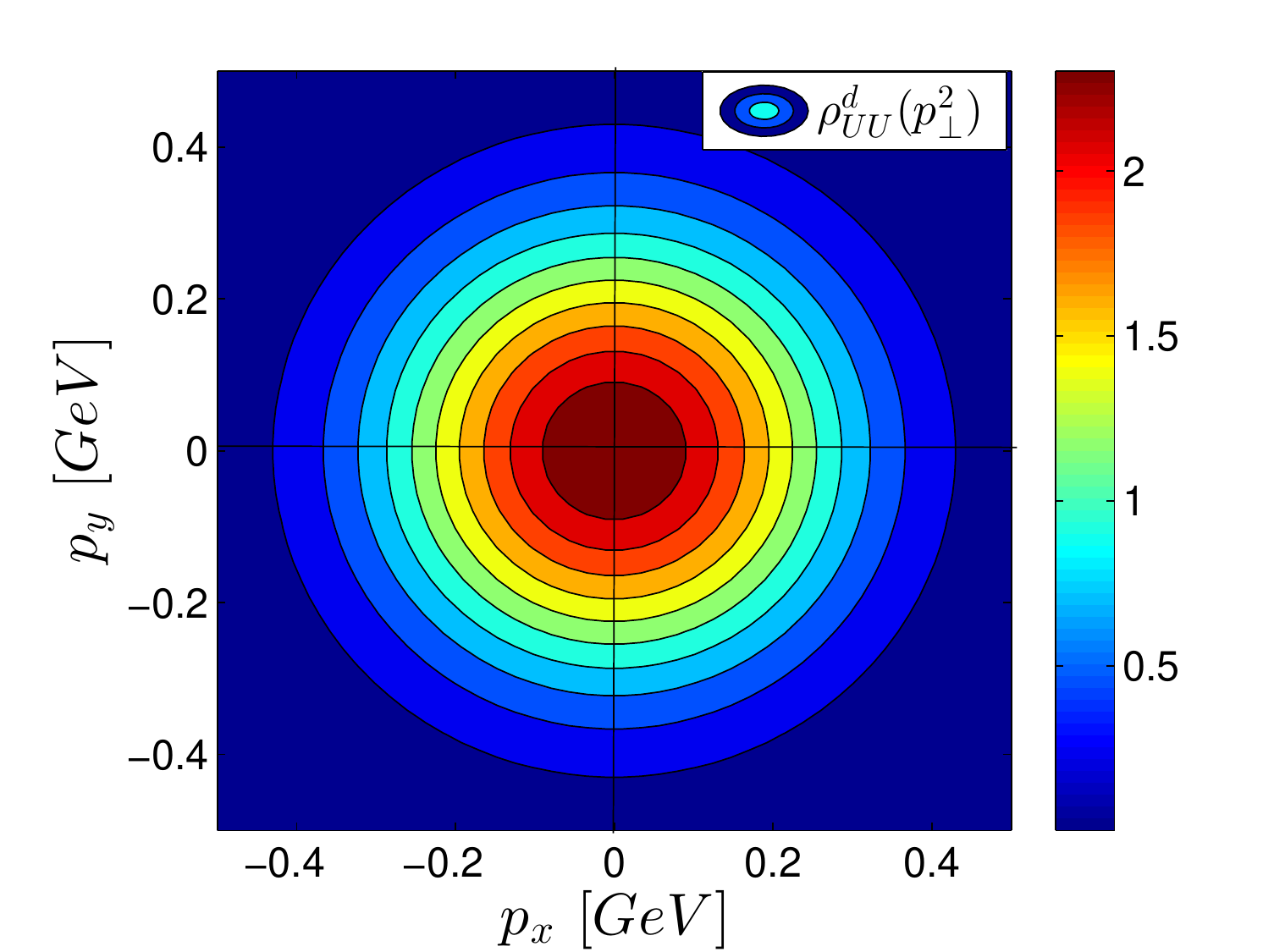}
\end{minipage}
\begin{minipage}[c]{0.98\textwidth}
\small{(c)}\includegraphics[width=7.5cm,clip]{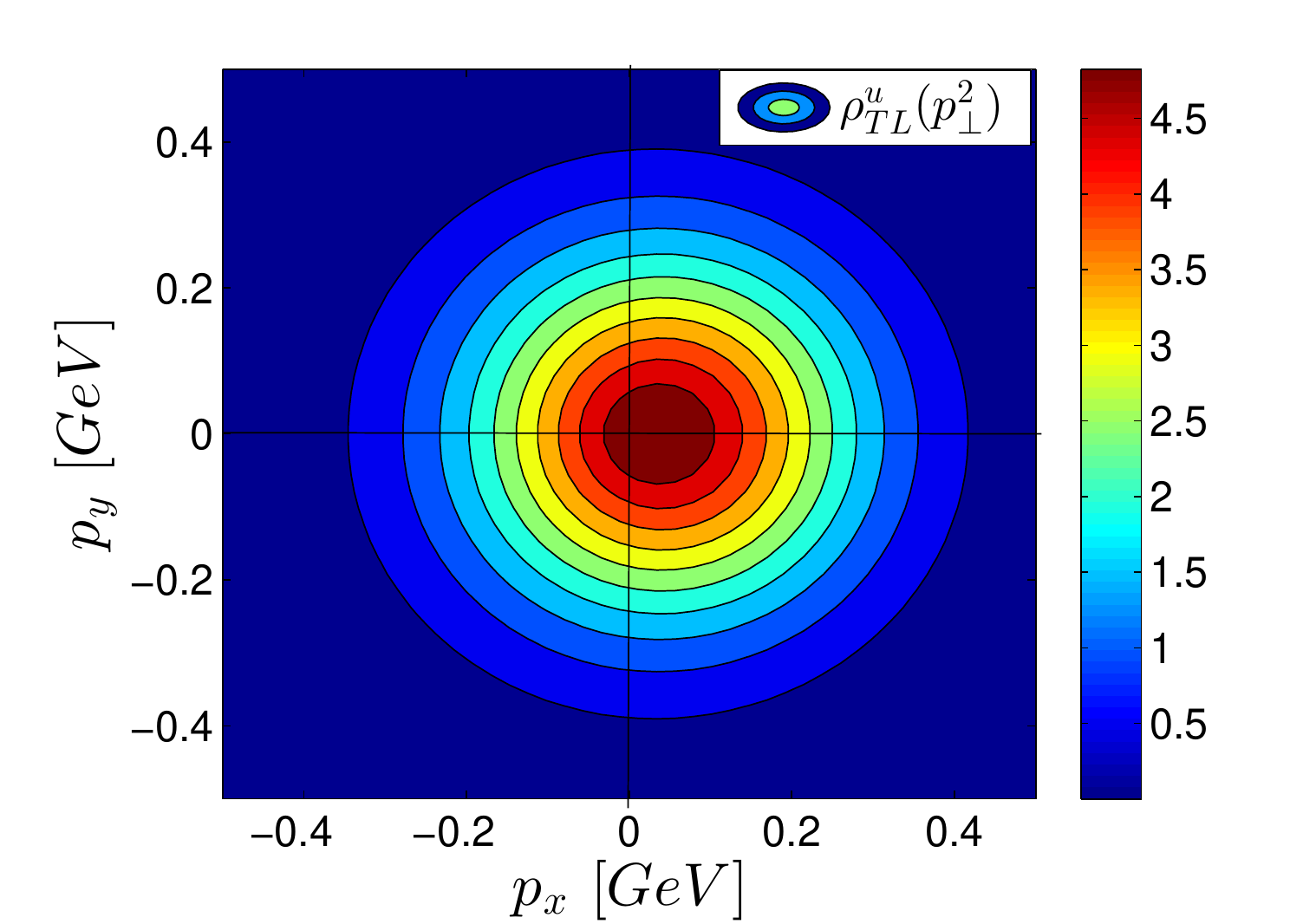}
\small{(d)}\includegraphics[width=7.5cm,clip]{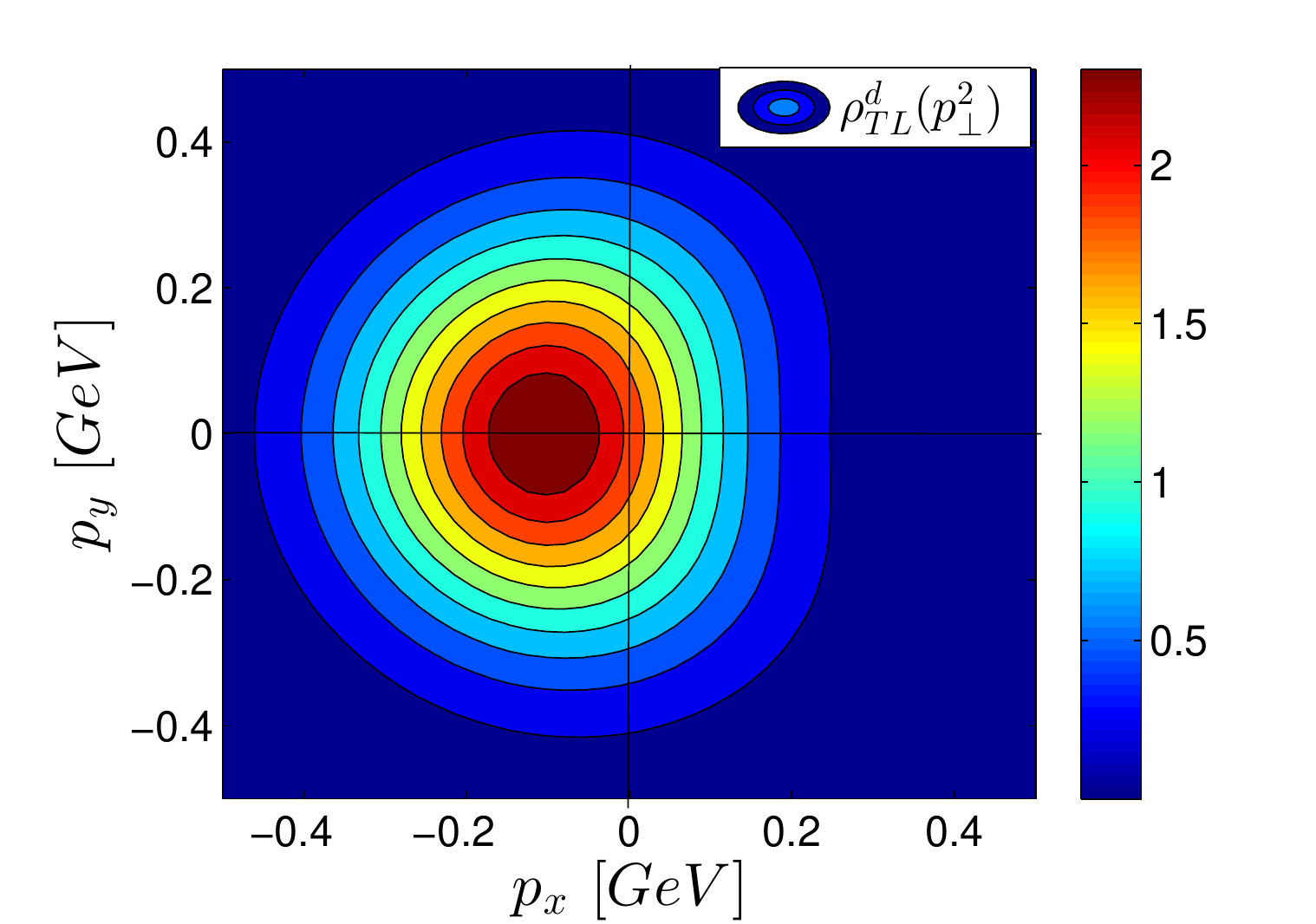}
\end{minipage}
\caption{\label{fig_rho} Quark densities $\rho^\nu_{UU}(\bfp^2)$ and $\rho^\nu_{TL}(\bfp^2)$ for u(left panel) and d(right panel) quarks.} 
\end{figure}

The TMDs can be interpreted as the quark densities inside a proton as:
\be
\rho^\nu_{UU}(\textbf{p}_\perp) &=& f^{\nu(1)}_1(\textbf{p}^2_\perp),\\ \label{rhoUU}
\rho^\nu_{TL}(\textbf{p}_\perp;\textbf{S}_\perp,\lambda) &=& \frac{1}{2} f^{\nu(1)}_1(\textbf{p}^2_\perp)+\frac{\lambda}{2}\frac{\textbf{p}_\perp.\textbf{S}_\perp}{M} g^{\nu(1)}_{1T}(\textbf{p}^2_\perp).\label{rhoTL}
\ee 
Where $f^{\nu(1)}_1(\textbf{p}^2_\perp)=\int dx f^{\nu}_1(x,\bfp^2)$ and the subscripts $XY$ in $\rho^\nu_{XY}(\textbf{p}_\perp)$ represent the proton polarization($X$) and quarks polarization($Y$) respectively i.e.,  $\rho^\nu_{UU}(\textbf{p}_\perp)$ gives the quark density when  both nucleon and quark  are unpolarized and  $\rho_{TL}(\textbf{p}_\perp)$ is the density of longitudinally polarized quark in a transversely polarized nucleon.  The quark densities in the transverse momentum plane are shown in Fig.(\ref{fig_rho})(a),(b) for u and d quarks. In this model the unpolarized distributions $\rho^\nu_{UU}(\textbf{p}_\perp)$ are circularly symmetric for both u and d quarks. Our result is consistent with Lattice data in \cite{Hagler09,Musch09}. 

For transversely polarized proton the quark densities $\rho^\nu_{TL}(\textbf{p}_\perp;\textbf{S}_\perp,\lambda)$, shown in Fig.\ref{fig_rho}(c),(d), are no longer axially symmetric for u and d quarks. This distortion is due  to  the non-zero values of $g^{(1)}_{1T}(x,\textbf{p}^2_\perp)$. Since the distribution $g_{1T}$ changes sign, shown in Fig.\ref{fig_TMD2}(a),(b), we find a sift towards the positive $p_x$ for u quark and towards the negative $p_x$ for d quark.  We consider the  quark spin pointing along $z$ direction and proton is polarized in transverse x-direction, $\textbf{S}_\perp = (1,0)$. The shift is larger for d quark, $\langle \textbf{p}_x\rangle_{\rho_{TL}} \approx - 105 ~ MeV $, compare to u quark $\langle \textbf{p}_x\rangle_{\rho_{TL}} \approx + 35~ MeV$.
 The deformation in $\rho_{TL}$ indicates that the transversely polarized nucleon is non-spherical, and the $u$ and $d$ quarks have opposite directional distributions.

\section{TMD evolutions}\label{evol}
\begin{figure}[htbp]
\begin{minipage}[c]{0.98\textwidth}
\small{(a)}\includegraphics[width=7.5cm,clip]{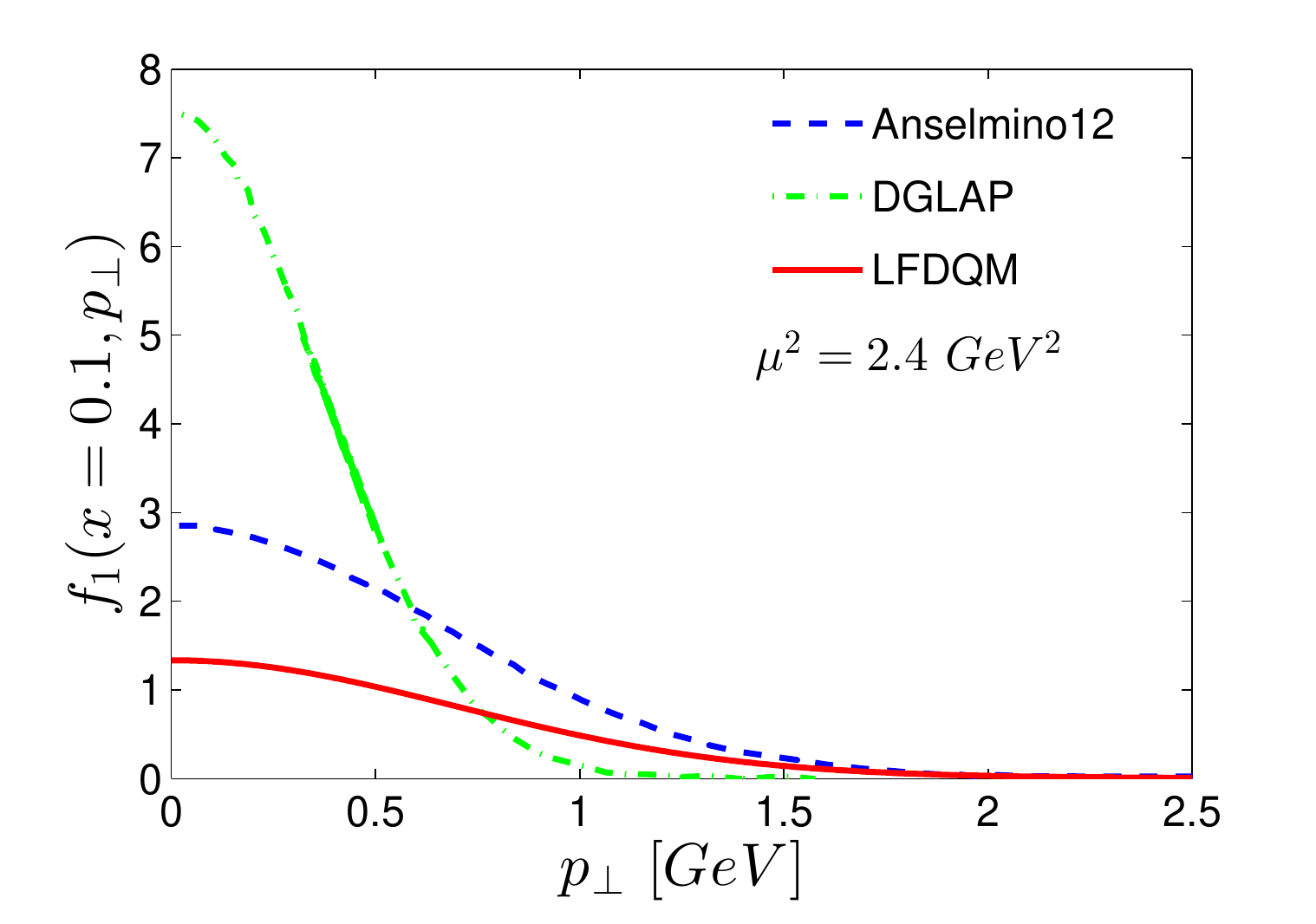}
\small{(b)}\includegraphics[width=7.5cm,clip]{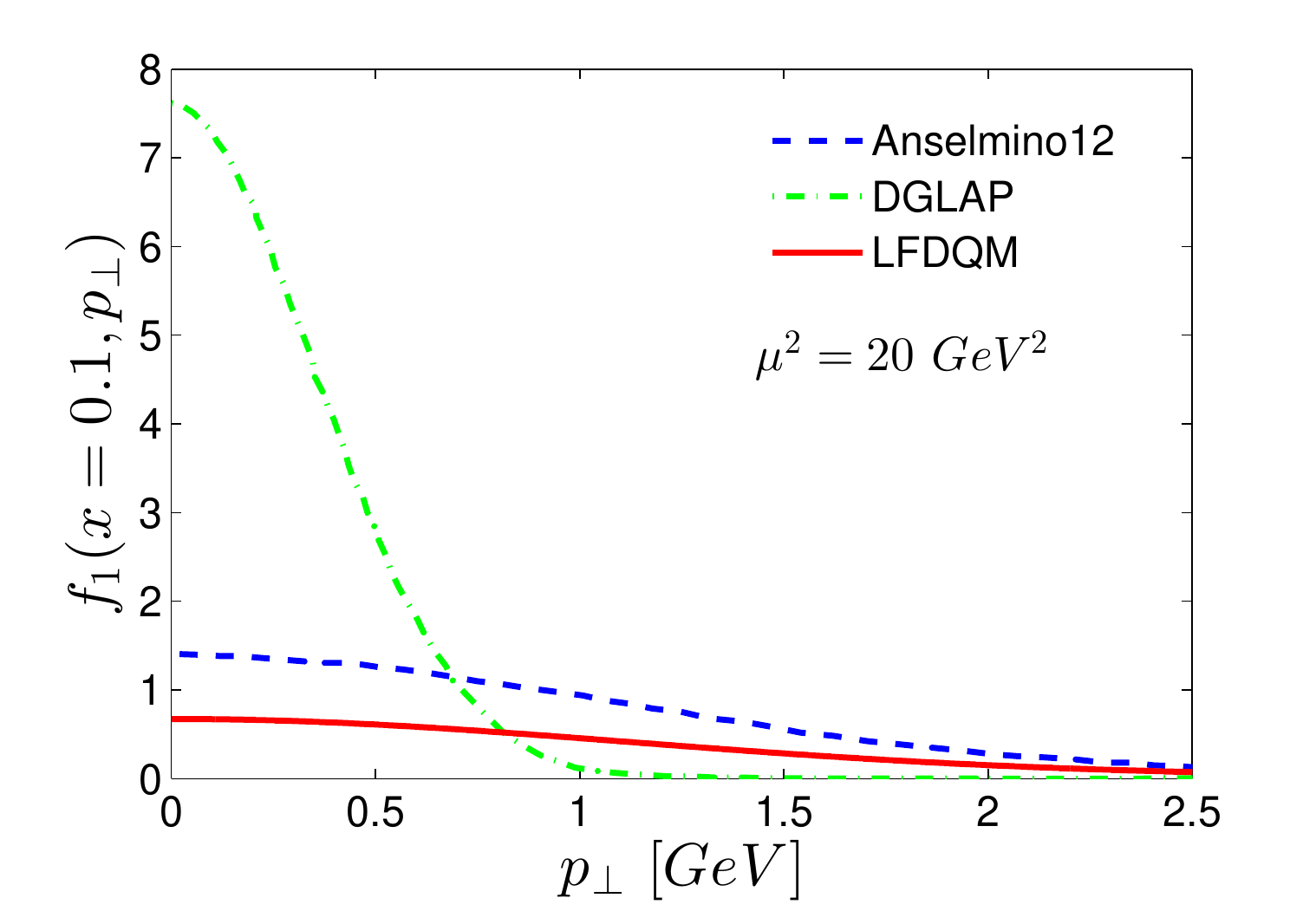}
\end{minipage}
\caption{\label{fig_f1evol} The unpolarized TMDs at $\mu^2=2.4~GeV^2$(average $\mu^2$ value for HERMES experiment) and at $\mu^2=20~GeV^2$(average $\mu^2$ value for COMPASS experiment) for u quark. A comparison is shown with DGLAP(green dash-doted line) and Anselmino\cite{Anselmino12}(blue dashed line). } 
\end{figure}

The scale evolution of TMDs in the coordinate space is defined\cite{Aybat11,Anselmino12} as
\be 
\tilde{F}(x,\bfb ;\mu)=\tilde{F}(x,\bfb ;\mu_0) \exp\bigg(\ln\frac{\mu}{\mu_0}\tilde{K}(b_\perp;\mu)+\int^\mu_{\mu_0}\frac{d\mu'}{\mu'}\gamma_F\big(\mu',\frac{\mu^2}{\mu'^2}\big)\bigg).
\ee
Where the $\tilde{F}(x,\bfb ;\mu)$ represents the T-even TMDs at scale $\mu$. $\tilde{K}(b_\perp;\mu)$ is given by\cite{Aybat12} 
\be 
\tilde{K}(b_\perp;\mu)=\tilde{K}(b_*;\mu_b)+\bigg[\int^{\mu_b}_\mu \frac{d\mu'}{\mu'}\gamma_K(\mu')\bigg]-g_K(b_T),
\ee
with 
\be 
\tilde{K}(b_*;\mu_b)=-\frac{\alpha_s C_F}{\pi}[\ln(b_*^2\mu^2_b)-\ln(4)+2\gamma_E],\\
b_*(b_T)=\frac{b_T}{\sqrt{1+\frac{b^2_T}{b^2_{max}}}} ~~~ ; ~~~ \mu_b=\frac{C_1}{b_*(b_T)} 
\ee
at $\mathcal{O}(\alpha_s)$\cite{Collins82,Collins85}.
$C_1$ is a constant, we adopt a particular choice $C_1=2 e^{-\gamma_E}$\cite{Aybat11,Aybat12}, where $\gamma_E=0.577$ is the Euler constant\cite{Collins85}.
Thus the evolution equation can be written as 
\be 
\tilde{F}(x,\bfb ;\mu)=\tilde{F}(x,\bfb ;\mu_0)\tilde{R}(\mu,\mu_0,b_T)\exp \bigg[-g_K(b_T)\ln(\frac{\mu}{\mu_0})\bigg],
\ee
with the kernel
\be  
\tilde{R}(\mu,\mu_0,b_T)=\exp\bigg[\ln\frac{\mu}{\mu_0}\int^{\mu_b}_\mu \frac{d\mu'}{\mu'}\gamma_K(\mu')+\int^\mu_{\mu_0}\frac{d\mu'}{\mu'}\gamma_F\big(\mu',\frac{\mu^2}{\mu'^2}\big)\bigg].
\ee 
The anomalous dimensions are given by
\be  
\gamma_F\big(\mu',\frac{\mu^2}{\mu'^2}\big)&=&\alpha_s(\mu')\frac{C_F}{\pi}\bigg(\frac{3}{2}-\ln\frac{\mu^2}{\mu'^2}\bigg),\\
\gamma_K(\mu')&=&\alpha_s(\mu')\frac{C_F}{\pi}.
\ee
Therefore, taking Fourier transformation the evolution of TMDs in momentum space is written as
\be 
F(x,\bfp ;\mu)=\int \frac{d^2\bfb}{(2\pi)^2} e^{i \bfb.\bfp} \tilde{F}(x,\bfb ;\mu) \label{TMDevol}
\ee
 The scale evolution of unpolarized TMDs are shown in Fig.\ref{fig_f1evol} and compared with the results of Anselmino et.al. \cite{Anselmino12}.

\section{Integrated TMDs}\label{integrated}
\begin{figure}[htbp]
\begin{minipage}[c]{0.98\textwidth}
\includegraphics[width=7.5cm,height=5.cm]{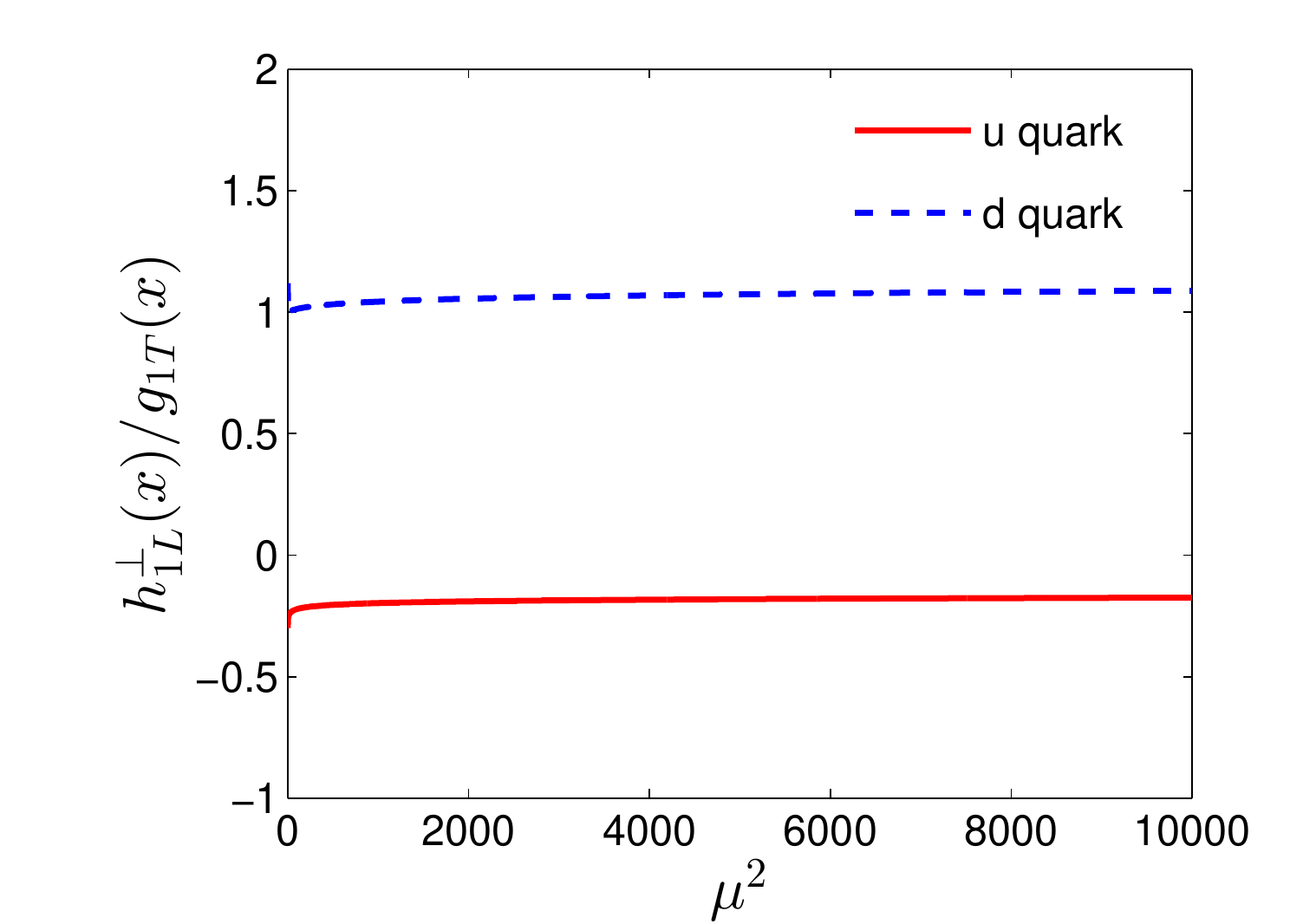}
\includegraphics[width=7.5cm,height=5.cm]{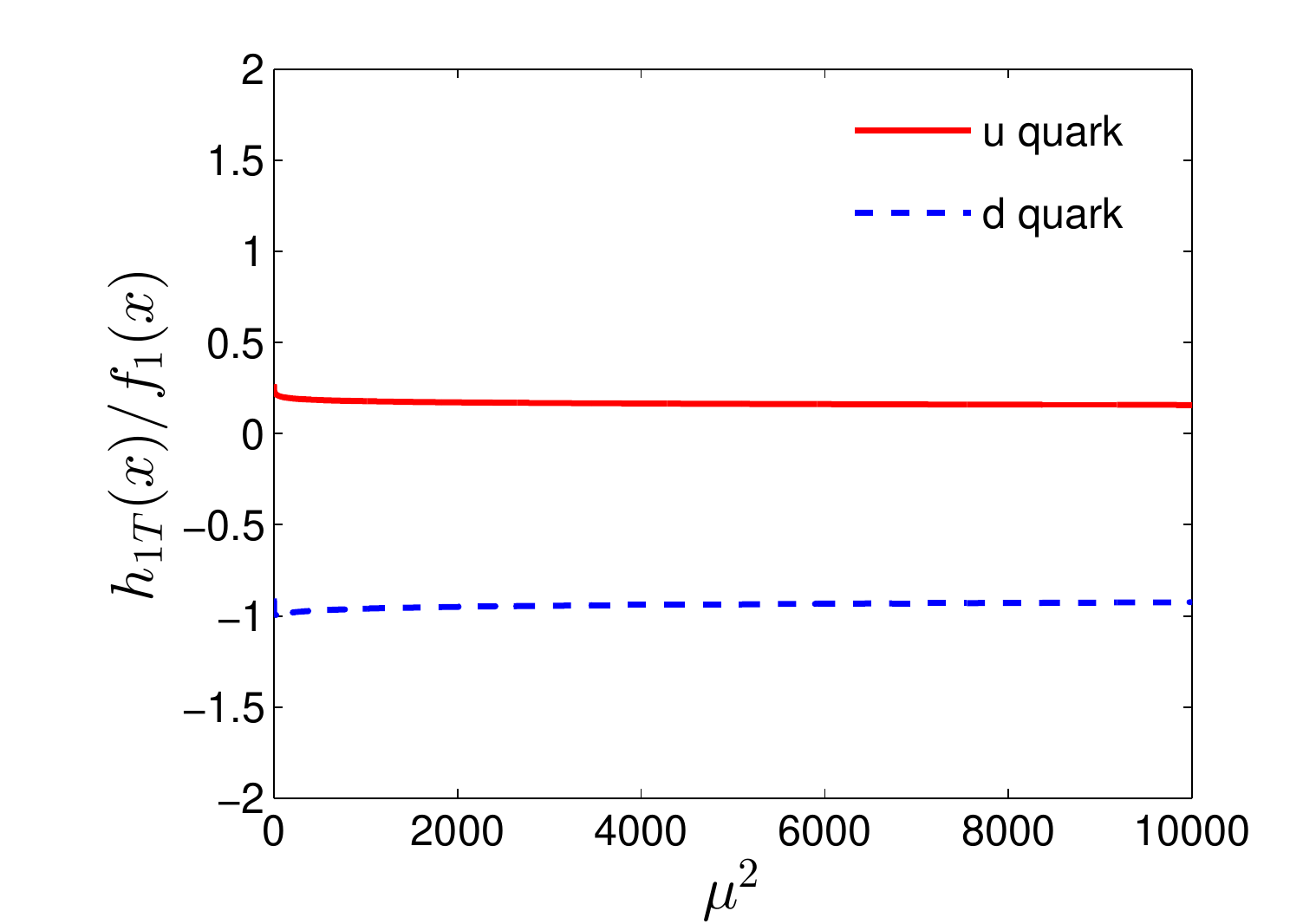}
\end{minipage}
\caption{\label{fig_ratioC} Scale evolution the ratios $\frac{g^{\nu}_{1T}(x)}{h^{\nu\perp}_{1L}(x)}$ and $\frac{h^{\nu}_{1T}(x)}{f^{\nu}_{1}(x)}$  are shown in (a) and (b)respectively.}
\end{figure}

The PDFs are found by integrating TMDs over transverse momentum $\bfp$. The PDF limit of the Eqs.(\ref{TMD_f1},\ref{TMD_g1L},\ref{TMD_h1}) give unpolarized PDF ($f^\nu_1(x)$), helicity distribution($g^\nu_1(x)$) and transversity distributions($h^\nu_1(x)$) respectively. At the leading twist, the integrated TMDs in this model read 
\be
f^{\nu}_1(x)&=&\bigg(C^2_SN^{\nu 2}_S+C^2_V\big(\frac{1}{3}N^{\nu 2}_0+\frac{2}{3}N^{\nu 2}_1\big)\bigg)(1-x)^2\frac{1}{\delta^\nu}\bigg[T^\nu_1(x) + \frac{1}{ M^2 R^\nu(x)}T^\nu_2(x) \bigg],\label{PDF_f1}\\
g^{\nu  }_1(x)&=&\bigg(C^2_SN^{\nu 2}_S+C^2_V\big(\frac{1}{3}N^{\nu 2}_0-\frac{2}{3}N^{\nu 2}_1\big)\bigg)(1-x)^2\frac{1}{\delta^\nu}\bigg[T^\nu_1(x) - \frac{1}{ M^2 R^\nu(x)}T^\nu_2(x) \bigg],\label{PDF_g1} \\
h^{\nu  }_1(x)&=&\bigg(C^2_SN^{\nu 2}_S-C^2_V\frac{1}{3}N^{\nu 2}_0\bigg)(1-x)^2\frac{1}{\delta^\nu} T^\nu_1(x), \label{PDF_h1}\\
g^{\nu}_{1T}(x)&=&\bigg(C^2_SN^{\nu 2}_S-C^2_V\frac{1}{3}N^{\nu 2}_0\bigg)(1-x)^2\frac{1}{\delta^\nu} 2 T^\nu_3(x),\label{PDF_g1T}\\
{h}^{\nu\perp  }_{1L}(x)&=& -\bigg(C^2_SN^{\nu 2}_S+C^2_V\big(\frac{1}{3}N^{\nu 2}_0 - \frac{2}{3}N^{\nu 2}_1\big)\bigg)(1-x)^2\frac{1}{\delta^\nu} 2 T^\nu_3(x),\label{PDF_h1Lp}\\
h^{\nu  }_{1T}(x)&=& \bigg(C^2_SN^{\nu 2}_S-C^2_V\frac{1}{3}N^{\nu 2}_0\bigg) (1-x)^2\frac{1}{\delta^\nu}\bigg[T^\nu_1(x) + \frac{1}{ M^2 R^\nu(x)}T^\nu_2(x) \bigg],\label{PDF_h1T}\\
{h}^{\nu\perp}_{1T}(x)&=& - \bigg(C^2_SN^{\nu 2}_S-C^2_V\frac{1}{3}N^{\nu 2}_0\bigg) (1-x)^2\frac{1}{\delta^\nu} 2 T^\nu_2(x).\label{PDF_h1Tp}
\ee
The distributions $f^\nu_1(x),~g^\nu_1(x)$ and $h^\nu_1(x)$ are discussed in \cite{TM_VD} in this model. Therefore, here we concentrate on the other four integrated TMDs i.e, $g^{\nu}_{1T}(x),h^{\nu\perp}_{1L}(x),h^{\nu  }_{1T}(x)$ and ${h}^{\nu\perp}_{1T}(x)$. 
Also we calculate the transverse moment of a distributions($f$) as 
\be 
f^{(1)\nu}=\int d^2\bfp \frac{\bfp^2}{2M^2}f^\nu(x,\bfp^2).
\ee
\begin{figure}[htbp]
\begin{minipage}[c]{0.98\textwidth}
\small{(a)}\includegraphics[width=7.5cm,height=5.cm]{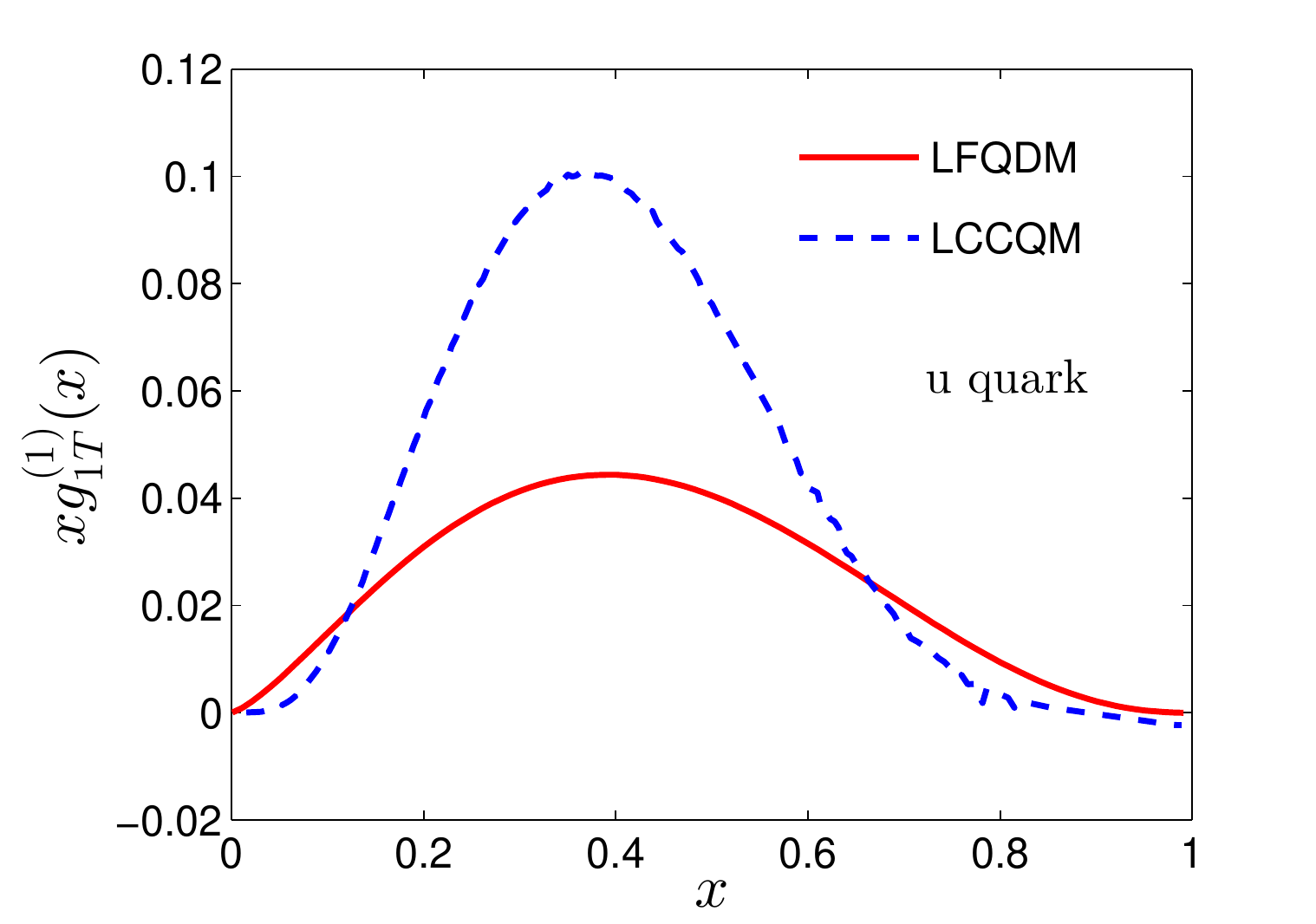}
\small{(b)}\includegraphics[width=7.5cm,height=5.cm]{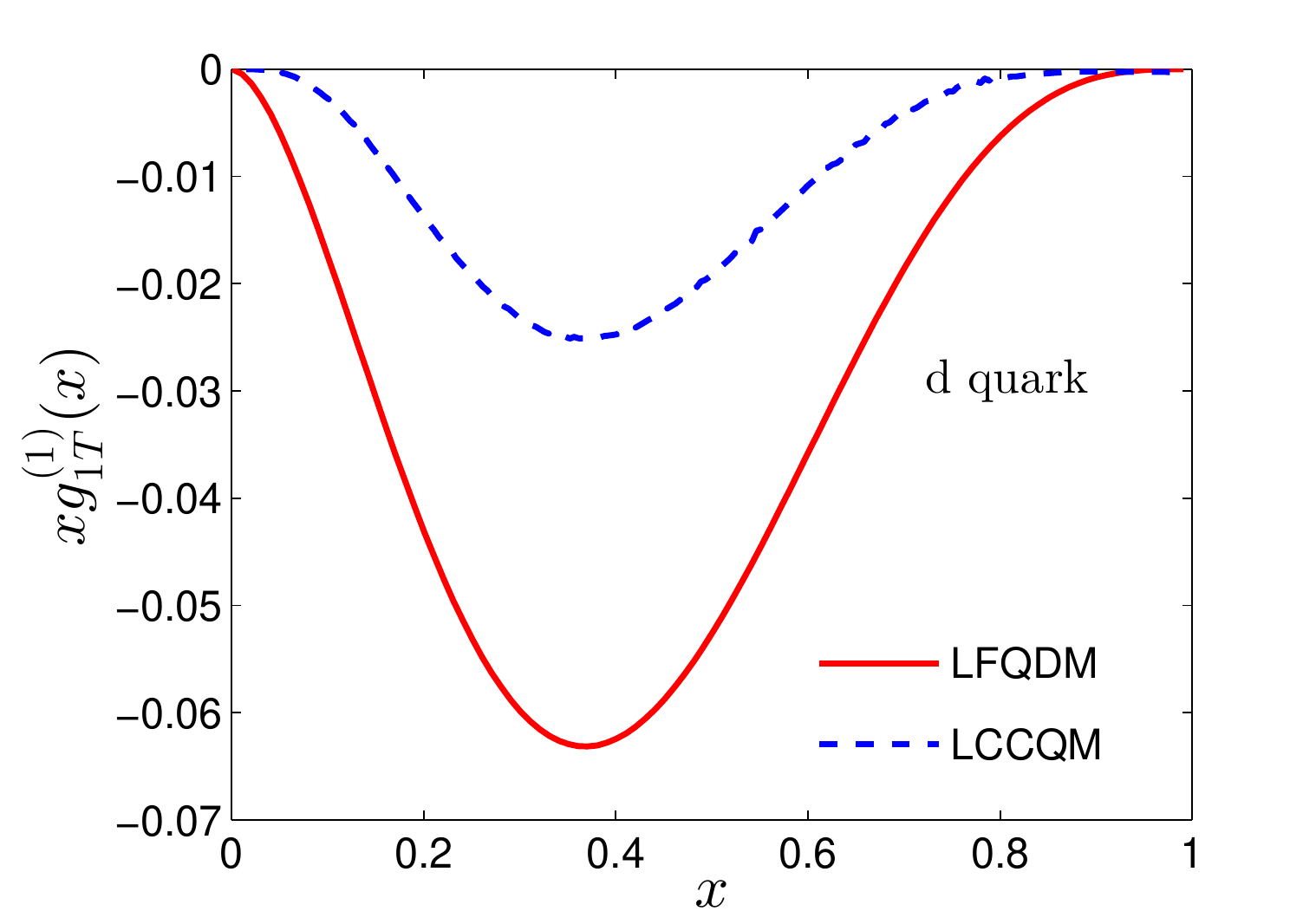}
\end{minipage}
\caption{\label{fig_PDFg1T}  The transverse moment $g^{(1)\nu}_{1T}(x)$ are shown for $u$ and $d$ quarks. The red continuous line represents our result(LFQDM) and dashed blue line is the result from light-cone constituent quark model(LCCQM)\cite{Pasquini09}.}
\end{figure}
The distribution $g^{\nu}_{1T}(x)$ is found for longitudinal quark in a transversely polarized proton. Its transverse moment $g^{(1)\nu}_{1T}(x)$ multiplied by $x$ is shown in Fig.\ref{fig_PDFg1T} at the initial scale. We observe a positive distribution for $u$ quarks and negative for $d$ quarks as found in light-cone constituent quark model(LCCQM)\cite{Pasquini09}. The difference in the magnitudes  of the distributions are expected as results in  two models are evaluated  at two different scales.
\begin{figure}[htbp]
\begin{minipage}[c]{0.98\textwidth}
\small{(a)}\includegraphics[width=7.5cm,height=5.cm]{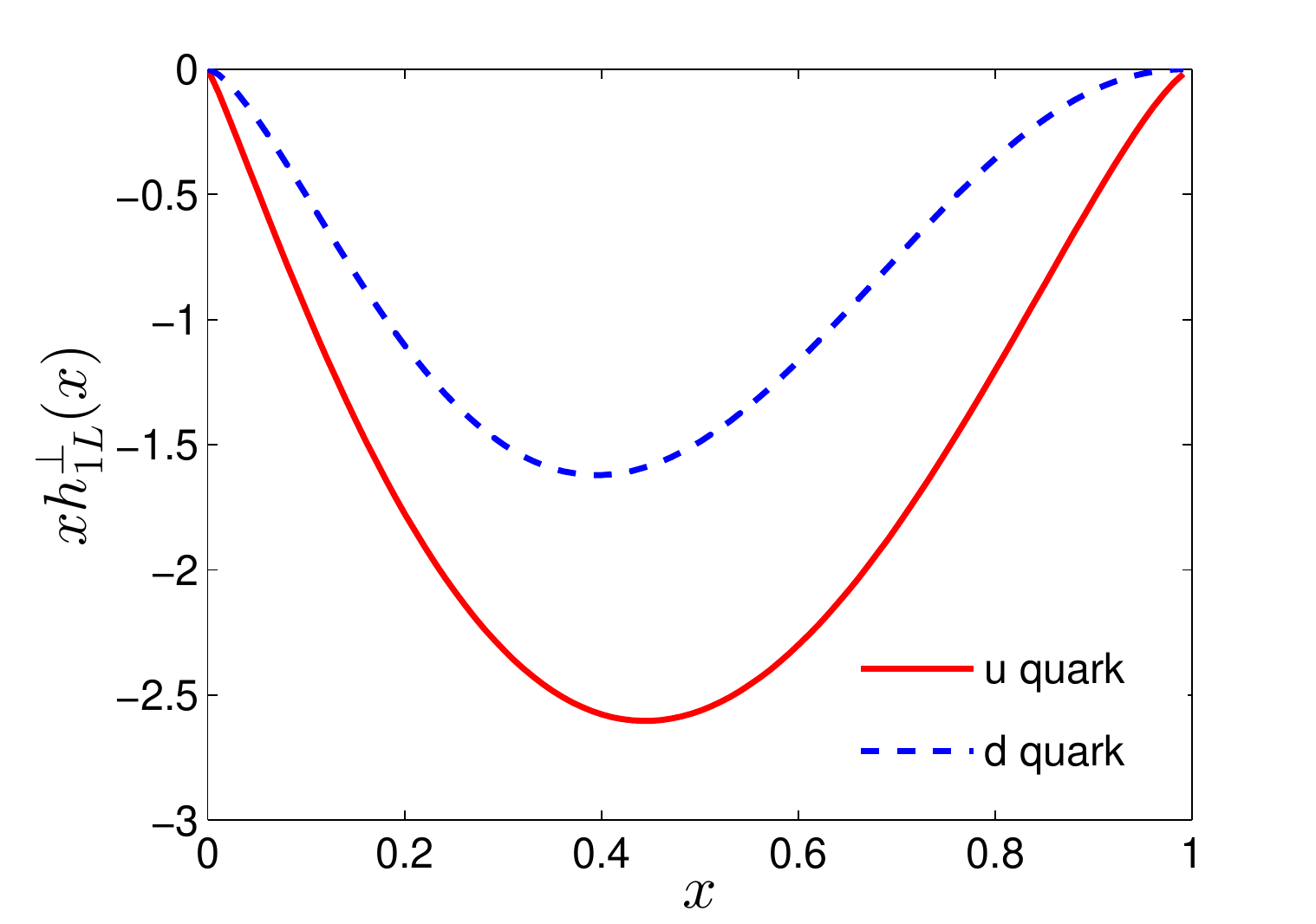}
\small{(b)}\includegraphics[width=7.5cm,height=5.cm]{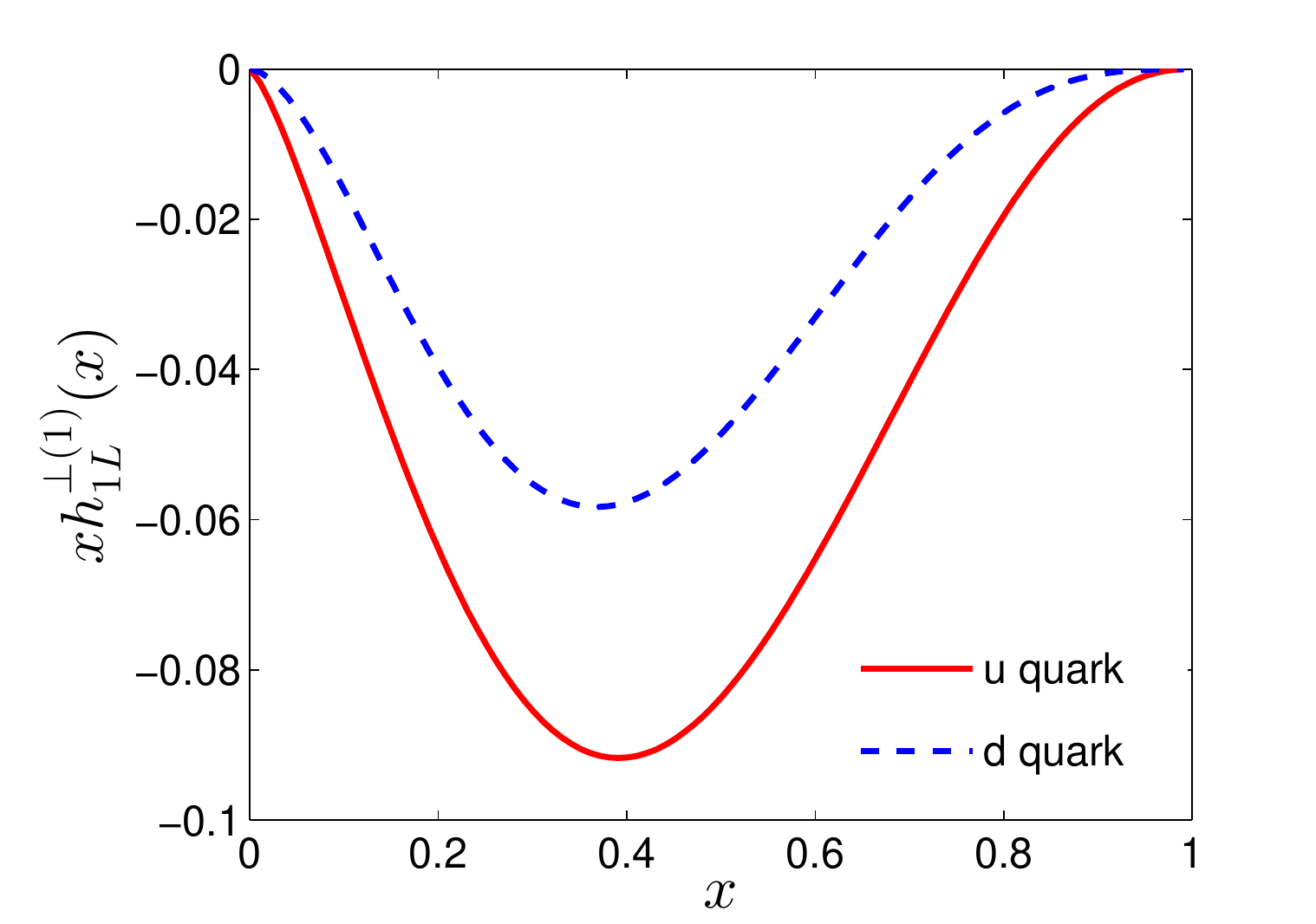}
\end{minipage}
\caption{\label{fig_PDFh1Lp}  $xh^{\perp\nu}_{1L}(x)$ and $xh^{(1)\nu}_{1L}(x)$ are shown for $u$ and $d$ quarks.}
\end{figure}
 The variation of $h^{\nu\perp}_{1L}(x)$ and its transverse moments with $x$ are shown in Fig.\ref{fig_PDFh1Lp}. A negative distribution is found for both the $u$ and $d$ quarks. Whereas a sign flip is observed in LCCQM. In this model the sign of the distribution is provided by the prefactor of that distribution and in Eq.(\ref{PDF_h1Lp}) the prefactor remains negative for $d$ quarks.  From Eq.(\ref{PDF_g1T}) and (\ref{PDF_h1Lp}) we notice $h^{\nu\perp}_{1L}(x)$ and $g^{\nu}_{1T}(x)$ differ from each other by the pre-factors only. Thus, in this model the ratio $\frac{g^{\nu}_{1T}(x)}{h^{\nu\perp}_{1L}(x)}=\mathcal{C}^\nu$ 
 depends on scale and flavour. We found $\mathcal{C}^u<0$ and $\mathcal{C}^d>0$. This ratio is almost constant at the higher scales as shown in Fig.\ref{fig_ratioC}(a). We perform the scale evolution using the model parameterization discussed in \cite{TM_VD}(also see Appendix-\ref{AppA}).
\begin{figure}[htbp]
\begin{minipage}[c]{0.98\textwidth}
\small{(a)}\includegraphics[width=7.5cm,height=5.cm]{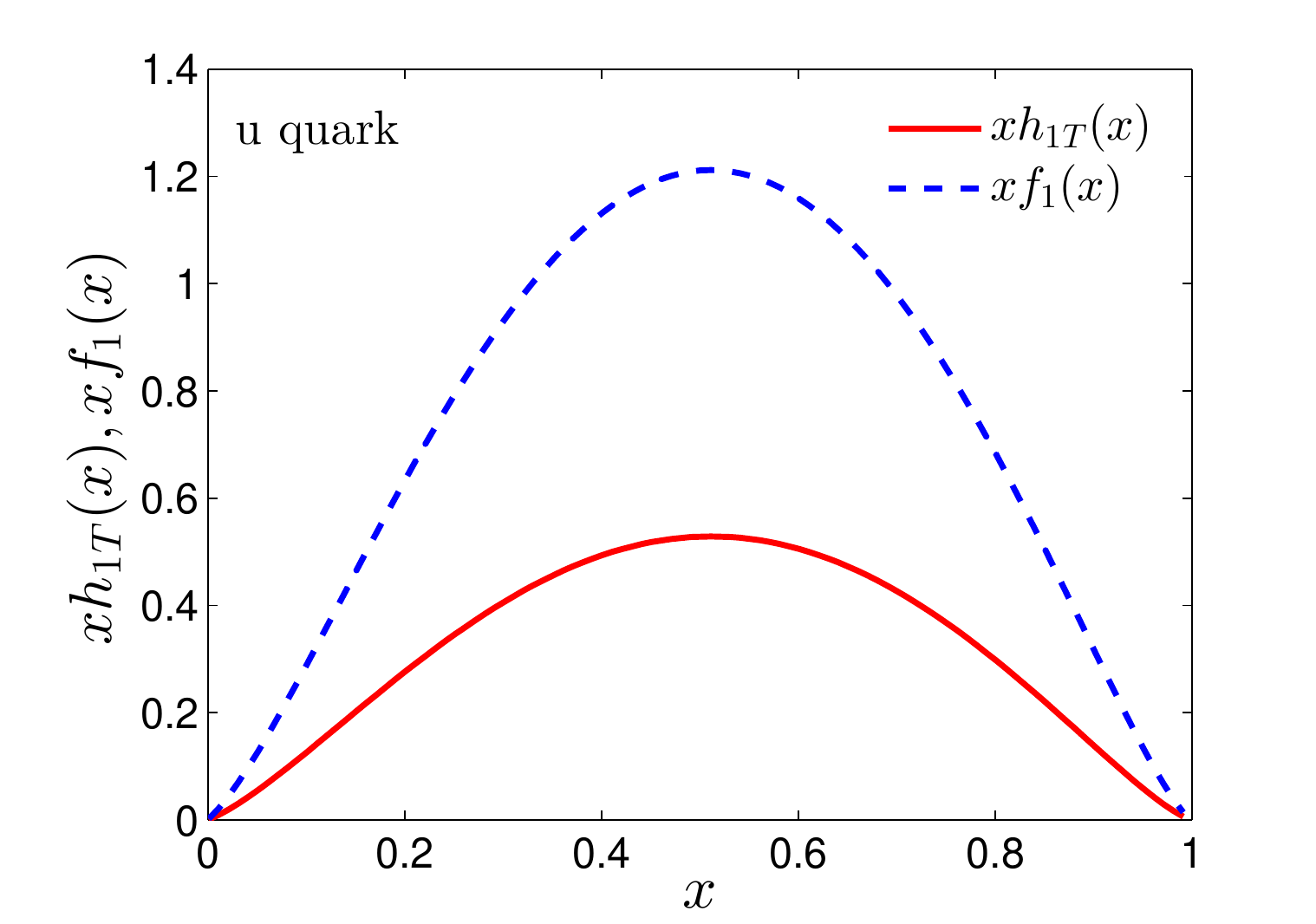}
\small{(b)}\includegraphics[width=7.5cm,height=5.cm]{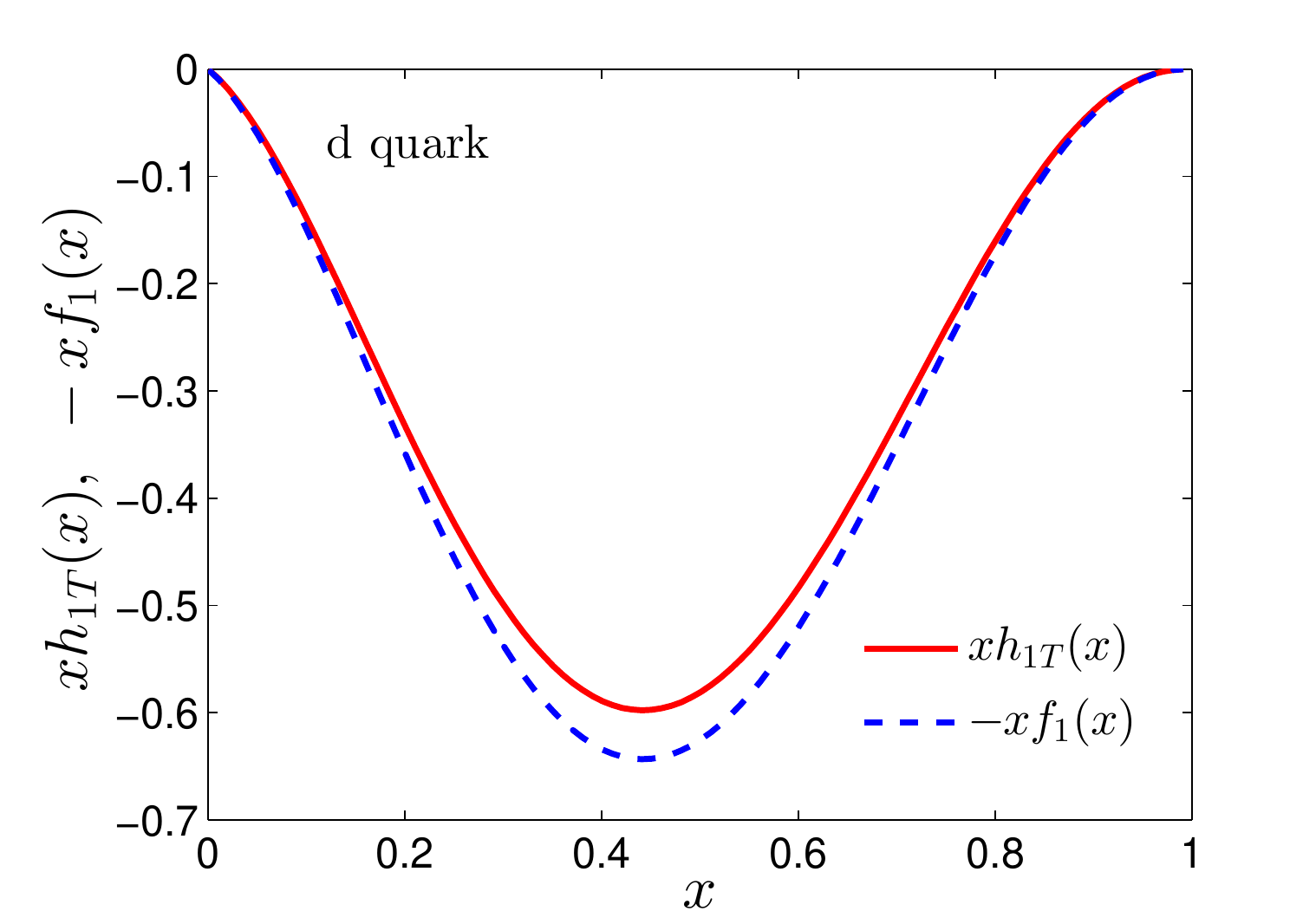}
\end{minipage}
\caption{\label{fig_PDFh1T}  The $x$ multiplied $h^{\nu}_{1T}(x)$ and $f^\nu_1(x)$ are shown for $u$ and $d$ quarks.}
\end{figure}
The distribution $h^{\nu}_{1T}(x)$ is shown in Fig.\ref{fig_PDFh1T} and compare with unpolarized PDF. Here $|h^{\nu}_{1T}(x)|<|f^{\nu}_{1}(x)|$. Again from Eq.(\ref{PDF_f1}) and (\ref{PDF_h1T}) we observe that 
 the $x$ dependent
  functional form of $h^{\nu}_{1T}(x)$ is the same as of $f^{\nu}_{1}(x)$, the only difference is in the  pre-factors. Thus the ratios $h^{\nu}_{1T}(x)/f^{\nu}_{1}(x)=\mathcal{C'}^\nu$ behaves like a constant at higher scales as shown in Fig.\ref{fig_ratioC}(b). The model predicts that  $\mathcal{C'}^u>0$ and $\mathcal{C'}^d<0$. 
\begin{figure}
\begin{minipage}[c]{0.98\textwidth}
\small{(a)}\includegraphics[width=7.5cm,height=5.cm]{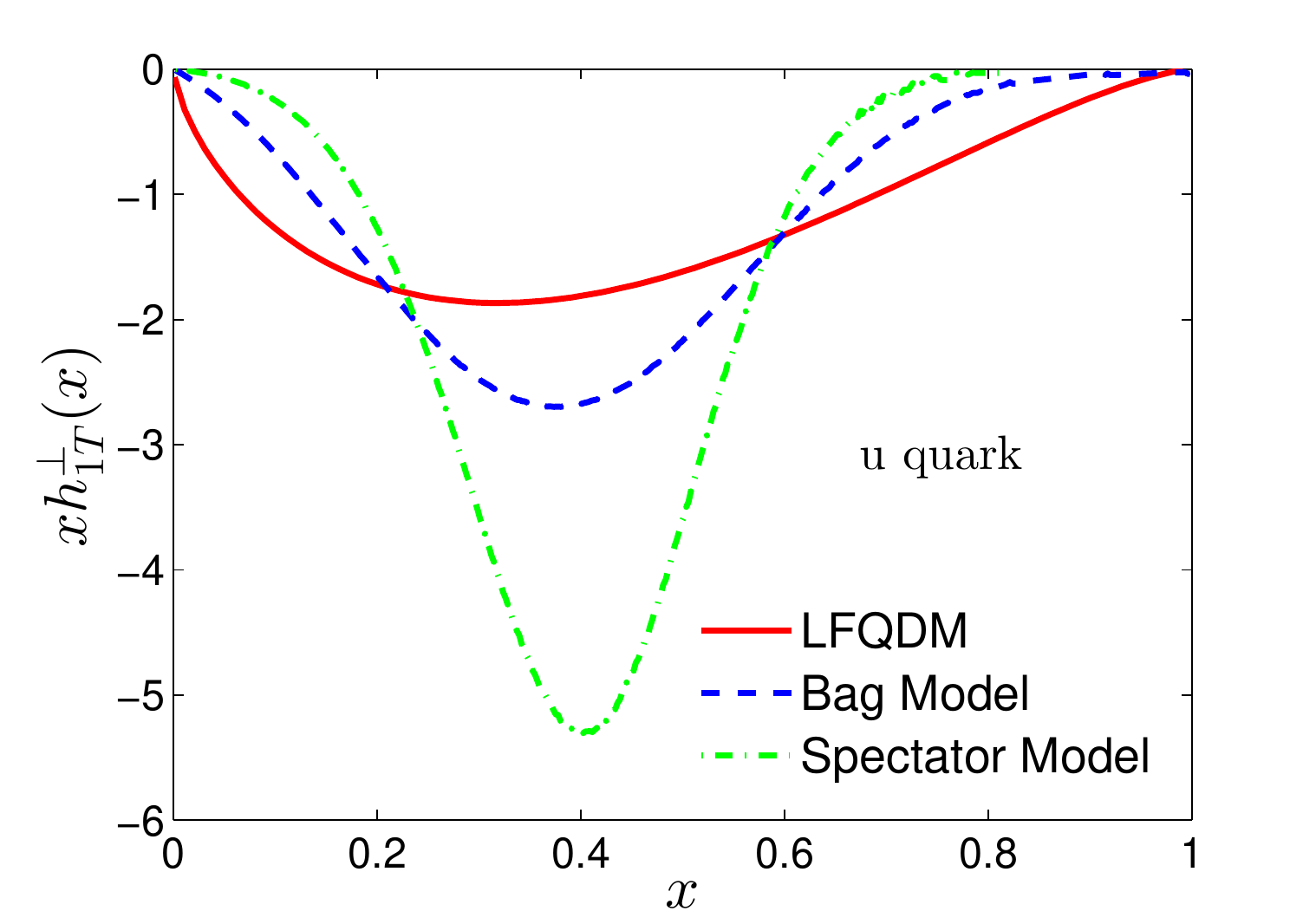}
\small{(b)}\includegraphics[width=7.5cm,height=5.cm]{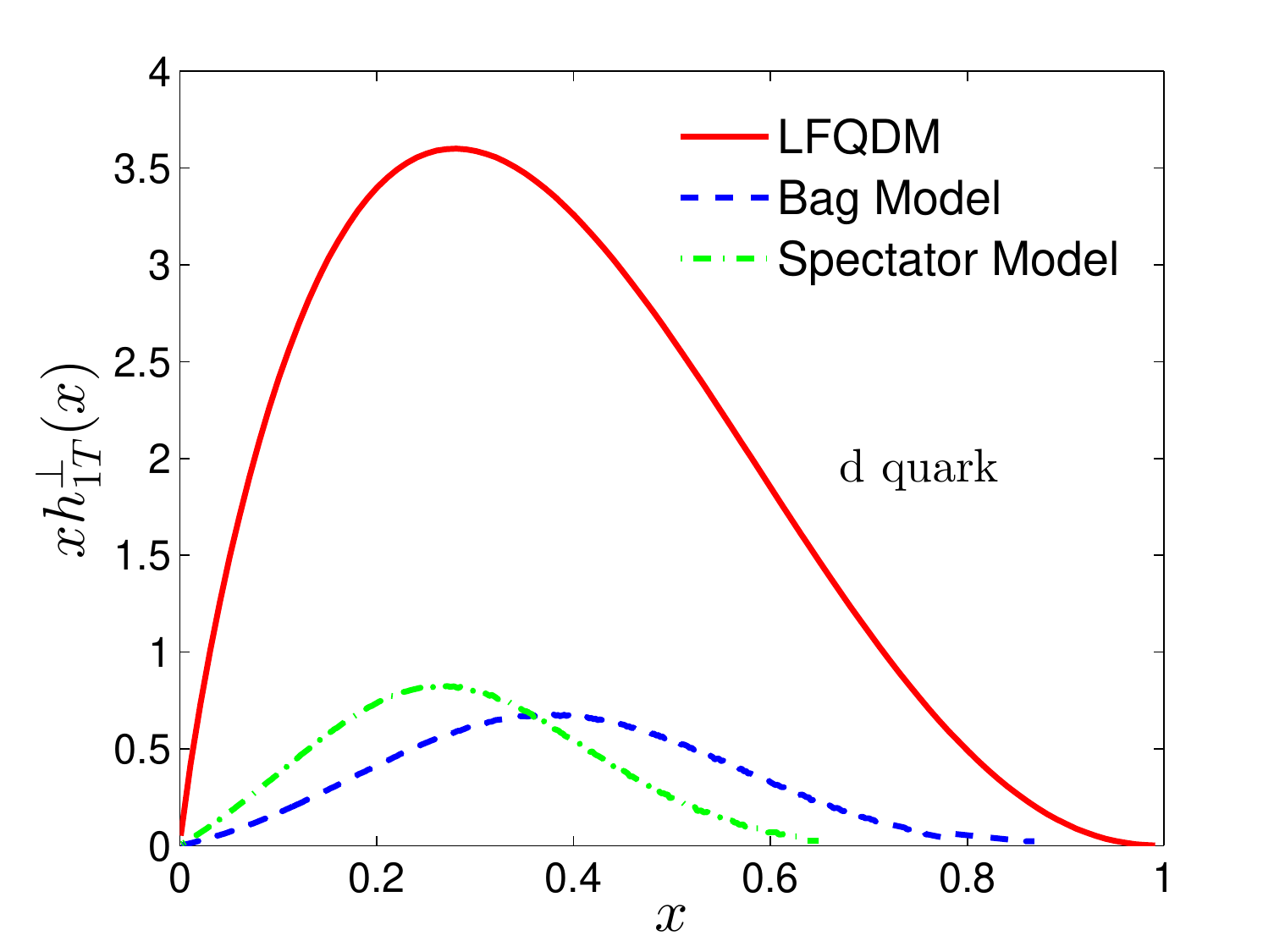}
\end{minipage}
\caption{\label{fig_PDFh1Tp} The pretzelosity distribution $h^{\perp\nu}_{1T}(x)$ multiplied by $x$ are shown for u and d quarks. The result in this model(LFQDM) is denoted by red continuous line, the dashed blue line represents Bag model\cite{bag} prediction and dot dashed green line is for spectator model\cite{Jakob97}.}
\end{figure}

\begin{figure}
\begin{minipage}[c]{0.98\textwidth}
\small{(a)}\includegraphics[width=7.5cm,height=5.cm]{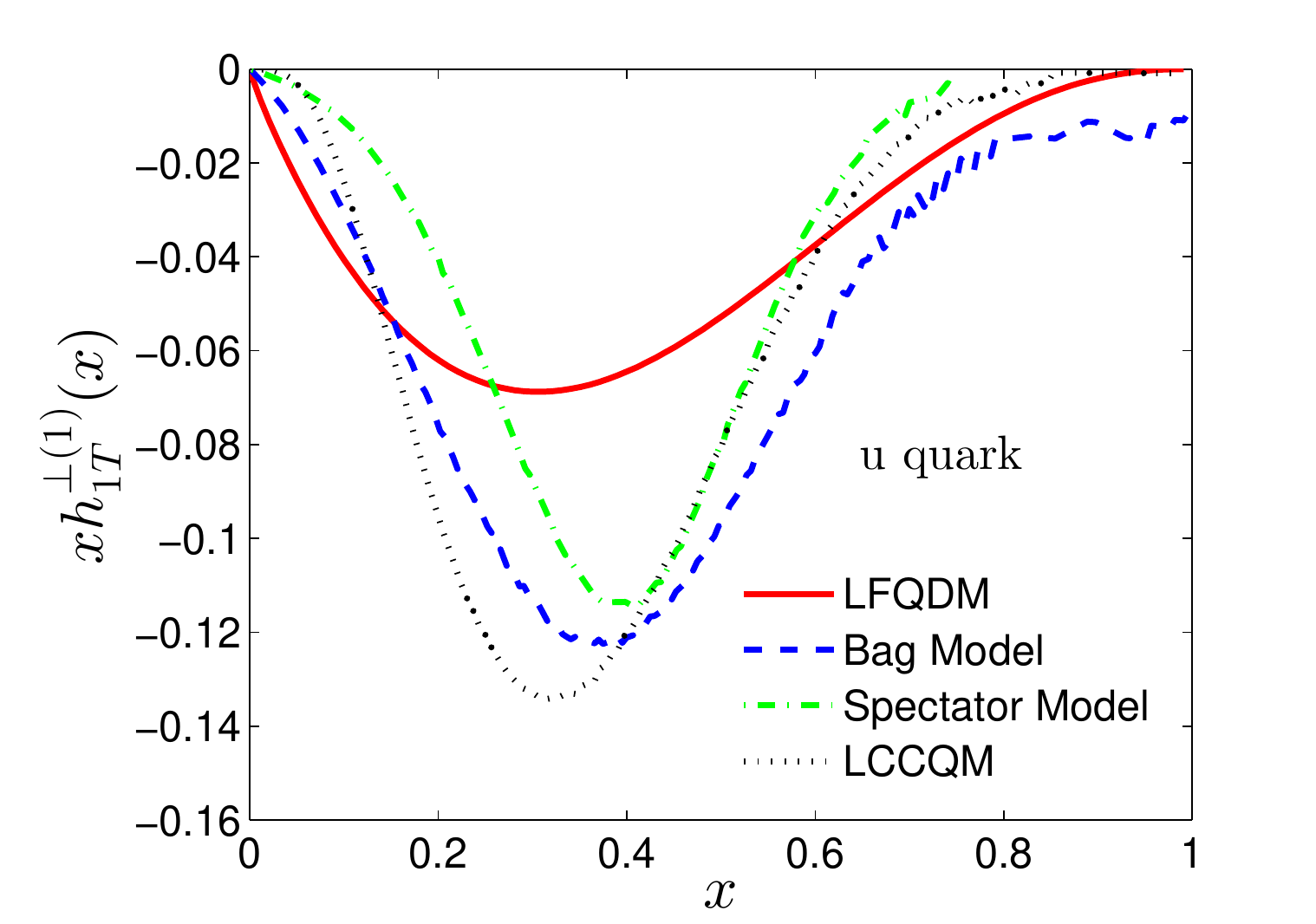}
\small{(b)}\includegraphics[width=7.5cm,height=5.cm]{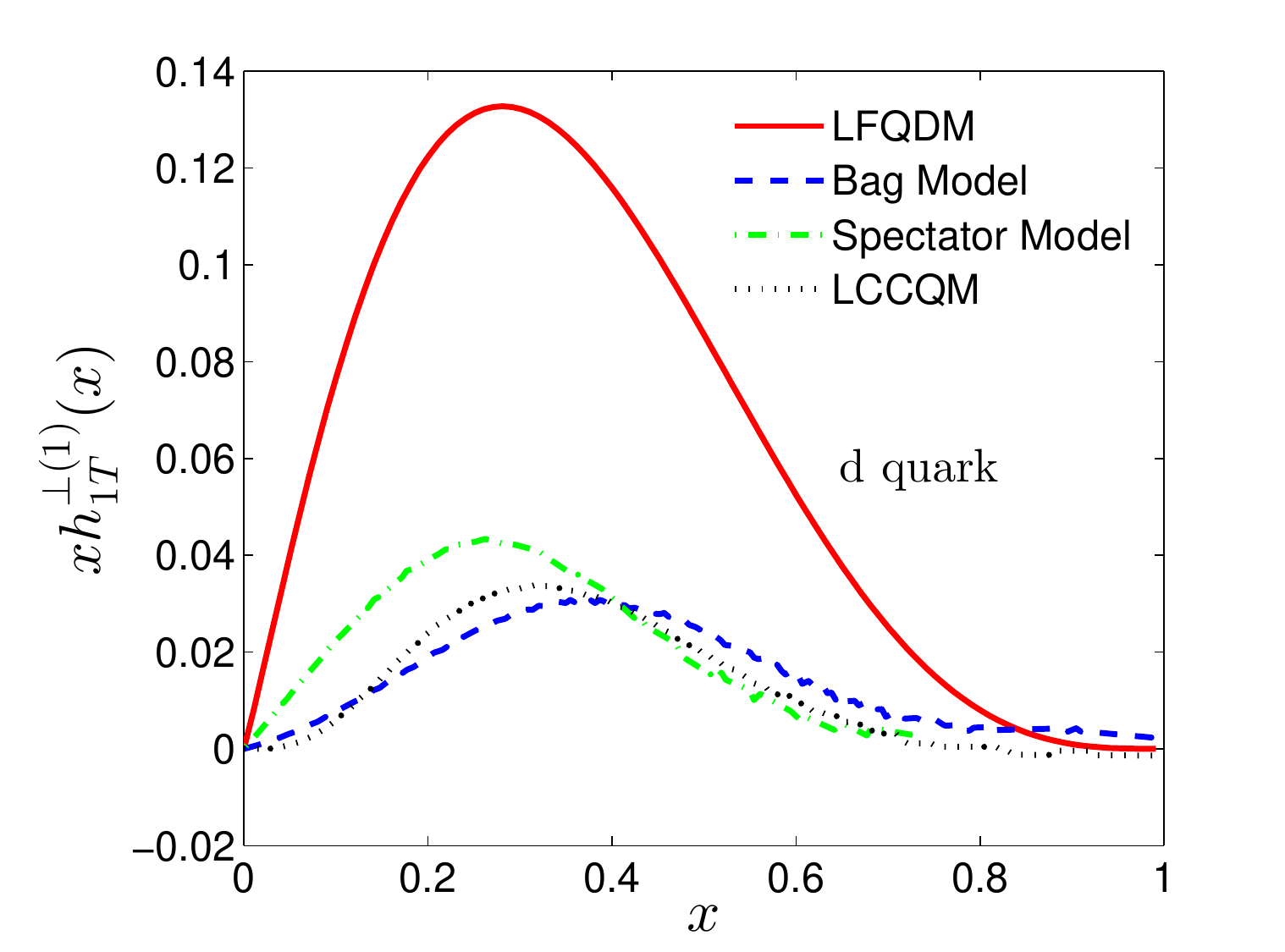}
\end{minipage}
\caption{\label{fig_PDFmh1Tp} The transverse moment of pretzelosity distribution, $h^{\perp(1)\nu}_{1T}(x)$ multiplied by $x$ are shown for u and d quarks. The result in this model is denoted by red continuous line. The dashed blue line, dot dashed green line and doted black line represent predictions of Bag model, spectator model and LCCQM\cite{Pasquini09} respectively.}. 
\end{figure}

Pretzelosity distribution $h^{\perp\nu}_{1T}(x)$ is shown in Fig.\ref{fig_PDFh1Tp} for $u$ and $d$ quarks. We compare our result with other models e.g, Bag model\cite{bag}, Spectator model\cite{Jakob97} for both the quarks. We observe a negative distribution for u quarks and a positive distribution for d quarks as found in most of the models. Whereas a opposite distribution is predicted by\cite{Proku15} with  big error corridor. In our model, the contribution coming from $d$ quark is much higher than that in other models. The difference in magnitudes in different model predictions  may be  due to  different energy scales used different models. The transverse moment of pretzelosity is shown in Fig.\ref{fig_PDFmh1Tp} and compared with other models.


\section{Transverse shape of Proton}\label{proton_shape}
It is interesting to study the contributions of transverse distributions e.g, $h_1$ and $ h^\perp_{1T}$ to the transverse shape of proton. Presence of  nonzero transversity and pretzelosity distribution causes a non-spherical shape of the proton. The transverse shape of proton\cite{Miller07} is defined as
\be 
\frac{\hat{\rho}_{\textrm{REL} T}(\bfp,\textbf{n})/M}{\tilde{f}_1(\bfp^2)}=1+\frac{\tilde{h}_1(\bfp^2)}{\tilde{f}_1(\bfp^2)}\cos\phi_n + \frac{\bfp^2}{2M^2}\cos(2\phi-\phi_n)\frac{\tilde{h}^\perp_{1T}(\bfp^2)}{\tilde{f}_1(\bfp^2)},\label{shape}
\ee
where the struck quark has a spin in an arbitrary fixed direction specified by $\textbf{n}$, the proton spin is denoted by $\textbf{S}_\perp$ and $\phi_n$ is the angle between $\textbf{n}$ and $\textbf{S}_\perp$. In the above equation $\phi$ is the angle between $\bfp$ and $\textbf{S}_\perp$. The tilde over a function is define as $\tilde{f}(\bfp^2)=\int dx f(x,\bfp^2)$. 
\begin{figure}[htbp]
\begin{minipage}[c]{0.98\textwidth}
\small{(a)}\includegraphics[width=7.5cm,clip]{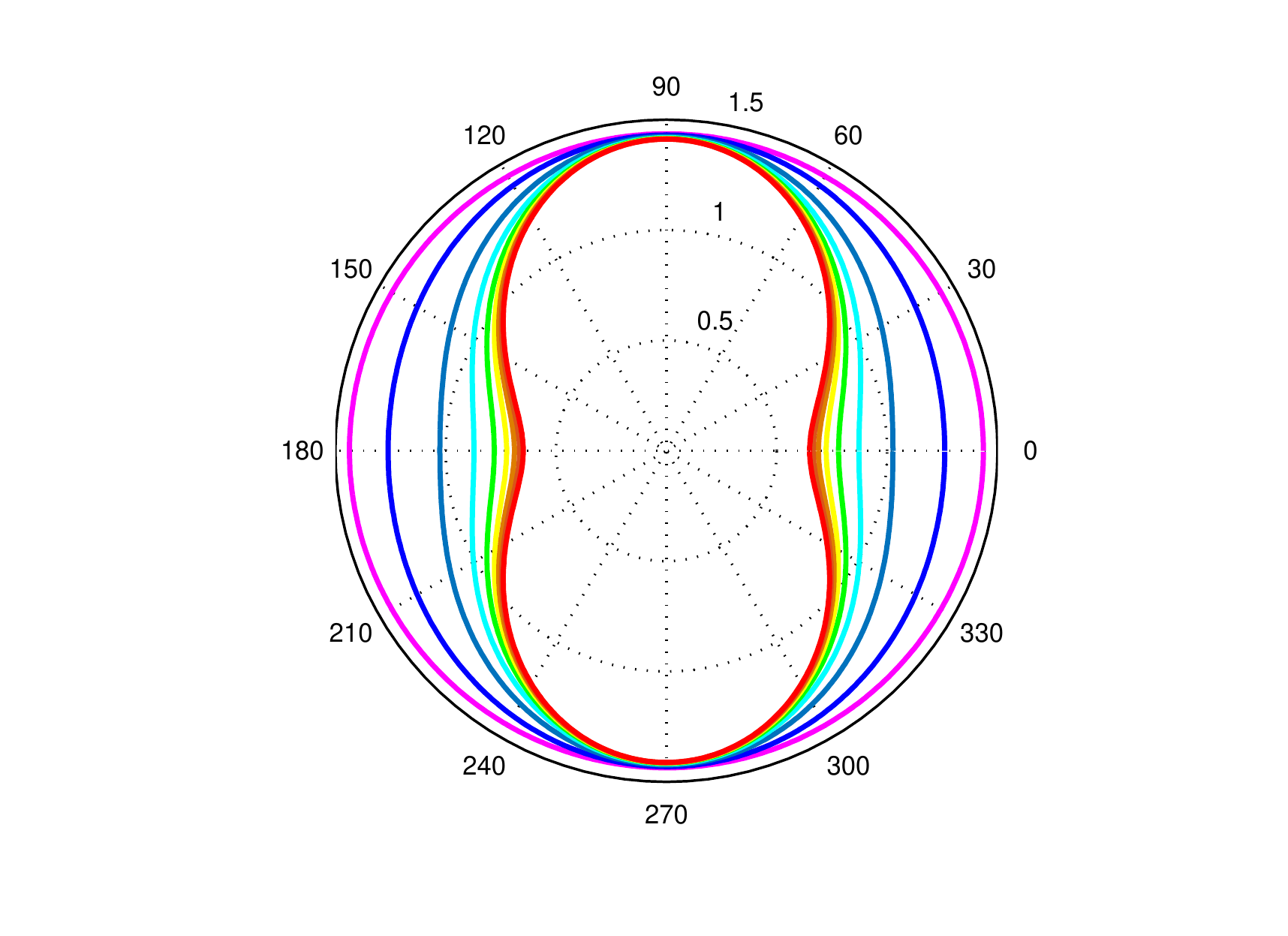}
\small{(b)}\includegraphics[width=7.5cm,clip]{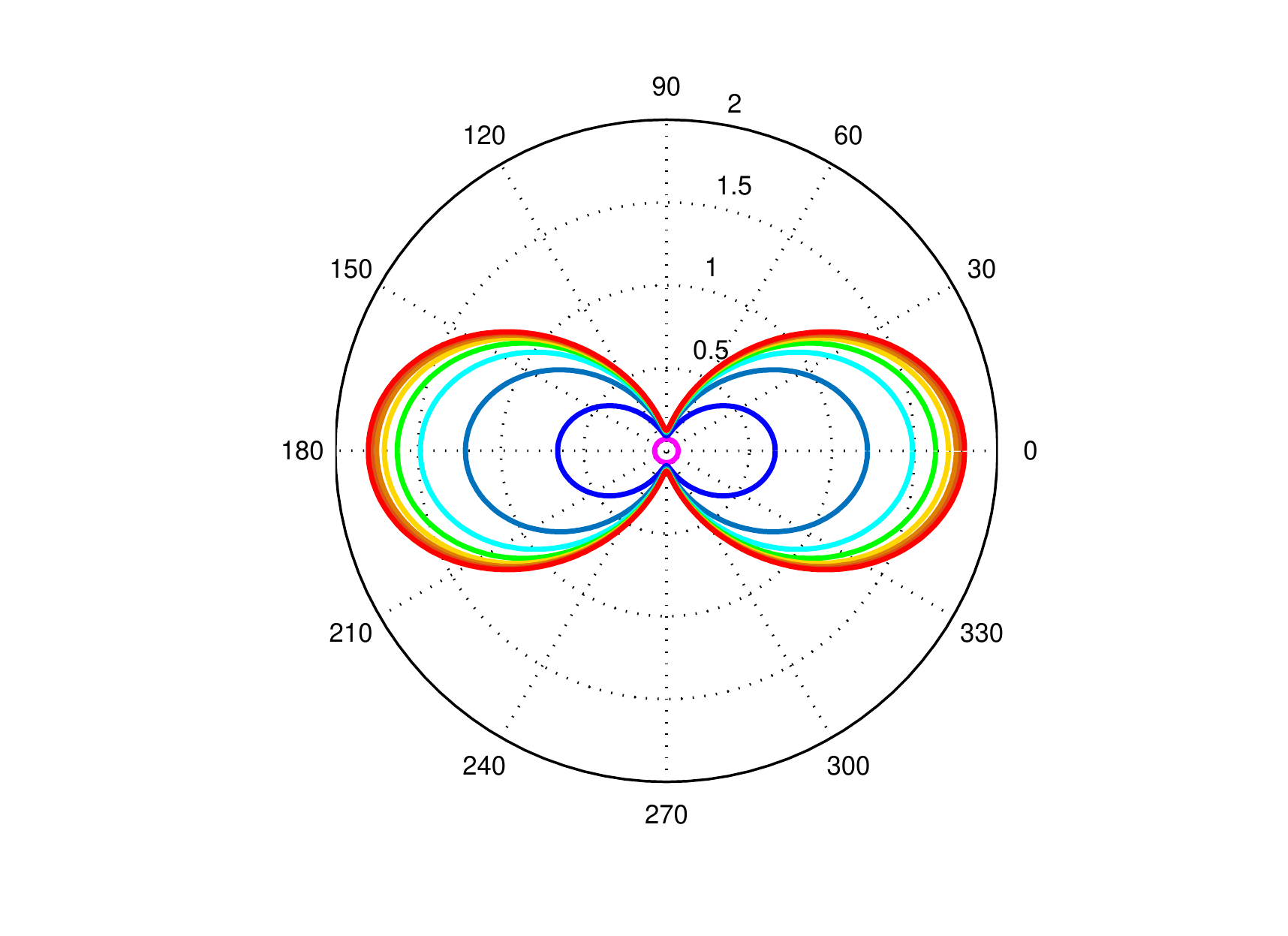}
\end{minipage}
\caption{\label{fig_phi0}  The transverse shape of proton, from Eq.(\ref{shape}), for u and d struck quarks shown in (a) and (b) respectively. The $\textbf{n}$ is parallel to $\bfS$ i.e, $\phi_n=0$. The shapes denoted by the different colors red$\to$ blue are corresponding to the values of $\bfp=0 \to 2$ GeV in steps of 0.25 GeV.} 
\end{figure}

\begin{figure}[htbp]
\begin{minipage}[c]{0.98\textwidth}
\small{(a)}\includegraphics[width=7.5cm,clip]{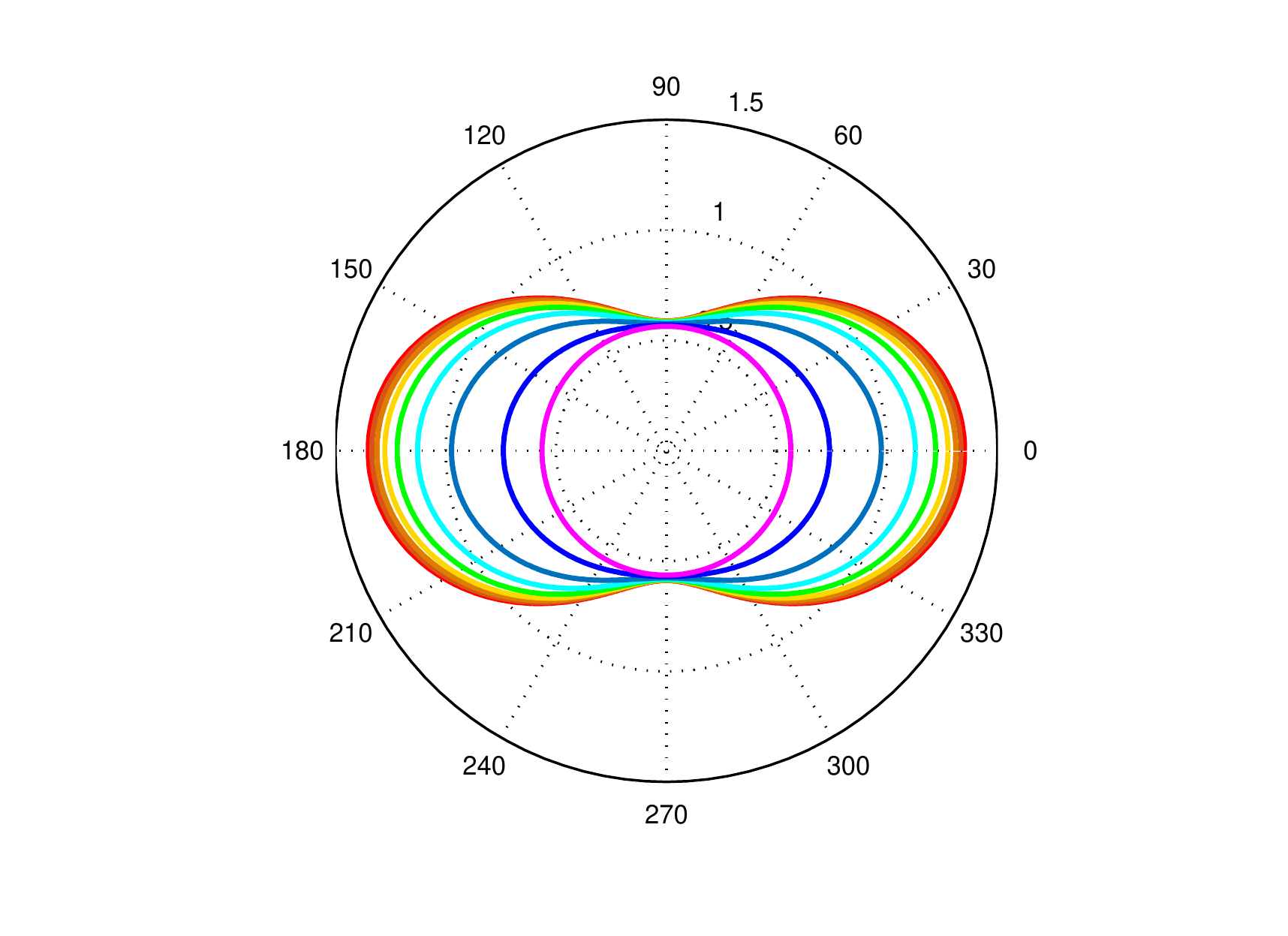}
\small{(b)}\includegraphics[width=7.5cm,clip]{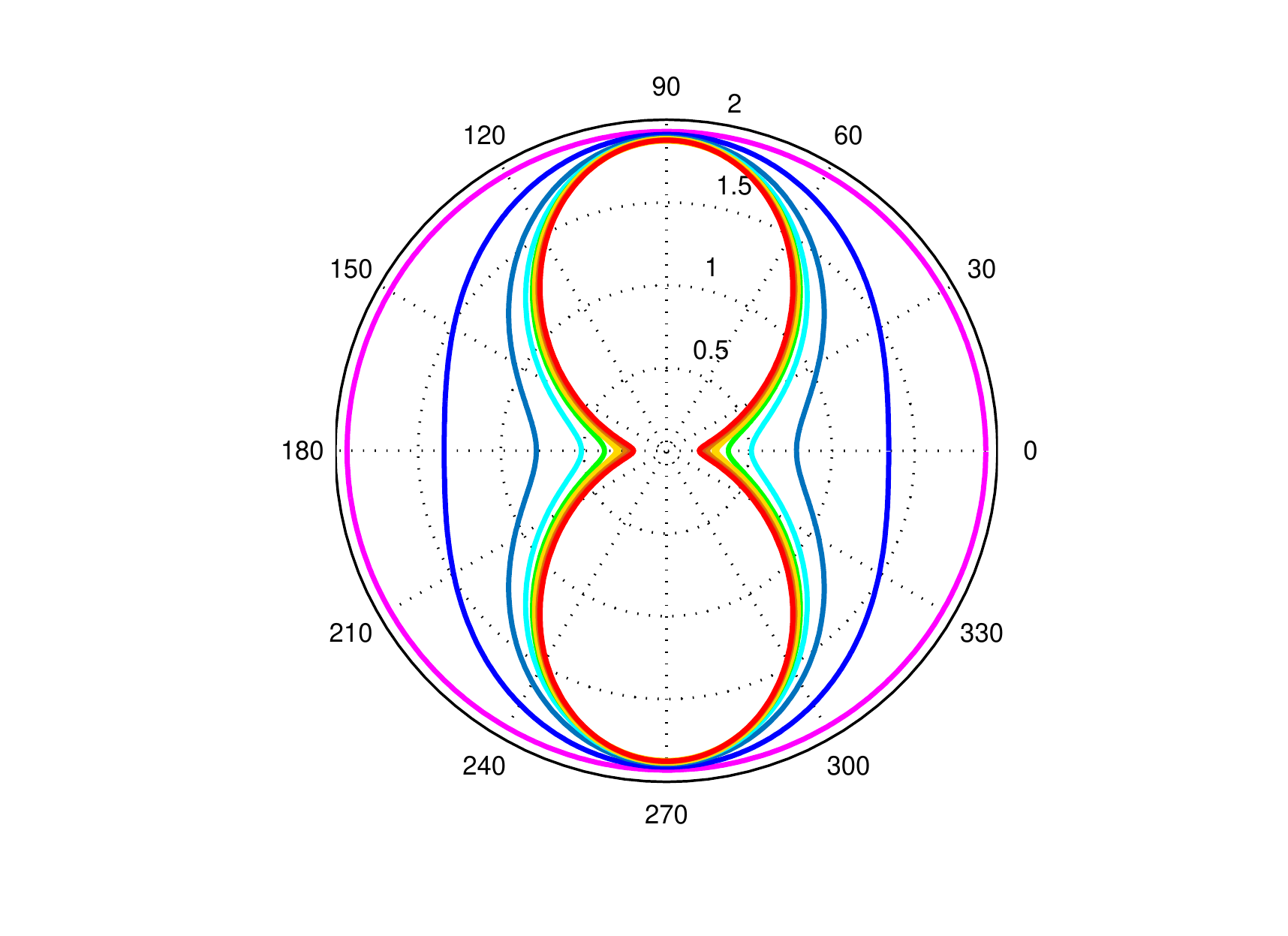}
\end{minipage}
\caption{\label{fig_phiPi} When the $\textbf{n}$ is anti-parallel to $\bfS$ i.e, $\phi_n=\pi$ the transverse shape of proton, from Eq.(\ref{shape}), for u and d struck quarks shown in (a) and (b) respectively. The shapes denoted by the different colors red$\to$ blue are corresponding to the values of $\bfp=0 \to 2$ GeV in steps of 0.25 GeV.} 
\end{figure}

The transverse shape of proton is shown in Fig.(\ref{fig_phi0}) for  $\textbf{n}$ lies parallel to $\bfS$, i.e, $\phi_n=0$.  Fig.\ref{fig_phi0}(a) represents the shapes for u struck quark and Fig.\ref{fig_phi0}(b) is for d struck quark. The shapes denoted by the different colors (pink$\to$blue$\to$red)  correspond to the values of $\bfp=0 \to 2$ GeV in steps of 0.25 GeV. The distributions with vanishing transverse moment i.e, $\bfp=0~$GeV do not contribute to the transverse shape and  spherical shapes are observed(denoted by pink colors) for both the quarks.
Similarly, for higher values of $\bfp$, the contribution from $h^\perp_{1T}$ becomes more significant and causes highly non-spherical transverse shape of proton. Similar deformations are found in other models e,g. CQM\cite{Pasquini08}, spectator model\cite{Miller07}.
 Fig.\ref{fig_phiPi} shows  the transverse shapes of proton, when  $\textbf{n}$ is anti-parallel to $\bfS$, i.e, $\phi_n=\pi$, for  up(a) and down(b) struck quarks respectively. Again the spherical shapes(denoted by pink color) are observed at $\bfp=0$ GeV for both the struck quarks and it gets distorted because of the significant values of $h^\perp_{1T}$ at $\bfp\neq 0$.

 In this model, $f_1(x,\bfp^2)$ and $h_1(x,\bfp^2)$ are of the same sign for $u$ struck quark, so the second term in Eq.(\ref{shape}) is negative(positive) for $\phi_n=\pi(0)$. Whereas,  because of the opposite sign between $f_1(x,\bfp^2)$ and $h^\perp_{1T}(x,\bfp^2)$ the third term is always  negative for both $\phi_n=0$ and $\pi$. Thus, the  sum of the three terms on the right side of Eq.(\ref{shape}) has a dominant contribution from $h^\perp_{1T}(x,\bfp^2)$ for $\phi_n=\pi$ and shows a larger distortion than for $\phi_n=0$.
 For $d$ quark, $f_1(x,\bfp^2)$ and $h_1(x,\bfp^2)$ are of the opposite sign, so the second term is negative(positive) for $\phi_n=0(\pi)$. Therefore, the  sum of the first two terms on the right side of Eq.(\ref{shape}) becomes small and effectively the dominating contribution comes from the third term which causes the distortion in the shape of proton. Since the the pretzelosity distribution $h^\perp_{1T}(x,\bfp^2)$, for d quarks, has same sign with $f_1(x,\bfp^2)$ the distortion is large for both $\phi_n=0,\pi$ cases.
 
\section{Summary and Conclusion}\label{concl}
The T-even TMDs are discussed in a light-front quark-diquark model of the proton. The model includes both scalar and vector diquarks where the light front wave functions are constructed from soft-wall AdS/QCD predictions.
The TMDs are found to satisfy different inequalities. Similar inequalities are also found in other models   and are generic to diquark models. The transversity TMD is found to  satisfy the Soffer bound. In  phenomenological models,  the TMD $f_1^\nu(x,p_\perp)$ is assumed to factorize in $x$ and $p_\perp$   where the $x$-dependence comes through the PDF $f_1^\nu(x)$ and a Gaussian ansatz is adopted for $p_\perp$ dependence. In our model, the TMDs $x-p_\perp^2$ factorization is not apparent. But,  interestingly, our numerical analysis indicates that TMDs in our model actually agree with the phenomenological ansatz.
 
 First moments in $x$  of TMDs $f_1^\nu(x,p_\perp^2)$ and $g_{1T}^\nu(x,p_\perp^2)$ are related with the quark densities for different polarization of the proton and quarks inside it. The quark densities for unpolarized and transversely polarized proton are presented in this paper. For transversely polarized proton the quark densities are found to be  non-spherical. On $p_\perp$-integration, $f_1^\nu(x,p_\perp^2),~ h_1^\nu(x,p_\perp^2)$ and $g_{1L}^\nu(x,p_\perp)$
give the PDFs $f_1^\nu(x),~h_1^\nu(x)$ and $g_1^\nu(x)$ while the other TMDs don't have such collinear interpretations. DGLAP evolution of the integrated TMDs in our model show that certain ratios like $g_{1T}^\nu(x)/h_{1L}^\nu(x)$ and $h_{1T}^\nu(x)/f_1^\nu(x)$ are independent of the evolution scale $\mu$ at high scales. The ratios are found to be positive for $u$ quark and negative for $d$ quark.
We have also presented the transverse shape of the  proton. For transversely polarized proton, the pretzelosity distribution causes a distortion in the spherical shape for nonzero transverse momentum. 

\textbf{Acknowledgment:} We thank Oleg Teryaev for many useful discussions.
\appendix

\section{parameters in the quark-diquark model} \label{AppA}
In \cite{TM_VD}, the initial scale was set to $\mu_0=0.313 $ GeV which is quite low for perturbative DGLAP evolution. So, here we set the initial scale to $\mu_0=0.8$ GeV  and reevaluate the parameters. Following the same strategy\cite{TM_VD} the new parameters are fitted to  reproduce  the DGLAP evolution of the unpolarized pdfs.  The parameters are listed in the Table.\ref{tab_evopar} and \ref{tab_DL}. 
\begin{table}[h]
\centering 
\begin{tabular}{|c c c c c|}
 \hline
 $P_i^\nu(\mu)$~~&~~$\alpha_i^\nu$~~&~~$\beta_i^\nu$~~ & ~~$\gamma_i^\nu$~~ & ~~$\chi^2/d.o.f$~~  \\ \hline
$A_1^u$ &~~ $-0.3143\pm0.0187$ ~~&~~ $-0.0113\pm0.0209$ ~~&~~ $0.1673\pm0.0937$ ~~&~~ 0.2\\
$B_1^u$ & $4.961^{+0.034}_{-0.035}$ & $0.0841\pm0.0057$ & $-0.7993\pm0.0242$ &0.02\\ 
$A_2^u$ & $-0.2413\pm0.0179$ & $0.0116\pm0.0245$ & $0.1606\pm0.1111$ & 0.16\\ 
$B_2^u$ & $3.255\pm0.320$ & $0.087\pm0.0605$ & $-0.9502^{+0.0241}_{-0.0251}$ &0.1\\ 
$A_1^d$ &~~ $0.0213\pm0.0098$ ~~&~~ $-0.1085\pm0.0257$ ~~&~~ $0.7663^{+0.246}_{-0.245}$ ~~&~~ 0.12\\
$B_1^d$ & $10.92\pm0.0193$ & $0.0306^{+0.021}_{-0.022}$ & $-0.4278\pm0.0972$ &0.10\\
$A_2^d$ & $-0.29\pm0.0456$ & $-0.0036\pm0.0408$ & $0.0489\pm0.1783$ & 0.23\\ 
$B_2^d$ & $0.9733\pm0.0737$ & $0.0661\pm0.1086$ & $-0.1307^{+0.0517}_{-0.0518}$ &0.17\\
 \hline
 \end{tabular} 
\caption{PDF evolution parameters with 95\% confidence bounds.} 
\label{tab_evopar} 
\end{table}
\begin{table}[h]
\centering  
\begin{tabular}{|c c c c |}
 \hline
 $\delta^\nu(\mu)$~~&~~$\delta_1^\nu$~~&~~$\delta_2^\nu$~~ & $\chi^2/d.o.f$~~  \\ \hline
$\delta^u$ ~~&~~ $0.0474\pm0.008$ ~~&~~ $1.252\pm0.032$ ~~&~~ 1.3\\
$\delta^d$ & $0.3271\pm0.0566$ & $0.3888\pm0.1504$ & 0.9 \\ \hline
\end{tabular} 
\caption{PDF evolution parameter $\delta^\nu_1$ and $\delta^\nu_2$ for $\nu=u,d$. }  
\label{tab_DL}  
\end{table}

The parameters vary with the scale as\cite{TM_VD}
\be 
a_i^\nu(\mu)&=&a_i^\nu(\mu_0) + A^\nu_{i}(\mu), \label{a_im}\\
b_i^\nu(\mu)&=&b_i^\nu(\mu_0) - B^\nu_{i}(\mu)\frac{4C_F}{\beta_0}\ln\bigg(\frac{\alpha_s(\mu^2)}{\alpha_s(\mu_0^2)}\bigg),\label{b_im}\\
\delta^\nu(\mu)&=& \exp\bigg[\delta^\nu_1\bigg(\ln(\mu^2/\mu_0^2)\bigg)^{\delta^\nu_2}\bigg],\label{DL}
\ee
Where, the scale dependent parts $A^\nu_{i}(\mu)$ and $B^\nu_{i}(\mu)$  evolve as 
\be 
P^\nu_{i}(\mu)&=&\alpha^\nu_{P,i} ~\mu^{2\beta^\nu_{P,i}}\bigg[\ln\bigg(\frac{\mu^2}{\mu_0^2}\bigg)\bigg]^{\gamma^\nu_{P,i}}\bigg|_{i=1,2} ,\label{Pi_evolu}
\ee

\begin{figure}[htbp]
\begin{minipage}[c]{0.98\textwidth}
\small{(a)}\includegraphics[width=7.5cm,clip]{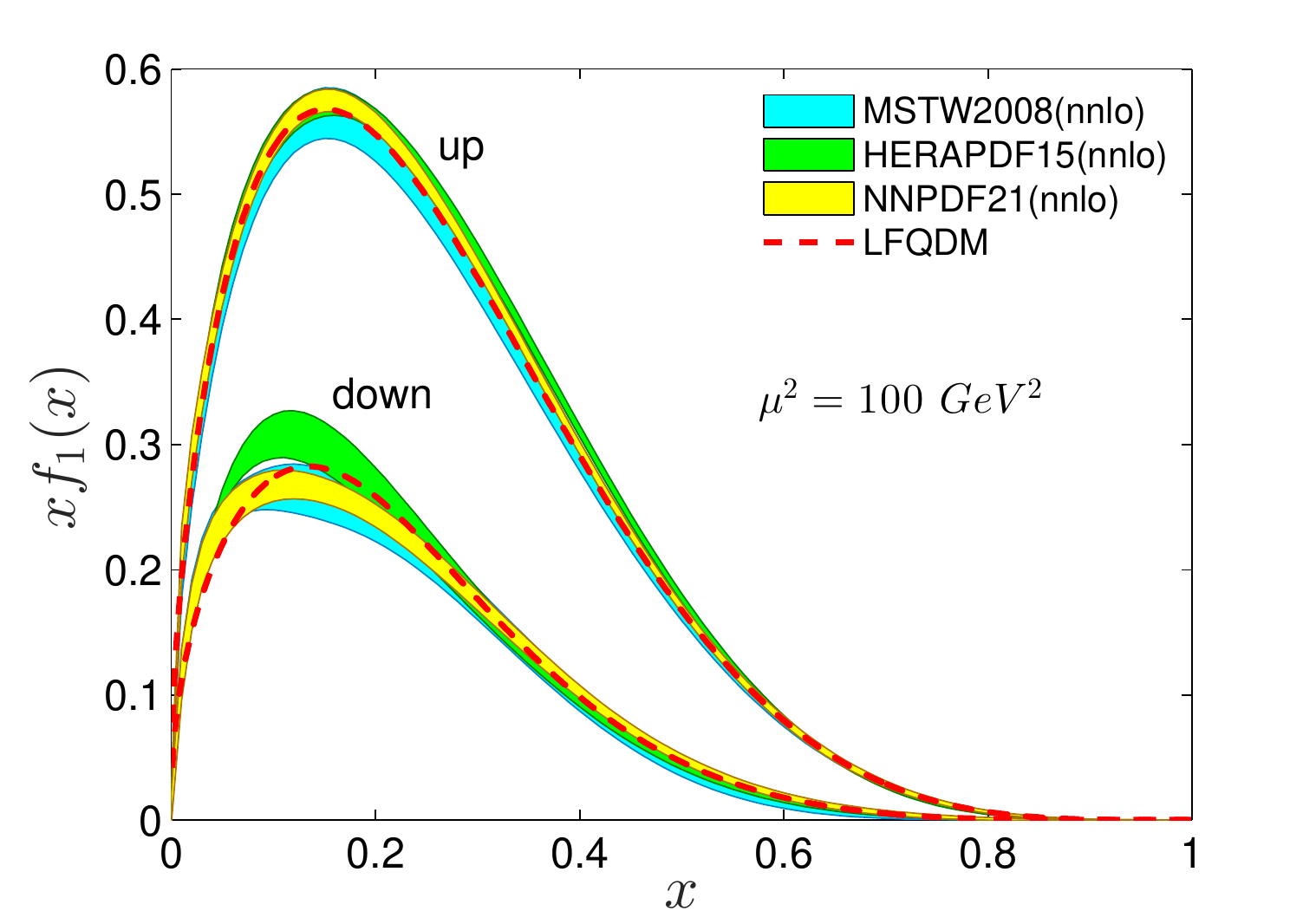}
\small{(b)}\includegraphics[width=7.5cm,clip]{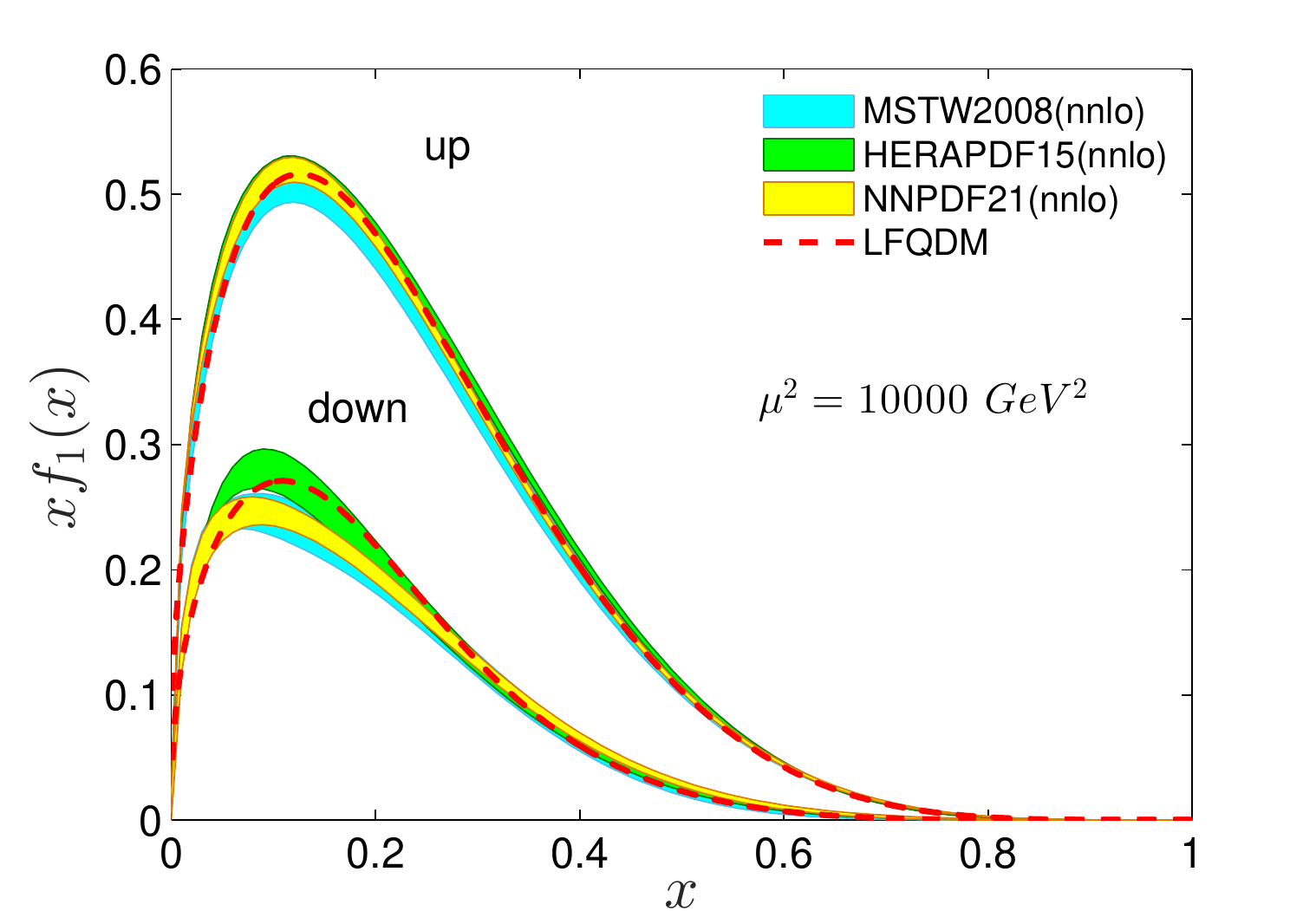}
\end{minipage}
\caption{\label{fig_unPDF}Evolution of unpolarized PDF in this model at  $\mu^2=100$ and $10000$ GeV$^2$ for both $u$ and $d$ quarks. Our model predictions are  compared with NNPDF21(NNLO)\cite{NNPDF}, HERAPDF15(NNLO)\cite{HERAPDF} and MSTW2008(NNLO)\cite{MSTW} results.} 
\end{figure}

For completeness, using the parameters of Table.\ref{tab_evopar} and \ref{tab_DL} in Eqs.(\ref{a_im},\ref{b_im},\ref{Pi_evolu}), we plot the unpolarized PDFs at two different scales $\mu^2=10^2,10^4~GeV^2$(shown in Fig.\ref{fig_unPDF}).

\section{quark-quark correlators} \label{AppB}
In this Appendix we list few steps to calculate the quark-quark correlators in this model for both the scalar diquark and the vector diquark.
\be 
\mathrm{Vector ~current:} &J^{\gamma^+}&=\bar{\psi}(0)\gamma^+\psi(z) 
= 2 \eta^\dagger(0)\eta(z)\\
\mathrm{Axial~ vector~ current:} 
&J^{\gamma^+\gamma^5}&=\bar{\psi}(0)\gamma^+\gamma^5\psi(z) 
= 2 \eta^\dagger(0)\sigma^3\eta(z) \\
\mathrm{Tensor~ current:}& J^{i\sigma^{j+}\gamma^5}&=\bar{\psi}(0)i\sigma^{j+}\gamma^5\psi(z) 
= 2 \eta^\dagger(0)(-i)\hat{\sigma}^j\sigma^3\eta(z) 
\ee
where the dynamical light-front component of the fermion field
\be
\psi^+=\frac{1}{2}\gamma^0\gamma^+\psi=\begin{pmatrix} \eta \\ 0\end{pmatrix}.
\ee
The two component light-front quark field $\eta(z)$ is given as
\be 
\eta(z)=\sum_\lambda \chi_\lambda \int\frac{dk^+d^2\bfk}{2(2\pi)^3\sqrt{k^+}}\big[b(k,\lambda)e^{-ik.z}+d^\dagger(k,\lambda)e^{ik.z}\big].
\ee
where $\chi_\lambda$ are the two component spinors. There is no antiquark in this model.

The two-particle basis states are normalized as
\be 
\langle \lambda',\Lambda';x'P^+,\bfq'|\lambda,\Lambda;xP^+,\bfq\rangle=\prod_{i=1}16\pi^3 q^+_i \delta(q^{\prime+}_i - q^+_i)\delta^{(2)}(\bf{q}'_{\perp i} - \bf{q}_{\perp i})\delta_{\lambda'\lambda}\delta_{\nu'\nu}\label{CR}
\ee
The transverse polarization states with momentum $P$  are defined as
\be
|P;\uparrow \downarrow \rangle=\frac{1}{\sqrt{2}}\bigg(|P;+ \rangle\pm|P;- \rangle\bigg)
\ee
Where the $|P;\pm \rangle$ represent the longitudinally polarized state along $+ve$ and $-ve$  axis receptively. 

The explicit expression for of the TMD correlator(Eq.(\ref{TMD_cor})) for vector current $J^{\gamma^+}$ in a longitudinally polarized proton reads\\
(i) for scalar diquark:
\be
&&\Phi^{\nu [\gamma^+]}_S(x,\bfp;+)\nonumber\\
&&=\frac{1}{2}\int \frac{dz^- d^2z_T}{2(2\pi)^3} e^{ip.z} \langle u ~S|\overline{\psi}^\nu (0)\gamma^+ \psi^\nu (z) |u~S\rangle\nonumber\\
\nonumber\\
&& = \frac{1}{2}\sum_{\lambda'\lambda} \chi^\dagger_{\lambda'}\chi_\lambda \int\frac{dk^{'+} d^2\bfk'}{2(2\pi)^3\sqrt{k^{'+}}}\int\frac{dk^+ d^2\bfk}{2(2\pi)^3\sqrt{k^+}} 2\delta(p^+-k^+)\delta^(2)(\bfp-\bfk)\nonumber\\
&&\times \int\frac{dx' d^2\bfq'}{2(2\pi)^3\sqrt{x'(1-x')}} \int\frac{dx d^2\bfq}{2(2\pi)^3\sqrt{x(1-x)}}\bigg[\bigg(\psi^{+\dagger\nu}_+(x',\bfq')\langle\frac{1}{2},0;x'P^+,\bfq'|\nonumber\\
&&+\psi^{+\dagger\nu}_-(x',\bfq')\langle-\frac{1}{2},0;x'P^+,\bfq'|\bigg)b^\dagger(k',\lambda')b(k,\lambda)\nonumber\\
&&\times\bigg(\psi^{+\nu}_+(x,\bfq)|\frac{1}{2},0;xP^+,\bfq\rangle+\psi^{+\nu}_-(x,\bfq)|-\frac{1}{2},0;xP^+,\bfq\rangle\bigg)\bigg]\nonumber\\
\nonumber\\
&&= \int\frac{dk^{'+} d^2\bfk'}{2(2\pi)^3\sqrt{k^{'+}}}\int\frac{dk^+ d^2\bfk}{2(2\pi)^3\sqrt{k^+}} \delta(p^+-k^+)\delta^(2)(\bfp-\bfk)\nonumber\\
&&\times \int\frac{dx' d^2\bfq'}{2(2\pi)^3\sqrt{x'(1-x')}} \int\frac{dx d^2\bfq}{2(2\pi)^3\sqrt{x(1-x)}}\nonumber\\
&&\times\bigg[\bigg(\langle\frac{1}{2},0;x'P^+,\bfq'|\psi^{+\dagger\nu}_+(x',\bfq') b^\dagger(k',1/2)b(k,1/2)\psi^{+\nu}_+(x,\bfq)|\frac{1}{2},0;xP^+,\bfq\rangle\bigg)\nonumber\\
&&+\bigg(\langle-\frac{1}{2},0;x'P^+,\bfq'|\psi^{+\dagger\nu}_-(x',\bfq') b^\dagger(k',-1/2)b(k,-1/2)\psi^{+\nu}_-(x,\bfq)|-\frac{1}{2},0;xP^+,\bfq\rangle\bigg)\bigg]\nonumber\\
\nonumber\\
&&=\frac{1}{16\pi^3}\bigg[|\psi^{+\nu}_+(x,\bfq)|^2+|\psi^{+\nu}_-(x,\bfq)|^2\bigg]\label{App_SD}
\ee

(ii) for vector diquark:
\be
&&\Phi^{\nu [\gamma^+]}_V(x,\bfp;+)=\frac{1}{2}\int \frac{dz^- d^2z_T}{2(2\pi)^3} e^{ip.z} \langle u ~A|\overline{\psi}^\nu (0)\gamma^+ \psi^\nu (z) |u~A\rangle\nonumber\\ 
&&= \frac{1}{2}\sum_{\lambda'\lambda} \chi^\dagger_{\lambda'}\chi_\lambda \int\frac{dk^{'+} d^2\bfk'}{2(2\pi)^3\sqrt{k^{'+}}}\int\frac{dk^+ d^2\bfk}{2(2\pi)^3\sqrt{k^+}} 2\delta(p^+-k^+)\delta^(2)(\bfp-\bfk)\nonumber\\
&&\times \int\frac{dx' d^2\bfq'}{2(2\pi)^3\sqrt{x'(1-x')}} \int\frac{dx d^2\bfq}{2(2\pi)^3\sqrt{x(1-x)}}\bigg[\bigg(\psi^{+\dagger\nu}_{++}(x',\bfq')\langle\frac{1}{2},+1;x'P^+,\bfq'|\nonumber\\
&&+\psi^{+\dagger\nu}_{-+}(x',\bfq')\langle-\frac{1}{2},+1;x'P^+,\bfq'|+\psi^{+\dagger\nu}_{+0}(x',\bfq')\langle\frac{1}{2},0;x'P^+,\bfq'|\nonumber\\
&&+\psi^{+\dagger\nu}_{-0}(x',\bfq')\langle-\frac{1}{2},0;x'P^+,\bfq'|+\psi^{+\dagger\nu}_{+-}(x',\bfq')\langle\frac{1}{2},-1;x'P^+,\bfq'|\nonumber\\
&&+\psi^{+\dagger\nu}_{--}(x',\bfq')\langle-\frac{1}{2},-1;x'P^+,\bfq'|\bigg)b^\dagger(k',\lambda')b(k,\lambda)\nonumber\\
&&\times\bigg(\psi^{+\nu}_{++}(x,\bfq)|\frac{1}{2},+1;xP^+,\bfq\rangle+\psi^{+\nu}_{-+}(x,\bfq)|-\frac{1}{2},+1;xP^+,\bfq\rangle \nonumber\\
&&+\psi^{+\nu}_{+0}(x,\bfq)|\frac{1}{2},0;xP^+,\bfq\rangle + \psi^{+\nu}_{-0}(x,\bfq)|-\frac{1}{2},0;xP^+,\bfq\rangle\bigg)\bigg]\nonumber\\
&&+\psi^{+\nu}_{+-}(x,\bfq)|\frac{1}{2},-1;xP^+,\bfq\rangle + \psi^{+\nu}_{--}(x,\bfq)|-\frac{1}{2},-1;xP^+,\bfq\rangle\bigg)\bigg]\nonumber\\
&&= \int\frac{dk^{'+} d^2\bfk'}{2(2\pi)^3\sqrt{k^{'+}}}\int\frac{dk^+ d^2\bfk}{2(2\pi)^3\sqrt{k^+}} \delta(p^+-k^+)\delta^(2)(\bfp-\bfk)\nonumber\\
&&\times \int\frac{dx' d^2\bfq'}{2(2\pi)^3\sqrt{x'(1-x')}} \int\frac{dx d^2\bfq}{2(2\pi)^3\sqrt{x(1-x)}}\nonumber\\
&&\times\bigg[\bigg(\langle\frac{1}{2},+1;x'P^+,\bfq'|\psi^{+\dagger\nu}_{++}(x',\bfq') b^\dagger(k',1/2)b(k,1/2)\psi^{+\nu}_{++}(x,\bfq)|\frac{1}{2},+1;xP^+,\bfq\rangle\bigg)\nonumber\\
&&+\bigg(\langle-\frac{1}{2},+1;x'P^+,\bfq'|\psi^{+\dagger\nu}_{-+}(x',\bfq') b^\dagger(k',-1/2)b(k,-1/2)\psi^{+\nu}_{-+}(x,\bfq)|-\frac{1}{2},+1;xP^+,\bfq\rangle\bigg)\nonumber\\
&&+\bigg(\langle\frac{1}{2},0;x'P^+,\bfq'|\psi^{+\dagger\nu}_{+0}(x',\bfq') b^\dagger(k',1/2)b(k,1/2)\psi^{+\nu}_{+0}(x,\bfq)|\frac{1}{2},0;xP^+,\bfq\rangle\bigg)\nonumber\\
&&+\bigg(\langle-\frac{1}{2},0;x'P^+,\bfq'|\psi^{+\dagger\nu}_{-0}(x',\bfq') b^\dagger(k',-1/2)b(k,-1/2)\psi^{+\nu}_{-0}(x,\bfq)|-\frac{1}{2},0;xP^+,\bfq\rangle\bigg)\nonumber\\
&&+\bigg(\langle\frac{1}{2},-1;x'P^+,\bfq'|\psi^{+\dagger\nu}_{+-}(x',\bfq') b^\dagger(k',1/2)b(k,1/2)\psi^{+\nu}_{+-}(x,\bfq)|\frac{1}{2},-1;xP^+,\bfq\rangle\bigg)\nonumber\\
&&+\bigg(\langle-\frac{1}{2},-1;x'P^+,\bfq'|\psi^{+\dagger\nu}_{--}(x',\bfq') b^\dagger(k',-1/2)b(k,-1/2)\psi^{+\nu}_{--}(x,\bfq)|-\frac{1}{2},-1;xP^+,\bfq\rangle\bigg)\bigg]\nonumber\\
&& =\frac{1}{16\pi^3}\bigg[|\psi^{+\nu}_{++}(x,\bfq)|^2+|\psi^{+\nu}_{-+}(x,\bfq)|^2+|\psi^{+\nu}_{+0}(x,\bfq)|^2\nonumber\\
&& \tab[2cm] +|\psi^{+\nu}_{-0}(x,\bfq)|^2+|\psi^{+\nu}_{+-}(x,\bfq)|^2+|\psi^{+\nu}_{--}(x,\bfq)|^2\bigg] \label{App_VD}
\ee
The last line in Eq.(\ref{App_SD}) and (\ref{App_VD}) is found using the commutation relation of Eq.(\ref{CR}).
Similarly, the TMD correlators can be calculated for axial vector current($\Gamma=\gamma^+\gamma^5$) and tensor current($\Gamma=i\sigma^{j +}\gamma^5$). The final expressions in terms of wave function are shown in Eq.(\ref{TMD_SD})and (\ref{TMD_VD}) for scalar sector and vector sector respectively. 


\end{document}